\documentclass[11pt,a4paper]{article}
\pdfoutput=1
\usepackage{jheppub}

\usepackage{verbatim}
\usepackage{listings}
\usepackage{fancyvrb}
\usepackage{xcolor}
\usepackage{fancyvrb}

\usepackage[T1]{fontenc}
\usepackage{graphicx}
\usepackage{slashed}
\usepackage{amsmath,amsfonts,bbm,latexsym,amssymb}
\usepackage{epsfig}
\usepackage{array}

\usepackage{booktabs}
\usepackage[utf8]{inputenc}
\usepackage{fancyvrb}
\usepackage{subfig}

\usepackage{graphicx}
\usepackage{caption}
\usepackage{feynmp-auto}

\usepackage[flushleft]{threeparttable}
\makeatletter
\def\hlinewd#1{%
	\noalign{\ifnum0=`}\fi\hrule \@height #1 %
	\futurelet\reserved@a\@xhline}
\makeatother
\usepackage{multirow}
\usepackage{multicol}
\usepackage{longtable}

\usepackage[b]{esvect}
 
\usepackage{hyperref}
\hypersetup{colorlinks=false}
\usepackage{array}

\usepackage{titlesec}
\usepackage{dashrule}
\usepackage{scrextend}
\usepackage{ragged2e}
\usepackage{cancel}
\usepackage{float,hypcap}
\usepackage{changepage}
\usepackage{sectsty}
\usepackage{soul}
\usepackage{bbold}
\usepackage{slashed}
\usepackage{tabularx}
\usepackage{blkarray}
\usepackage{latexsym}
\usepackage{braket}
\usepackage{enumerate}
\usepackage{bm}
\usepackage{booktabs,hyperref}
\usepackage{bigints}
\usepackage{amsmath}
\usepackage{amssymb}
\usepackage{empheq}
\usepackage{stackengine}
\stackMath

\usepackage{color}
\definecolor{orange}{rgb}{1,0.7294,0.1843}
\definecolor{uglyblue}{RGB}{95,158,160}
\definecolor{mygray}{RGB}{129,129,129}
\definecolor{mywhite}{RGB}{255,250,240}
\usepackage[most]{tcolorbox}

\allowdisplaybreaks

\def\be{\begin{equation}}
\def\ee{\end{equation}}
\def\ba{\begin{alignedat}}
\def\ea{\end{alignedat}}
\def\bea{\begin{eqnarray}}
\def\eea{\end{eqnarray}}
\newcommand{\bs}{\begin{subequations}}
\newcommand{\es}{\end{subequations}}
\newcommand{\no}{\nonumber \\}
\def\bd{\begin{dmath}} 
\def\ed{\end{dmath}}

\def\vs{\vspace}
\def\hs{\hspace}

\definecolor{mycolor}{RGB}{165,111,29}
\definecolor{mycolor2}{RGB}{128,0,0}

\def\n{\noindent}
\def\fn{\footnote}

\def\t{\text}

\def\t{\texttt}
\def\ts{\textsc}
\def\FM{\textsc{FeynMaster}}
\def\FMS{\textsc{FeynMaster} }
\def\FMT{\textsc{FeynMaster} 2}
\def\FMTS{\textsc{FeynMaster} 2 }
\def\FR{\textsc{FeynRules}}
\def\FC{\textsc{FeynCalc}}

\begin{document}

\begin{flushright}
CFTP/21-006
\end{flushright}

\vspace{-5mm}

\title{\boldmath Renormalization of the C2HDM with FeynMaster 2}

\author[a]{Duarte Fontes,}
\author[a]{Jorge C. Rom\~ao}
\affiliation[a]{Departamento de F\'{\i}sica and CFTP,
Instituto Superior T\'{e}cnico, Universidade de Lisboa, \\
Avenida Rovisco Pais 1, 1049-001 Lisboa, Portugal}
\emailAdd{duartefontes@tecnico.ulisboa.pt}
\emailAdd{jorge.romao@tecnico.ulisboa.pt}

\vspace{0.5cm}
\noindent\abstract{
We present the one-loop electroweak renormalization of the CP-violating 2-Higgs-Doublet Model with softly broken $\mathbb{Z}_2$ symmetry (C2HDM). The existence of CP violation in the scalar sector of the model leads to a quite unique process of renormalization, since it requires the introduction of several non-physical parameters. The C2HDM will thus have more independent counterterms than independent renormalized parameters. As a consequence, different combinations of counterterms can be taken as independent for the same set of independent renormalized parameters. We compare the behaviour of selected combinations in specific NLO processes, which are assured to be gauge independent via a simple prescription. \FMTS is used to derive the Feynman rules, counterterms and one-loop processes in a simultaneously automatic and flexible way. This illustrates its use as an ideal tool to renormalize models such as the C2HDM and investigate them at NLO.
}

\maketitle
\noindent

\section{Introduction}
\label{Sect:intro}

\n The discovery of a scalar particle at the Large Hadron Collider (LHC) in 2012~\cite{Aad:2012tfa, Chatrchyan:2012ufa} marked the beginning of the unveiling of the scalar sector of particle physics.
Since then, several properties of the newfound particle have been ascertained and they are so far compatible with those of the Higgs boson predicted by the Standard Model (SM) \cite{Khachatryan:2016vau, ATLAS:2017ovn, CMS:2018lkl}.
But it is possible that such particle belongs to a theory with an extended scalar sector.
Actually, since the SM is unable to explain open problems like dark matter or baryogenesis, New Physics beyond the SM (BSM) is expected.
One of the most promising directions of BSM physics is precisely the possibility of extra scalars.
In recent years, the LHC has been actively searching for extra Higgs bosons \cite{Morvaj:2019ldy,Tao:2019hpy}, while plans for future colliders clearly aim at a better understanding of the scalar sector~\cite{deBlas:2019rxi,DiMicco:2019ngk}.
Among the various classes of models with an extended scalar sector, the simplest one that can provide a new source of CP violation required by the three Sakharov criteria for baryogenesis~\cite{Sakharov:1967dj} is the 2-Higgs-Doublet Model (2HDM)~\cite{Lee:1973iz} (for reviews of the 2HDM, cf. refs.~\cite{Gunion:1989we,Branco:2011iw,Ivanov:2017dad}).

\n One of the simplest versions of 2HDM with a CP-violating scalar sector is the so-called complex 2HDM (C2HDM). This model, characterized by a softly broken $\mathbb{Z}_2$ symmetry and complex parameters in the scalar potential, was originally discussed in ref.~\cite{Ginzburg:2002wt}, having been later developed and used by many authors~\cite{Khater:2003wq, ElKaffas:2006gdt, WahabElKaffas:2007xd, ElKaffas:2007rq, Osland:2008aw, Grzadkowski:2009iz, Arhrib:2010ju, Barroso:2012wz, Inoue:2014nva, Cheung:2014oaa, Fontes:2014xva, Fontes:2015mea, Chen:2015gaa, Fontes:2015xva, Belusca-Maito:2017iob, Fontes:2017zfn, Basler:2017uxn, Aoki:2018zgq}.
It describes a rich phenomenology, featuring five physical scalars: a charged Higgs pair and three neutral scalars
which are in general a mixture of a CP-even and a CP-odd component. 
The model becomes especially relevant in face of the very recent suggestion \cite{Fontes:2021znm} that the so-called ``real 2HDM''---a widely known variant of 2HDM where CP conservation is enforced in the scalar sector---is theoretically inconsistent. It turns out that the C2HDM is the simplest consistent version of such variant which can account for CP violation.

\n In the last few months, the C2HDM has aroused particular interest; for example, a 2-loop renormalization group evolution of the model was discussed in ref.~\cite{Oredsson:2019mni}; ref.~\cite{Wang:2019pet} used the model to study phase transition dynamics and gravitational wave signals; ref.~\cite{Boto:2020wyf} investigated a basis-independent treatment of the model; studies on electric dipole moments (EDM) of the C2HDM have been put forward in refs.~\cite{Cheung:2020ugr, Altmannshofer:2020shb};
ref.~\cite{Azevedo:2020vfw} discussed the impact of the discovery of a new scalar particle on the parameter space of the model; ref.~\cite{Huang:2020zde} considered CP-violating gauge-scalar interactions; ref.~\cite{Low:2020iua} investigated Higgs alignment and signatures of CP violation in the context of the model; and ref. \cite{Azevedo:2020fdl} derived constraints on the model from the phenomenology of light scalars.

\n However, precise predictions in the C2HDM have not yet been systematically considered. Precise predictions are the key to probe the SM and BSM models, since they are indispensable for a) a sound interpretation of the observed results, b) a correct determination of the parameter space of models and c) a proper distinction between different BSM models.
Precise predictions require the inclusion of one-loop electroweak corrections, which in turn require the one-loop electroweak renormalization of the model.%
\fn{Hereafter, and unless stated otherwise, all references to renormalization mean the one-loop electroweak renormalization; higher order terms will always be neglected.}
The literature devoted to the renormalization of models with an extended Higgs sector is vast (for the SM, reviews can be found in refs. \cite{Aoki:1982ed, Hollik:1988ii,Denner:1991kt,Denner:2019vbn,Freitas:2020kcn});
the ``real 2HDM'' has been subject to several studies~\cite{Santos:1996vt, Kanemura:2004mg, Kanemura:2015mxa, Krause:2016oke, Altenkamp:2017ldc, Denner:2016etu, Denner:2017vms, Denner:2018opp};
the next-to-minimal 2HDM has been considered in ref.~\cite{Krause:2017mal}, and
the scalar sector of a variant of the Minimal Supersymmetric extension of the SM with complex parameters (cMSSM) was explored for instance in refs.~\cite{Pilaftsis:1996ac,Pilaftsis:1997dr,Pilaftsis:1998pe,Pilaftsis:1998dd,Pilaftsis:1999qt,Carena:2000ks,Choi:2000wz,Frank:2006yh,Fowler:2009ay,Heinemeyer:2010mm,Fritzsche:2011nr,Fritzsche:2013fta}.

\n In this paper, we present the renormalization of the C2HDM. To our knowledge, it is the first time that the renormalization of a model with explicit CP violation in the scalar sector is put forward. This leads to a unique process of renormalization, as one is forced to introduce several parameters that can be rephased away in the context of the renormalized parameters, but must be considered anyway in order to assure the generation of all the necessary counterterms. As a consequence, there will be more independent counterterms than independent renormalized parameters. In particular, different combinations of counterterms can be taken as independent for the same set of independent renormalized parameters.

\n A crucial element in all of this is \FM~\cite{Fontes:2019wqh}, a multi-tasking software for particle physics studies. Combining \ts{FeynRules}~\cite{Christensen:2008py,Alloul:2013bka}, \ts{QGRAF}~\cite{Nogueira:1991ex} and \ts{FeynCalc}~\cite{Mertig:1990an,Shtabovenko:2016sxi,Shtabovenko:2020gxv}, \FMS is able to perform the totality of the following list of tasks in a consistent, automatic and flexible way:
%
\begin{small}
\begin{center}
\quad a) generation and drawing of Feynman rules (for both tree-level interactions and counterterms);
b) generation of amplitudes; \hs{3mm} c) generation and drawing of Feynman diagrams;\\ \hs{3mm} d) loop calculations; \hs{3mm} e) algebraic calculations; \hs{3mm} f) renormalization.
\end{center}
\end{small}
%
%
A new version of the program---\FMT---was recently made publicly available at
\begin{center}
\url{https://porthos.tecnico.ulisboa.pt/FeynMaster/},
\end{center}
containing several improvements over the first version. \FMTS turns out to be an ideal tool for building and investigating models, and especially for renormalizing them. In this paper, we apply it to the renormalization of the C2HDM and illustrate the advantages of doing so.

\n The paper is organized as follows. 
In section \ref{sec:model}, we describe the C2HDM in such a way that we aim at the one-loop renormalization of the model, presenting different combinations of independent parameters.
Section \ref{sec:UTOL} introduces the treatment of the theory when considered at up to one-loop level, clarifying how the theory can contain counterterms for quantities that do not show up in the set of renormalized quantities.
Sections \ref{sec:FJTS-C2HDM} and \ref{sec:calculation-CTs} proceed with the aforementioned treatment: the former describes the selection of the true vacuum expectation value (vev), whereas the latter is devoted to the counterterms of the C2HDM.
%
Finally, we present our results in section \ref{sec:num-res-main}.

\n After the conclusions in section \ref{sec:conclu}, some appendices are included.
Appendix \ref{app:OSS} provides a simple description of the on-shell subtraction scheme and investigates the scenarios where the mass of a particle is a dependent parameter.
Appendix \ref{sec:apD} is devoted to CP violation in fermionic 2-point functions and its influence on counterterms.
Then, in appendix \ref{app:Sym}, we show how counterterms can be fixed through symmetry relations.
Finally, we dedicate appendix \ref{app:FM} to \FMTS and to its application to the C2HDM.

\section{The C2HDM}
\label{sec:model}

\n The complete Lagrangian of the C2HDM can be written as a sum of the partial Lagrangians for the different sectors of the theory:
\be
\mathcal{L}_{\text{C2HDM}}
=
\mathcal{L}_{\text{Gauge}}
+
\mathcal{L}_{\text{Fermion}}
+
\mathcal{L}_{\text{Higgs}}
+
\mathcal{L}_{\text{Yukawa}}
+
\mathcal{L}_{\text{GF}}
+
\mathcal{L}_{\text{Ghost}}.
\label{eq:full-Lag}
\ee
The terms $\mathcal{L}_{\text{Gauge}}$ and $ \mathcal{L}_{\text{Fermion}}$ are just those of the SM, while $\mathcal{L}_{\text{GF}}$ and $\mathcal{L}_{\text{Ghost}}$ can be easily derived from the SM ones (cf. e.g. ref.~\cite{Romao:2012pq}).
Here, we study in detail the Higgs and the Yukawa partial Lagrangians. The former can be split into kinetic terms and potential,
\be
\mathcal{L}_{\text{Higgs}}
=
\mathcal{L}_{\text{Higgs}}^{\text{kin}}
- V.
\label{eq:LHiggs}
\ee
We start by studying the potential $V$ in section \ref{section:pot}, 
then the scalar kinetic terms in section \ref{section:kin}, and finally the Yukawa Lagrangian in section \ref{section:Yukawa}.
We shall use a general parameterization, which implies that we will introduce several parameters that are absent in the usual tree-level description of the model (cf. e.g. ref.~\cite{Fontes:2017zfn}).%
\fn{By the usual tree-level description, we mean the tree-level description that does not aim at the one-loop renormalization.}
As it will become clear, this is required in order to assure the one-loop renormalization of the theory.

\subsection{The potential}
\label{section:pot}

\n Assuming a softly broken $\mathbb{Z}_2$ symmetry, under which $\Phi_1 \to \Phi_1$, $\Phi_2 \to - \Phi_2$, the potential can be written as \cite{Fontes:2017zfn}
\bea
\label{eq:mypot}
V &=& m_{11}^2 |\Phi_1|^2 + m_{22}^2 |\Phi_2|^2
- \left(m_{12}^2 \, \Phi_1^\dagger \Phi_2 + \text{h.c.}\right)
+ \frac{\lambda_1}{2} (\Phi_1^\dagger \Phi_1)^2 +
\frac{\lambda_2}{2} (\Phi_2^\dagger \Phi_2)^2 \nonumber \\
&& + \lambda_3
(\Phi_1^\dagger \Phi_1) (\Phi_2^\dagger \Phi_2) + \lambda_4
(\Phi_1^\dagger \Phi_2) (\Phi_2^\dagger \Phi_1) +
\left[\frac{\lambda_5}{2} (\Phi_1^\dagger \Phi_2)^2 + \text{h.c.}\right] \;.
\eea
The hermiticity of the Lagrangian obliges all parameters to be real except $m_{12}^2$ and $\lambda_5$, which are in general complex.
In what follows, we define:
\be
m_{12 \mathrm{R}}^2 \equiv \mathrm{Re}\left(m_{12}^2\right),
\qquad
m_{12 \text{I}}^2 \equiv \mathrm{Im}\left(m_{12}^2\right),
\qquad
\lambda_{5 \mathrm{R}} \equiv \mathrm{Re}\left(\lambda_{5}\right),
\qquad
\lambda_{5 \text{I}} \equiv \mathrm{Im}\left(\lambda_{5}\right).
\label{eq:my-re-and-im}
\ee
%


\n After spontaneous symmetry breaking (SSB), each of the Higgs doublets
acquires a vev. These are in general complex, and in general relatively complex when compared to the scalar fields. We thus parameterize the Higgs doublets as:
\be
\Phi_1 =
\left(
\begin{array}{c}
\phi_1^+\\
\tfrac{1}{\sqrt{2}} (v_1 \, e^{i \zeta_1}+ \rho_1 + i \eta_1)
\end{array}
\right),
\hspace{5ex}
\Phi_2 =
\left(
\begin{array}{c}
\phi_2^+\\
\tfrac{1}{\sqrt{2}} (v_2 \, e^{i \zeta_2}+ \rho_2 + i \eta_2)
\end{array}
\right),
\label{eq:doublets-first}
\ee
where $v_i$ are real parameters, $\zeta_i$ (real) phases, $\phi_i^+$ complex fields, and $\rho_i$ and $\eta_i$ real fields ($i=1,2$).
The writing of the vevs with modulus and phase allows us to define the real parameters $v$ and $\beta$ such that
\be
v^2 \equiv v_1^2 + v_2^2,
\qquad
\tan(\beta) \equiv \dfrac{v_2}{v_1},
\qquad
v_1 = v \, c_{\beta},
\qquad
v_2 = v \, s_{\beta},
\label{eq:preliminar}
\ee
with $c_{\beta} = \cos \beta$, $s_{\beta} = \sin \beta$.

\subsubsection{Neutral linear terms: minimum equations}

\n One must start by assuring that the true minimum of the theory at tree-level is selected. This corresponds to the requirement that
no tree-level tadpoles $t_i$ for the neutral scalar fields $\phi_{n,i}$ exist, i.e. that the terms of the potential which are linear in the neutral scalar fields vanish.%
\fn{The tadpoles for the charged fields $\phi_{c,i}$ are trivially zero, due to the parameterization of eq. \ref{eq:doublets-first} (whose validity is proved throughout the paper).}
Hence,
\be
t_i \equiv \left\langle \frac{\partial V}{\partial \phi_{n,i}} \right\rangle
=
0.
\ee
This leads to three conditions, the so-called minimum equations:
\bs
\bea
{{m}_{11}^{2}} &=&  \dfrac{1}{4} \bigg[  \left( -{\lambda_1} + \lambda_{345} \right)  \, {v}^2 \, \cos(2 \, {\beta}) + 4 \, {{m}_{12 \mathrm{R}}^{2}} \, \tan({\beta}) \, \sec({\zeta_1} - {\zeta_2}) \nonumber \\[-4mm]
&& \hs{30mm} - {v}^2 \,  \left( {\lambda_1} + \lambda_{345} + 2 \, {\lambda_{5 \text{I}}} \, \sin({\beta})^2 \, \tan({\zeta_1} - {\zeta_2}) \right)  \bigg], \\[1mm]
{{m}_{22}^{2}} &=&  \dfrac{1}{4} \bigg[  \left( {\lambda_2} -  \lambda_{345} \right)  \, {v}^2 \, \cos(2 \, {\beta}) + 4 \, {{m}_{12 \mathrm{R}}^{2}} \, \cot({\beta}) \, \sec({\zeta_1} - {\zeta_2}) \nonumber \\[-4mm]
&& \hs{30mm} - {v}^2 \,  \left( {\lambda_2} +  \lambda_{345} + 2 \, {\lambda_{5 \text{I}}} \, \cos({\beta})^2 \, \tan({\zeta_1} - {\zeta_2}) \right) \bigg], \\[1mm]
{{m}_{12 \text{I}}^{2}} &=&  {{m}_{12 \mathrm{R}}^{2}} \, \tan({\zeta_1} - {\zeta_2}) + \dfrac{1}{4} \bigg[{v}^2 \, \sec({\zeta_1} - {\zeta_2}) \, \sin(2 \, {\beta}) \,  \Big( {\lambda_{5 \text{I}}} \, \cos(2 \,  \left( {\zeta_1} - {\zeta_2} \right) ) \nonumber \\[-4mm]
&& \hs{65mm} - {\lambda_{5 \mathrm{R}}} \, \sin(2 \,  \left( {\zeta_1} - {\zeta_2} \right) ) \Big) \bigg],
\eea
\label{eq:min}
\es
where we define $\lambda_{345} \equiv \lambda_3 + \lambda_4 + \lambda_{5 \mathrm{R}}$.

\subsubsection{Charged bilinear terms}
\label{section:charged}

The mass matrix of the charged scalars, defined by
\be
({\cal M}_c^{2})_{ij} = \left\langle \frac{\partial^2 V}{\partial \phi_{c,i}^{*} \,
  \partial \phi_{c,j}} \right\rangle \;,
\label{eq:chargedmassmat}
\ee
is hermitian, so that one needs in general a unitary matrix to diagonalize it. We then define the unitary matrix $X$ such that:%
\fn{\label{note:overall}The most general parameterization of $X$ includes an overall phase, which we ignore (in doing so, we force $\det X$ to be 1). It can be shown that such phase is not necessary; for details, cf. appendix \ref{app:Sym}.}
\be
S_c
=
X \, \phi_c
\quad
\Leftrightarrow
\quad
\begin{pmatrix}
G^+ \\
H^+
\end{pmatrix}
=
\begin{pmatrix}
e^{i(-\zeta_a - \zeta_b)} \cos \chi &
- e^{i(-\zeta_a + \zeta_b)} \sin \chi \\
e^{i(\zeta_a - \zeta_b)} \sin \chi &
e^{i(\zeta_a + \zeta_b)} \cos \chi
\end{pmatrix}
\begin{pmatrix}
\phi_1^+ \\
\phi_2^+
\end{pmatrix},
\label{eq:charged-param-original}
\ee
where $\zeta_a$ and $\zeta_b$ are (real) phases, $\chi$ a (real) angle, and the fields $S_c$ (with $S_c=(G^+ \, \, H^+)^{\mathrm{T}}$) are the charged states in the mass basis, with $G^+$ corresponding to the (massless) charged would-be Goldstone boson.
By definition, $X$ is such that:
\be
X {\cal M}_c^{2} X^{\dagger} = {\cal D}_c^2 \equiv 
\text{diag}(0,m_{\mathrm{H}^{+}}^2),
\label{eq:charged-mass-diag}
\ee
which in turn implies
\be
{\cal M}_c^{2} = X^{\dagger}  {\cal D}_c^2 X.
\label{eq:identity-charged}
\ee
It turns out that there are two non-trivial relations between the elements of $\mathcal{M}_c^{2}$, namely,
\be
\dfrac{\left(\mathcal{M}_c^{2}\right)_{11}}{\left(\mathcal{M}_c^{2}\right)_{22}} = \tan(\beta)^2,
\qquad
\dfrac{\left(\mathcal{M}_c^{2}\right)_{11}}{\left(\mathcal{M}_c^{2}\right)_{12}} = - \tan(\beta) e^{i(\zeta_2 - \zeta_1)},
\label{eq:charged-rels}
\ee
which must also be verified for the right-hand side of eq. \ref{eq:identity-charged}, thus leading to two identities:
\be
\chi = - \beta,
\qquad
\zeta_b = \dfrac{\zeta_1 - \zeta_2}{2}.
\label{eq:charged-conditions}
\ee

\subsubsection{Neutral bilinear terms}
\label{sec:neutral-bi}

The mass matrix of the neutral scalars, defined by
\be
({\cal M}_n^{2})_{ij}
=
\left\langle \frac{\partial^2 V}{\partial \phi_{n,i} \,
  \partial \phi_{n,j}} \right\rangle ,
\label{eq:c2hdmmassmat}
\ee
is symmetric, which means that one needs an orthogonal matrix to diagonalize it. Hence, we define the orthogonal matrix $Q$ such that:
\be
S_n = Q \, \phi_n
\quad
\Leftrightarrow
\quad
\begin{pmatrix}
h_1\\
h_2\\
h_3\\
G_0
\end{pmatrix}
= Q
\begin{pmatrix}
\rho_1\\
\rho_2\\
\eta_1\\
\eta_2
\end{pmatrix},
\label{eq:newmain}
\ee
where $S_n$ (with $S_n = (h_1 \,\, h_2 \,\, h_3 \,\, G_0)^{\mathrm{T}}$) are the neutral states in the mass basis, with $G^0$ corresponding to the (massless) neutral would-be Goldstone boson. Concerning $Q$, since it is a $4 \times 4$ orthogonal matrix, one in general needs 6 angles to parameterize it, which we take to be $\alpha_0, \alpha_1, \alpha_2, \alpha_3, \alpha_4, \alpha_5$, such that:
\be
Q
=
Q_5 \, Q_4 \, Q_3 \, Q_2 \, Q_1 \, Q_0,
\label{eq:Q1}
\ee
with
\bea
&&
Q_5
=
\begin{pmatrix}
1 & 0 & 0 & 0\\
0 & c_{5} & 0 & s_{5}\\
0 & 0 & 1 & 0\\
0 & -s_{5} & 0 & c_{5}
\end{pmatrix},
\ \
Q_4
=
\begin{pmatrix}
c_{4} & 0 & 0 & s_{4}\\
0 & 1 & 0 & 0\\
0 & 0 & 1 & 0\\
-s_{4} & 0 & 0 & c_{4}
\end{pmatrix}
\ \
Q_3
=
\begin{pmatrix}
1 & 0 & 0 & 0\\
0 & c_{3} & s_{3} & 0\\
0& -s_{3} & c_{3} & 0 \\
0 & 0 & 0 & 1
\end{pmatrix},
\nonumber \\[3mm]
&&
Q_2
=
\begin{pmatrix}
 c_{2} & 0 & s_{2} & 0\\
0 & 1 & 0 & 0\\
-s_{2} & 0 & c_{2} & 0 \\
0 & 0 & 0 & 1
\end{pmatrix},
\ \
Q_1
=
\begin{pmatrix}
c_{1} & s_{1} & 0 & 0\\
-s_{1} & c_{1} & 0 & 0\\
0 & 0 & 1 & 0\\
0 & 0 & 0 & 1
\end{pmatrix}
\ \
Q_0
=
\begin{pmatrix}
1 & 0 & 0 & 0\\
0 & 1 & 0 & 0\\
0 & 0 & -s_{0} & c_{0}\\
0 & 0& c_{0} & s_{0}
\end{pmatrix},
\label{eq:Q2}
\eea
with $s_i = \sin \alpha_i$, $c_i = \cos \alpha_i$ ($i = \{0,1,2,3,4,5\}$).
%
By definition, $Q$ is such that:
\be
Q {\cal M}_n^{2} Q^{\mathrm{T}} = {\cal D}_n^2 \equiv
\text{diag}(m_1^2,m_2^2,m_3^2,0),
\label{eq:neutral-mass-diag}
\ee
where $m_i$ represents the mass of $h_i$, such that $m_1 < m_2 < m_3$.
In order to find relations in this case, it is convenient to rewrite the doublets as:
\be
\Phi_1 = e^{i \zeta_1}
\left(
\begin{array}{c}
\phi_1^{+ \, \prime}\\
\tfrac{1}{\sqrt{2}} (v_1 + \rho_1^{\prime} + i \eta_1^{\prime})
\end{array}
\right),
\hspace{5ex}
\Phi_2 = e^{i \zeta_2}
\left(
\begin{array}{c}
\phi_2^{+ \, \prime}\\
\tfrac{1}{\sqrt{2}} (v_2 + \rho_2^{\prime} + i \eta_2^{\prime})
\end{array}
\right),
\label{eq:alternativebasis}
\ee
where the neutral fields with primes are related to the original ones (in eq. \ref{eq:doublets-first}) via:
\be
\phi_n = Z \, \phi_n^{\prime}
\quad
\Leftrightarrow
\quad
\begin{pmatrix}
\rho_1\\
\rho_2\\
\eta_1\\
\eta_2
\end{pmatrix}
=
\begin{pmatrix}
c_{\zeta_1} & 0 & -s_{\zeta_1} & 0\\
0 & c_{\zeta_2} & 0 & -s_{\zeta_2}\\
s_{\zeta_1} & 0 & c_{\zeta_1} & 0\\
0 & s_{\zeta_2} & 0 & c_{\zeta_2}
\end{pmatrix}
\begin{pmatrix}
\rho_1^{\prime}\\
\rho_2^{\prime}\\
\eta_1^{\prime}\\
\eta_2^{\prime}
\end{pmatrix}.
\ee
Then,
\begin{gather}
({\cal M}_n^{2 })_{ij}
=
\left\langle \frac{\partial^2 V}{\partial \phi_{n,i} \,
  \partial \phi_{n,j}} \right\rangle
=
({\cal M}_n^{2\prime})_{kl}
\dfrac{\partial \phi_{n,k}'}{\phi_{n,i}}
\dfrac{\partial \phi_{n,l}'}{\phi_{n,j}}
=
({\cal M}_n^{2 \prime})_{kl}
Z^{\mathrm{T}}_{ki} Z^{\mathrm{T}}_{lj}
=
(Z {\cal M}_n^{2 \prime} Z^{\mathrm{T}})_{ij},
\label{eq:trick-neutral}
\end{gather}
where we defined:
\be
({\cal M}_n^{2\prime})_{ij}
=
\left\langle \frac{\partial^2 V}{\partial \phi_{n,i}^{\prime} \,
  \partial \phi_{n,j}^{\prime}} \right\rangle.
\label{eq:neutral-mass-matrix-primes}
\ee
%
%
There are simple relations between the elements of ${\cal M}_n^{2\prime}$:
\be
\begin{pmatrix}
\dfrac{\left(\mathcal{M}_n^{2\prime}\right)_{13}}{\left(\mathcal{M}_n^{2\prime}\right)_{23}}
&
\dfrac{\left(\mathcal{M}_n^{2\prime}\right)_{14}}{\left(\mathcal{M}_n^{2\prime}\right)_{23}}
&
\dfrac{\left(\mathcal{M}_n^{2\prime}\right)_{23}}{\left(\mathcal{M}_n^{2\prime}\right)_{24}}
\\[5mm]
\dfrac{\left(\mathcal{M}_n^{2\prime}\right)_{13}}{\left(\mathcal{M}_n^{2\prime}\right)_{14}}
&
\dfrac{\left(\mathcal{M}_n^{2\prime}\right)_{13}}{\left(\mathcal{M}_n^{2\prime}\right)_{24}}
&
\dfrac{\left(\mathcal{M}_n^{2\prime}\right)_{14}}{\left(\mathcal{M}_n^{2\prime}\right)_{24}}
\\[5mm]
\dfrac{\left(\mathcal{M}_n^{2\prime}\right)_{33}}{\left(\mathcal{M}_n^{2\prime}\right)_{34}}
&
\dfrac{\left(\mathcal{M}_n^{2\prime}\right)_{33}}{\left(\mathcal{M}_n^{2\prime}\right)_{44}}
&
\dfrac{\left(\mathcal{M}_n^{2\prime}\right)_{34}}{\left(\mathcal{M}_n^{2\prime}\right)_{44}}
\end{pmatrix}
=
\begin{pmatrix}
\tan(\beta)
&
-1
&
-\tan(\beta)
\\
-\tan(\beta)
&
-\tan^2(\beta)
&
\tan(\beta)
\\
-\tan(\beta)
&
\tan^2(\beta)
&
-\tan(\beta)
\label{eq:neutral-rels}
\end{pmatrix}.
\ee
On the other hand, eqs. \ref{eq:neutral-mass-diag} and \ref{eq:trick-neutral} imply:
\be
{\cal M}_n^{2 \prime} = (Q Z)^{\mathrm{T}} {\cal D}_n^2 \, Q Z,
\label{eq:identity-neutral}
\ee
which means that the elements of the right-hand side of this equation must obey the same nine relations of eq. \ref{eq:neutral-rels}.
We thus have nine relations, involving in general the parameters:
\be
\alpha_0, \, \alpha_1, \, \alpha_2, \, \alpha_3, \, \alpha_4, \, \alpha_5, \, \zeta_1, \, \zeta_2, \, \beta, \, m_1^2, \, m_2^2, \, m_3^2.
\label{eq:params-aux}
\ee
It turns out that only four relations are independent, which means that one can only fix four of the parameters of eq. \ref{eq:params-aux}.
Obviously, there are many choices (or combinations) for the possible set of dependent parameters.
In this work, we consider the combinations $C_i$ ($i=1,2,3,4$) described in table \ref{tab:combos}.
\begin{table}[h!]
\centering
\begin{tabular}
{
@{\hspace{-0.8mm}}
>{\centering}p{3cm}
>{\centering}p{0.5cm}
>{\centering}p{0.5cm}
>{\centering}p{0.5cm}
>{\centering\arraybackslash}p{0.5cm}@{\hspace{3mm}}
}
\hlinewd{1.1pt}
Combination & \multicolumn{4}{c}{Dependent parameters}\\
\hline\\[-5mm]
$C_1$ & $m_3^2,$ & $\zeta_1,$ & $\alpha_0,$ & $\alpha_4$\\[1mm]
$C_2$ & $m_3^2,$ & $\zeta_1,$ & $\alpha_0,$ & $\alpha_5$\\[1mm]
$C_3$ & $m_3^2,$ & $\zeta_1,$ & $\alpha_4,$ & $\alpha_5$\\[1mm]
$C_4$ & $m_3^2,$ & $\alpha_0,$ & $\alpha_4,$ & $\alpha_5$\\[0.5mm]
\hlinewd{1.1pt}
\end{tabular}
\vspace{-3mm}
\caption{The dependent parameters associated to the different combinations $C_i$.}
\label{tab:combos}
\end{table}
\normalsize
Some notes are in order here.

\n First, $m_3^2$ is chosen as a dependent parameter in all combinations; this is not accidental, and can be justified as follows. When considering the model up to one-loop level, we shall find that there can be a clear relation between the renormalized parameters, on the one hand, and the parameters of the usual tree-level description, on the other. For the latter, we use ref. \cite{Fontes:2017zfn} as a guiding reference;
as it turns out, $m_3^2$ is taken as a dependent parameter in that reference. Hence, in order for this parameter to show up as dependent in the renormalized parameters of the model considered up to one-loop level, it must necessarily be taken as dependent at tree-level.

\n This also allows to explain the four combinations chosen in Table \ref{tab:combos}. Indeed, if we wish to obtain the aforementioned clear relation, the parameters taken as independent in ref.~\cite{Fontes:2017zfn} must also be taken as independent here; in particular, $\alpha_1, \alpha_2, \alpha_3, \beta, m_1^2, m_2^2$ must always be taken as independent. As a consequence, besides $m_3^2$ (which is always dependent), the only parameters available to be dependent are $\alpha_0$, $\alpha_4$, $\alpha_5$, $\zeta_1$ and $\zeta_2$.
We decide to take $\zeta_2$ as independent, which we will justify below. Therefore, since there are only four dependence relations, there are only four possible combinations of dependent counterterms---precisely those of Table \ref{tab:combos}.

\n Finally, the system of equations at stake is non-linear, and a rather complex one. We shall only solve it when we consider the model up to one-loop level.%
\fn{By then, the parameters involved in the system of equations are identified as bare parameters, and are split in renormalized parameters and counterterms (cf. section \ref{sec:UTOL}). The equations are then solved separately for the former and the latter.} 
For now, the dependent parameters must be understood simply as functions of the independent parameters.

\subsubsection{Combined sectors}

\n We have already derived some dependence conditions that resulted from relations we found within the squared mass matrices. Specifically, we found relations between the elements of $\mathcal{M}_c^{2}$, as well as relations between the elements of $\mathcal{M}_n^{2\prime}$. The former lead to the two con- ditions in eq. \ref{eq:charged-conditions}, whereas the latter to the dependence relations implied in table \ref{tab:combos}.

\n But there are still relations we have not yet used, which come from two equalities: that of eq. \ref{eq:identity-charged}
and that of eq. \ref{eq:identity-neutral}.
%
These lead to six independent relations, that we may use to rewrite the six $\lambda_i$ ($i=1,2,3,4,5\mathrm{R},5\text{I})$ in terms of other parameters.%
\fn{These expressions, though, are too long to be written here, but we have checked that they coincide with those of ref. \cite{Fontes:2014xva} in the appropriate limit.}

\subsection{Scalar kinetic sector}
\label{section:kin}

\n We now consider the first term of the right-hand side of eq. \ref{eq:LHiggs},
\be
\mathcal{L}_{\text{scalar}}^{\text{kin}}
=
\left( D_{\mu} \Phi_1\right)^{\dagger} \left( D_{\mu} \Phi_1\right)
+
\left( D_{\mu} \Phi_2\right)^{\dagger} \left( D_{\mu} \Phi_2\right),
\label{eq:kin}
\ee
where the covariant derivative is defined by
\be
D_{\mu}
=
\partial_{\mu}
+
i g_2 \dfrac{\tau_a}{2} W^a_{\mu}
+
i g_1 Y B_{\mu}.
\ee
Here, $g_1$ and $g_2$ are the gauge couplings of the $\mathrm{U(1)}$ and $\mathrm{SU(2)}$ gauge groups, respectively, with $B_{\mu}$ and $W^a_{\mu}$ ($a = 1,2,3$) being the corresponding gauge fields, and $Y$ and $\tau_a$ the corresponding group generators, respectively 
(we are following the conventions of ref.~\cite{Romao:2012pq} with all $\eta$'s positive).%
\fn{However, we use $g_1$ and $g_2$ instead of $g'$ and $g$, respectively.}
After SSB, the physical gauge fields $W^{\pm}_{\mu}$, $A_{\mu}$ and $Z_{\mu}$ are obtained from the original fields through the relations:
\be
W^{\pm}_{\mu} = \dfrac{W^1_{\mu} \mp i W^2_{\mu}}{\sqrt{2}},
\qquad
\begin{pmatrix}
A_{\mu} \\
Z_{\mu}
\end{pmatrix}
=
\begin{pmatrix}
c_{\text{w}} & s_{\text{w}} \\
-s_{\text{w}} & c_{\text{w}} 
\end{pmatrix}
\begin{pmatrix}
B_{\mu} \\
W^3_{\mu}
\end{pmatrix},
\label{eq:gauge-rot-tree}
\ee
with $s_{\text{w}} = \sin\theta_{\text{w}}$, $c_{\text{w}} = \cos\theta_{\text{w}}$, where $\theta_{\text{w}}$ is the weak mixing parameter, which is such that:
\be
c_{\text{w}} = \dfrac{g_2}{\sqrt{g_1^2 + g_2^2}}.
\label{eq:mycw}
\ee
Expanding the bilinear terms in the gauge fields in eq. \ref{eq:kin}, one easily finds that the photon $A_{\mu}$ is massless, whereas the squared masses of the $W^{+}_{\mu}$ and $Z_{\mu}$ bosons are, respectively,
\be
m_{\mathrm{W}}^2 = \dfrac{1}{4} \, g_2^2 \, v^2,
\qquad
m_{\mathrm{Z}}^2 = \dfrac{1}{4} \left(g_1^2 + g_2^2 \right) v^2.
\label{eq:gauge-masses}
\ee
Finally, defining the electric charge as
\be
e = \dfrac{g_1 g_2}{\sqrt{g_1^2 + g_2^2}},
\label{eq:charge}
\ee
we can take $v$, $g_1$ and $g_2$ as dependent parameters, which are then written as:%
\fn{The weak mixing parameter will be a dependent parameter itself, through the relation $c_{\text{w}} = m_{\mathrm{W}}/m_{\mathrm{Z}}$ (which follows from eqs. \ref{eq:mycw} and \ref{eq:gauge-masses}), since $m_{\mathrm{W}}$ and $m_{\mathrm{Z}}$ will be taken as independent.}
%
\be
v = \dfrac{2 \, m_{\mathrm{W}} \, s_{\text{w}}}{e},
\qquad
g_1 = \dfrac{e}{c_{\text{w}}},
\qquad
g_2 = \dfrac{e}{s_{\text{w}}}.
\label{eq:dep-params-gauge}
\ee

\subsection{Yukawa sector}
\label{section:Yukawa}


\n The Yukawa Lagrangian in general leads to flavour-changing neutral currents at tree-level. A simple way to avoid this is to assure that each right-handed fermionic singlet couples to only one Higgs doublet. This in turn can be accomplished if the $\mathbb{Z}_2$ symmetry is extended to the fermion fields, such that:
\begin{gather}
\bar{q}_{\mathrm{L}} \to (-1)^a \, \bar{q}_{\mathrm{L}},
\qquad
\bar{n}_{\mathrm{R}} \to (-1)^b \, \bar{n}_{\mathrm{R}},
\qquad
p_{\mathrm{R}} \to (-1)^c \, p_{\mathrm{R}}, \no
\bar{L}_{\mathrm{L}} \to (-1)^d \, \bar{L}_{\mathrm{L}},
\qquad
l_{\mathrm{R}} \to (-1)^e \, l_{\mathrm{R}}.
\label{eq:coeffs-a-to-e}
\end{gather}
Here, $a, b, c, d, e$ are general powers, $q_{\mathrm{L}} = (p_{\mathrm{L}} \, n_{\mathrm{L}})^{\mathrm{T}}$  and $L_{\mathrm{L}} = (\nu_{\mathrm{L}} \, l_{\mathrm{L}})^{\mathrm{T}}$ are the quark and lepton left-handed $\mathrm{SU(2)}$ doublets, respectively, and $p_{\mathrm{R}}$, $n_{\mathrm{R}}$ and $l_{\mathrm{R}}$ are the up-type quark, down-type quark and lepton right-handed singlets, respectively. There are four different combinations of the powers $a$ to $e$, each combination corresponding to a different type of C2HDM, as can be seen in Table \ref{tab:coeffs_models}.
\begin{table}[!h]
\begin{normalsize}
\normalsize
\begin{center}
\begin{tabular}
{@{\hspace{3mm}} >{\raggedright\arraybackslash}p{2.9cm} >{\raggedleft\arraybackslash}p{1cm} >{\raggedleft\arraybackslash}p{1cm}
>{\raggedleft\arraybackslash}p{1cm}
>{\raggedleft\arraybackslash}p{1cm} >{\raggedleft\arraybackslash}p{1cm}@{\hspace{3mm}}}
\hlinewd{1.1pt}
Type of C2HDM & $a$ & $b$ & $c$ & $d$ & $e$\\
\hline
Type I & $0$ & $1$ & $1$ & $0$ & $1$ \\
Type II & $0$ & $0$ & $1$ & $0$ & $0$ \\
Lepton-specific & $0$ & $1$ & $1$ & $0$ & $0$ \\
Flipped & $0$ & $0$ & $1$ & $0$ & $1$ \\
\hlinewd{1.1pt}
\end{tabular}
\end{center}
\vspace{-5mm}
\end{normalsize}
\caption{The powers $a, b, c, d, e$ of eq. \ref{eq:coeffs-a-to-e} for each type of C2HDM.}
\label{tab:coeffs_models}
\end{table}
\normalsize
In this article, we restrict ourselves to the Type II model.
To write the Yukawa Lagrangian, it is convenient to parameterize the Higgs doublets according to:
\be
\Phi_i =
\begin{pmatrix}
\phi_i^+ \\
\phi_i^0
\end{pmatrix}
\,
, 
\quad
\Phi_i^* =
\begin{pmatrix}
\phi_i^- \\
\phi_i^{0*}
\end{pmatrix},
\ee
in which case we have, in the Type II C2HDM:
\be
- \mathcal{L}_{\text{Yukawa}}
=
\begin{pmatrix}
\bar{p}_{\mathrm{L}} & \bar{n}_{\mathrm{L}}
\end{pmatrix}
Y_d
\begin{pmatrix}
\phi_1^+ \\
\phi_1^0
\end{pmatrix}
n_{\mathrm{R}}
\, + \,
\begin{pmatrix}
\bar{p}_{\mathrm{L}} & \bar{n}_{\mathrm{L}}
\end{pmatrix}
Y_u 
\begin{pmatrix}
\phi_2^{0*} \\
-\phi_2^-
\end{pmatrix}
p_{\mathrm{R}}
\, + \,
\begin{pmatrix}
\bar{\nu}_{\mathrm{L}} & \bar{l}_{\mathrm{L}}
\end{pmatrix}
Y_l 
\begin{pmatrix}
\phi_1^+ \\
\phi_1^0
\end{pmatrix}
l_{\mathrm{R}}
\, + \, 
\text{h.c.},
\label{eq:LYukawa-original}
\ee
where $Y_d$, $Y_u$ and $Y_l$ are the Yukawa matrices for the down-type quarks, up-type quarks and leptons, respectively.%
\fn{In this equation, the $\mathrm{SU(2)}$ product is shown explicitly, but the sum over fermion generations is left implicit by the matrix notation. Had we written it explicitly, the quantities  $\bar{p}_{\mathrm{L}}, \bar{n}_{\mathrm{L}}, Y_d, n_{\mathrm{R}}$ in the first term on the right-hand side of eq. \ref{eq:LYukawa-original}, for example, would have been $\bar{p}_{L_i}, \bar{n}_{L_i}, Y_d^{ij}, n_{R_j}$, respectively.}
We can, however, rewrite this equation in a more meaningful way.
We start by noting that the quarks are rotated to the mass basis through unitary transformations
\be
\bar{p}_{\mathrm{L}} = \bar{u}_{\mathrm{L}} U_{u_{\mathrm{L}}}^\dagger,
\hspace{8mm}
\bar{n}_{\mathrm{L}} = \bar{d}_{\mathrm{L}} U_{d_{\mathrm{L}}}^\dagger,
\hspace{8mm}
p_{\mathrm{R}} = U_{u_{\mathrm{R}}} u_{\mathrm{R}},
\hspace{8mm}
n_{\mathrm{R}} = U_{d_{\mathrm{R}}} d_{\mathrm{R}} ,
\label{eq:quarks-rot}
\ee
in such a way that the interaction with the vevs of the Higgs doublets generates the mass terms, that is,
\be
-\mathcal{L}_{\text{Yukawa}}^{\text{mass}}
=
\bar{d}_{\mathrm{L}} \, U_{d_{\mathrm{L}}}^{\dagger} Y_d U_{d_{\mathrm{R}}} \,  d_{\mathrm{R}} \, \langle \phi_1^0 \rangle
+ \bar{u}_{\mathrm{L}} \, U_{u_{\mathrm{L}}}^{\dagger} Y_u U_{u_{\mathrm{R}}} \, u_{\mathrm{R}} \, \langle \phi_2^{0*} \rangle
+ \bar{l}_{\mathrm{L}} \, Y_l \, l_{\mathrm{R}}  \, \langle \phi_1^0 \rangle
\, + \, 
\text{h.c.} \, .
\ee
On the other hand, the mass terms must also obey:
\be
- \mathcal{L}_{\text{Yukawa}}^{\text{mass}}
=
\bar{d}_{\mathrm{L}} \, M_d \, d_{\mathrm{R}}
+ \bar{u}_{\mathrm{L}} \, M_u \, u_{\mathrm{R}}
+ \bar{l}_{\mathrm{L}} \, M_l \, l_{\mathrm{R}}
\, + \, 
\text{h.c.} \, ,
\ee
with
$M_d = \text{diag}(m_d, m_s, m_b)$,
$M_u = \text{diag}(m_u, m_c, m_t)$,
$M_l = \text{diag}(m_e, m_{\mu}, m_{\tau})$.
Then, noting that
\be
\langle \phi_1^0 \rangle = \dfrac{v}{\sqrt{2}} c_{\beta} \, e^{i \zeta_1},
\qquad
\langle \phi_2^{0*} \rangle = \dfrac{v}{\sqrt{2}} s_{\beta} \, e^{-i \zeta_2},
\ee
we find:%
\fn{We assume there are no right-handed neutrinos, which implies we can take $Y_l$ to be diagonal without loss of generality.}.
\be
Y_d = \dfrac{\sqrt{2}}{v \, c_{\beta} \, e^{i \zeta_1}} U_{d_{\mathrm{L}}} M_d U_{d_{\mathrm{R}}}^{\dagger},
\qquad
Y_u = \dfrac{\sqrt{2}}{v \, s_{\beta} \, e^{-i \zeta_2}} U_{u_{\mathrm{L}}} M_u U_{u_{\mathrm{R}}}^{\dagger},
\qquad
Y_l = \dfrac{\sqrt{2}}{v \, c_{\beta} \, e^{i \zeta_1}} M_l,
\ee
which allows us to write eq. \ref{eq:LYukawa-original} as
\begin{align}
&
-\mathcal{L}_{\text{Yukawa}}
=
\dfrac{\sqrt{2}}{v}
\Bigg[
\dfrac{1}{c_{\beta} \, e^{i \zeta_1}}
\begin{pmatrix}
\bar{u}_{\mathrm{L}} V & \bar{d}_{\mathrm{L}}
\end{pmatrix}
M_d
\begin{pmatrix}
\phi_1^+ \\
\phi_1^0
\end{pmatrix}
d_{\mathrm{R}}
\, + \,
\dfrac{1}{s_{\beta} \, e^{-i \zeta_2}}
\begin{pmatrix}
\bar{u}_{\mathrm{L}} & \bar{d}_{\mathrm{L}} V^{\dagger}
\end{pmatrix}
M_u
\begin{pmatrix}
\phi_2^{0*} \\
-\phi_2^-
\end{pmatrix}
u_{\mathrm{R}}
\nonumber
\\[1mm]
&
\hspace{40mm}
+
\dfrac{1}{c_{\beta} \, e^{i \zeta_1}}
\begin{pmatrix}
\bar{\nu}_{\mathrm{L}} & \bar{l}_{\mathrm{L}}
\end{pmatrix}
M_l 
\begin{pmatrix}
\phi_1^+ \\
\phi_1^0
\end{pmatrix}
l_{\mathrm{R}}
\Bigg]
\, + \, 
\text{h.c.} \, ,
\end{align}
where $V= U_{u_{\mathrm{L}}}^{\dagger} U_{d_{\mathrm{L}}}$ is the Cabibbo–Kobayashi–Maskawa (CKM) matrix.

\n Finally, note that the fermion masses $m_f$ are in general complex parameters, since the Yukawa matrices are general complex matrices.%
\fn{$f$ represents the physical down-type quarks, up-type quarks and leptons, whose masses are respectively contained in $M_d$, $M_u$ and $M_l$.}
Usually, one performs a chiral rotation (that is, a rotation of the Weyl spinors) in order to render the masses real.%
\fn{Cf. e.g. section 29.3.2 of ref.~\cite{Schwartz:2013pla}.}
But the circumstance that the masses are in general complex means that, when the theory is considered up to one-loop level, the counterterms for the masses will also be in general complex. And although this is an irrelevant detail in most models, it becomes most relevant whenever there is CP violation in fermionic 2-point functions, as in the present model at one-loop level.
We discuss in detail complex mass counterterms---as well as their relation with CP violation in fermionic 2-point functions---in appendix \ref{sec:apD}.
For now, we assume $m_f$ to be in general complex.

\subsection{Parameters}

\n The relations introduced in the previous sections allow us to replace the original sets of parameters in the Yukawa and Higgs sectors, respectively given by
\bs
\label{eq:original-params}
\bea
\{p^{\text{Y}}_{\text{ori.}}\}
&=&
\{ 
Y_d, Y_u, Y_l
\},
\\
\{p^{\mathrm{H}}_{\text{ori.}}\}
&=&
\{ 
g_1, \, g_2, \, m_{11}^2, \, m_{22}^2, \, m_{12 \mathrm{R}}^2, \, m_{12 \text{I}}^2, \,
\lambda_1, \, \lambda_2, \, \lambda_3, \, \lambda_4, \, \lambda_{5\mathrm{R}}, \, \lambda_{5\text{I}}
\},
\eea
\es
by new sets of parameters:
\bs
\label{eq:new-params}
\bea
\label{eq:new-params-Y}
\{p^{\text{Y}}_{\text{new}}\}
&=&
\{ 
m_f, V
\},
\\[1mm]
\label{eq:new-params-H}
\begin{blockarray}{c}
\\[-1.4mm]
\{p^{\mathrm{H}}_{C_1}\}\\[1.5mm]
\{p^{\mathrm{H}}_{C_2}\}\\[1.5mm]
\{p^{\mathrm{H}}_{C_3}\}\\[1.5mm]
\{p^{\mathrm{H}}_{C_4}\}
\end{blockarray}
\hspace{-1.5mm}
&=&
\{e, \, m_{\mathrm{W}}, \, m_{\mathrm{Z}},
\, \alpha_1, \, \alpha_2, \, \alpha_3,
\, \beta, \, m_1, \, m_2, \, m_{\mathrm{H}^{+}},
\, \mu^2, \, \zeta_2, \, \zeta_a,
\begin{blockarray}{c}
\\[-1.4mm]
\alpha_5\},\\[1.5mm]
\alpha_4\},\\[1.5mm]
\alpha_0\},\\[1.5mm]
\zeta_1\},
\end{blockarray}
\eea
\es
where we define:
\be
\mu^2 \equiv
\dfrac{m_{12 \mathrm{R}}^2}{c_{\beta} \, s_{\beta}}.
\label{eq:mu}
\ee
We thus see that, in the Higgs sector, there are four possibilities, $\{p^{\mathrm{H}}_{C_i}\}$ ($i=1,2,3,4$), respectively associated to the combinations $C_i$ introduced in section \ref{sec:neutral-bi}.
For a given $C_i$, the parameters in the new total set (including parameters from the Yukawa and Higgs sectors) are all independent, and are more convenient than the ones in eqs. \ref{eq:original-params}.
One can then express the first four terms of the right-hand side of eq. \ref{eq:full-Lag}
in terms of the parameters of eqs. \ref{eq:new-params}.

%
%
%

\section{Going up to one-loop level}
\label{sec:UTOL}

\n So far, we have been discussing the theory at tree-level, i.e. in lowest order (LO) in perturbation theory. When we include the next order---that is, when we consider the theory up to one-loop level, or at the next-to-leading order (NLO)---, we must start by modifying the notation.%
\fn{\label{note:utol}The notion `NLO' does not refer to one-loop level only, but to both tree-level and one-loop level. For the sake of clarity, we 
will often use the notion `up to one-loop' instead. Both notions are equivalent and shall be used indifferently in what follows.
What we identify here as up-to-one-loop theory is then part (namely, the part which includes the tree-level and one-loop level) of what is usually identified as the effective theory (cf. e.g. ref. \cite{Denner:2019vbn}).
}
The theory we have been considering, indeed, is to be identified with a \textit{bare} theory, and the parameters and fields therein contained as \textit{bare} parameters and fields. Moreover, we make sure to select the true vev of the theory up to one-loop level; such selection is discussed in section \ref{sec:FJTS-C2HDM}.

\n The bare parameters and fields are identified with an index ``$(0)$'' and they can \textit{renormalized}. To do so, one starts by splitting them into renormalized quantities and their corresponding counterterms; for example, for a generic bare parameter $p_{(0)}$ and a generic bare field $\psi_{(0)}$,
\be
p_{(0)} = p + \delta p,
\qquad
\psi_{(0)} = \psi + \dfrac{1}{2} \delta Z_{\psi} \, \psi.
\label{eq:generic-expansion}
\ee
Here, $p$ represents the renormalized parameter and $\delta p$ the corresponding counterterm, and $\psi$ the renormalized field and $\delta Z_{\psi}$ the corresponding counterterm.%
\fn{The renormalized quantities in the context of the theory up to one-loop level (which have no index whatsoever) should not be confused with the bare quantities in the context of the theory at tree-level (which, in that context, did not have the index ``$(0)$'').
Note also that $\delta p$ is dubbed counterterm for the parameter (or parameter counterterm) and $\delta Z_{\psi}$ is dubbed counterterm for the field (or field counterterm).
}
The renormalization is completed by the calculation of the counterterms, which is performed in section \ref{sec:calculation-CTs}.%
\fn{\label{note:LSZ}In general, the LSZ factors (present in the LSZ reduction formula) must also be calculated in order to obtain correct $S$-matrix elements \cite{Bohm:2001yx,Denner:2019vbn}. However, they become trivial when fields are renormalized in the on-shell subtraction scheme, so that they can be ignored. For details, cf. ref. \cite{Fontes:PhD}.}

\n Now, in order to assure that all $S$-matrix elements are ultraviolet (UV) finite, one \textit{must} renormalize the independent parameters. The first step is choosing an independent set of parameters.\fn{Dependent parameters need not be renormalized, although they can be renormalized for convenience; for details, see ref. \cite{Fontes:PhD}.} As independent set of parameters, besides those of eq. \ref{eq:new-params-Y} we will consider the four possibilities of independent sets of eq. \ref{eq:new-params-H}. All these parameters are taken as bare parameters in the context of the up-to-one-loop theory, and thus read:
\begingroup\makeatletter\def\f@size{9.5}\check@mathfonts
\def\maketag@@@#1{\hbox{\m@th\normalsize\normalfont#1}}
\makeatother
\bs
\label{eq:bare-params}
\begin{flalign}
\label{eq:bare-params-Y}
&\{p^{\text{Y}}_{\text{new}(0)}\}
=
\{ 
m_{f(0)}, V_{(0)}
\},
\\[2mm]
&
\label{eq:bare-params-H}
\hs{-2mm}
\begin{blockarray}{c}
\\[-1.2mm]
\{p^{\mathrm{H}}_{C_1(0)}\}\\[1.5mm]
\{p^{\mathrm{H}}_{C_2(0)}\}\\[1.5mm]
\{p^{\mathrm{H}}_{C_3(0)}\}\\[1.5mm]
\{p^{\mathrm{H}}_{C_4(0)}\}
\end{blockarray}
\hspace{-3mm}
=
\hspace{-1mm}
\{e_{(0)}, m_{\mathrm{W}(0)},  m_{\mathrm{Z}(0)},
 \alpha_{1(0)},  \alpha_{2(0)},  \alpha_{3(0)},
 \beta_{(0)},  m_{1(0)},  m_{2(0)},  m_{\mathrm{H}^{+}(0)},
 \mu^2_{(0)},  \zeta_{2(0)}, \zeta_{a(0)},
\begin{blockarray}{c}
\\[-1.2mm]
\alpha_{5(0)}\},\\[1.5mm]
\alpha_{4(0)}\},\\[1.5mm]
\alpha_{0(0)}\},\\[1.5mm]
\zeta_{1(0)}\}.
\end{blockarray}
\end{flalign}
\es
\endgroup
Afterwards, and as suggested above, these bare parameters are separated into renormalized parameter and parameter counterterm, and the latter must be calculated.

\n On the other hand, if one intends to obtain finite Green's functions (GFs) and not only finite $S$-matrix elements, fields must be renormalized as well. This can be done in the symmetric form of the theory \cite{Bohm:1986rj,Hollik:1988ii}---i.e. before the SSB of the symmetry---, in which case it is enough to consider one field counterterm for each multiplet \cite{Denner:2019vbn}. As an alternative, one can renormalize the fields in the mass basis \cite{Aoki:1982ed,Denner:1991kt,Denner:2019vbn}. Here, counterterms for the mixing of fields (mixed fields counterterms) must in general be considered whenever fields mix at loop-level.

\n In the present section, we start by identifying the bare quantities of the theory with the sum of renormalized ones and counterterms (we do not do this for gauge parameters
nor ghost fields, since it would be irrelevant for the calculation of $S$-matrix elements~\cite{Denner:2019vbn}).%
\fn{Actually, in linear gauges (as the Feynman gauge), it can be shown that one can simply restrain from applying any renormalization transformation whatsoever to the Gauge Fixing Lagrangian
\cite{tHooft:1972qbu,Ross:1973fp,Baulieu:1983tg}.
That is, all the parameters and fields therein are assumed to be already renormalized.}
Then, in section \ref{sec:simple}, we review the need to consider a general parameterization, and show that the renormalized parameters can be described by a simpler parameterization.

\subsection{Expansion of the bare quantities}
\label{sec:expa}

\n In eq. \ref{eq:generic-expansion} above, we considered generic expansions. We now need to apply them to the specific content of the C2HDM. 
Concerning the parameters, we include not only those of eqs. \ref{eq:bare-params}, but also some dependent parameters (for convenience).
Some renormalized parameters will be set equal to non-obvious quantities, which will be justified below.
As for the fields, we choose those in the mass basis.


\n We start with the gauge boson masses and fields. For the masses, we have:
\be
m_{{\mathrm{W}}(0)}^2 = m_{\mathrm{W}}^2 + \delta m_{\mathrm{W}}^2,
\qquad
m_{\mathrm{Z}(0)}^2 = m_{\mathrm{Z}}^2 + \delta m_{\mathrm{Z}}^2,
\label{eq:mass-gauge-expa}
\ee
while for the fields:
\be
W_{\mu(0)}^{+} =
\Big( 1 + \dfrac{1}{2} \delta Z_W \Big) W_{\mu}^{+},
\qquad
\begin{pmatrix}
A_{\mu(0)} \\ Z_{\mu(0)}
\end{pmatrix}
=
\begin{pmatrix}
1 + \dfrac{1}{2} \delta Z_{AA} & \dfrac{1}{2} \delta Z_{AZ} \\
\dfrac{1}{2} \delta Z_{ZA} & 1 + \dfrac{1}{2} \delta Z_{ZZ}
\end{pmatrix}
\begin{pmatrix}
A_{\mu}
\\
Z_{\mu}
\end{pmatrix}.
\label{eq:field-gauge-expa}
\ee
%


\n In the fermions, the masses obey
\be
m_{f,i(0)} = m_{f,i} + \delta m_{f,i},
\label{eq:mass-fermion-expa}
\ee
in such a way that $m_{f,i}$ are real, and we define:%
\fn{For details, cf. appendix \ref{sec:apD}.}
\be
\delta m_{f,i}^{\mathrm{R}} \equiv \delta m_{f,i},
\qquad
\delta m_{f,i}^{\mathrm{L}} \equiv \delta m_{f,i}^*.
\label{eq:mass-fermion-expa-2}
\ee
The fields obey:
\be
f_{i(0)}^{\mathrm{L}} =
\sum_j \Big( \delta_{ij} + \dfrac{1}{2} \delta Z_{ij}^{f,\mathrm{L}} \Big) f_j^{\mathrm{L}},
\qquad
f_{i(0)}^{\mathrm{R}} =
\sum_j \Big( \delta_{ij} + \dfrac{1}{2} \delta Z_{ij}^{f,\mathrm{R}} \Big) f_j^{\mathrm{R}}.
\label{eq:field-fermion-expa}
\ee


\n Considering now the scalar sector, we have for the masses:
\be
m_{\mathrm{H}^{+}(0)}^2 = m_{\mathrm{H}^{+}}^2 + \delta m_{\mathrm{H}^{+}}^2,
\qquad
m_{1(0)}^2 = m_1^2 + \delta m_1^2,
\qquad
m_{2(0)}^2 = m_2^2 + \delta m_2^2.
\label{eq:mass-scalar-expa}
\ee
We renormalize $m_{3(0)}^2$ for convenience:%
\be
m_{3(0)}^2 = m_{3\mathrm{R}}^2 + \delta m_{3}^2.
\label{eq:mass-scalar-expa-2}
\ee
The renormalized squared mass in this case explicitly includes an index $\mathrm{R}$ (from `renormalized').%
\fn{Not to be confused with the index $\mathrm{R}$ expressing `real', as in eq. \ref{eq:my-re-and-im}.}
The reason why this is necessary in this case is that, since $m_{3(0)}$ is a dependent parameter, it is fixed by other parameters. As a consequence, when the bare quantities are identified with a renormalized quantity plus a counterterm, the renormalized squared mass $m_{3\mathrm{R}}^2$ will also be fixed by other parameters, which means that it cannot be set equal to the \textit{pole} squared mass---which we identify in the following as $m_{3\mathrm{P}}^2$. Moreover, $\delta m_{3}^2$ will be a dependent counterterm.
All this implies that $h_3$ plays a special role in theory, which is described detail in appendix \ref{app:OSS}.

\n For the scalar fields,
\bs
\label{eq:field-scalar-expa}
\bea
\label{eq:charged-diag-bare-exp}
\begin{pmatrix}
G_{(0)}^{+} \\ H_{(0)}^{+}
\end{pmatrix}
&=&
\begin{pmatrix}
1 + \dfrac{1}{2} \delta Z_{G^{+}G^{+}} & \dfrac{1}{2} \delta Z_{G^{+}H^{+}} \\
\dfrac{1}{2} \delta Z_{H^{+}G^{+}} & 1 + \dfrac{1}{2} \delta Z_{H^{+}H^{+}}
\end{pmatrix}
\begin{pmatrix}
G^{+} \\ H^{+}
\end{pmatrix},
\\
\begin{pmatrix}
h_{1(0)} \\ h_{2(0)} \\ h_{3(0)} \\ G_{0(0)}
\end{pmatrix}
&=&
\begin{pmatrix}
1 + \frac{1}{2} \delta Z_{h_1h_1} & \frac{1}{2} \delta Z_{h_1h_2} & \frac{1}{2} \delta Z_{h_1h_3} & \frac{1}{2} \delta Z_{h_1G_0}\\
\frac{1}{2} \delta Z_{h_2h_1} & 1 + \frac{1}{2} \delta Z_{h_2h_2} & \frac{1}{2} \delta Z_{h_2h_3} & \frac{1}{2} \delta Z_{h_2G_0}\\
\frac{1}{2} \delta Z_{h_3h_1} & \frac{1}{2} \delta Z_{h_3h_2} & 1 + \frac{1}{2} \delta Z_{h_3h_3}  & \frac{1}{2} \delta Z_{h_3G_0}\\
\frac{1}{2} \delta Z_{G_0h_1} & \frac{1}{2} \delta Z_{G_0h_2} & \frac{1}{2} \delta Z_{G_0h_3} & 1 + \frac{1}{2} \delta Z_{G_0G_0}
\end{pmatrix}
\begin{pmatrix}
h_{1} \\ h_{2} \\ h_{3} \\ G_{0}
\end{pmatrix}.
\eea
\es
The mixing parameters are such that:%
\fn{Recall that $\alpha_{0(0)}$, $\alpha_{4(0)}$ and $\alpha_{5(0)}$ will be independent according to the combination $C_i$ chosen.}
\bs
\label{eq:mix-angles-bare-exp}
\begin{gather}
\label{eq:my-dza}
\beta_{(0)} = \beta + \delta \beta,
\qquad
\zeta_{a(0)} = \delta \zeta_a,
\\
\alpha_{1(0)} = \alpha_1 + \delta \alpha_{1},
\qquad
\alpha_{2(0)} = \alpha_2 + \delta \alpha_{2},
\qquad
\alpha_{3(0)} = \alpha_3 + \delta \alpha_{3},
\\
\alpha_{0(0)} = \beta + \delta \alpha_{0},
\qquad
\alpha_{4(0)} = \delta \alpha_{4},
\qquad
\alpha_{5(0)} = \delta \alpha_{5}.
\label{eq:consoante}
\end{gather}
\es
The electric charge and the CKM matrix obey:
\be
e_{(0)}
= e + \delta Z_e e,
\qquad
V_{ij(0)} = V_{ij} + \delta V_{ij}.
\ee
As for the remaining parameters that are taken as independent ($\zeta_{1(0)}$ will be independent in the combination $C_4$), we have:
\be
\mu_{(0)}^2 = \mu^2 + \delta \mu^2,
\qquad
\zeta_{1(0)} = \delta \zeta_1,
\qquad
\zeta_{2(0)} = \delta \zeta_2.
\label{eq:remaining-bare-exp}
\ee

\n For convenience, we also renormalize several dependent parameters:%
\fn{
\label{eq:note-dep-CT}
In the combinations $C_i$ for which the parameters $\alpha_{0(0)}$, $\alpha_{4(0)}$, $\alpha_{5(0)}$ and $\zeta_{1(0)}$ are dependent, we also renormalize them for convenience (according to eqs. \ref{eq:consoante} and \ref{eq:remaining-bare-exp}).
We do not renormalize $\chi$, $\zeta_b$, $g_1$, $g_2$ or $v$. This means that, before renormalization, they must be replaced by the corresponding expressions (cf. eqs. \ref{eq:charged-conditions} and \ref{eq:dep-params-gauge}), and the latter must be renormalized. More details can be found in ref. \cite{Fontes:PhD}.
}
\begin{gather}
s_{{\text{w}}(0)} = s_{\text{w}} + \delta s_{\text{w}}, 
\quad
c_{{\text{w}}(0)} = c_{\text{w}} + \delta c_{\text{w}},
\nonumber
\\
m_{11(0)}^2 = m_{11}^2 + \delta m_{11}^2,
\quad
m_{22(0)}^2 = m_{22}^2 + \delta m_{22}^2,
\qquad
m_{12\text{I}(0)}^2 = m_{12\text{I}}^2 + \delta m_{12\text{I}}^2,
\nonumber
\\
\lambda_{i(0)} = \lambda_i + \delta \lambda_i,
\qquad
X^{\dagger}_{ij(0)} = X^{\dagger}_{ij} + \delta X^{\dagger}_{ij},
\qquad
Q_{ij(0)} = Q_{ij} + \delta Q_{ij}.
\label{eq:unnec}
\end{gather}

\subsection{General \textit{vs.} simple parameterization}
\label{sec:simple}

\n When studying the theory at tree-level, we considered the most general parameterization of the Higgs doublets, as well as of the diagonalization of the scalar states---recall section \ref{section:pot}.
This general information was required in order to have general counterterms (which are necessary to absorb the UV divergences at one-loop). But it is \textit{not} required to describe the renormalized parameters. Indeed, if the only reason we considered such a general parameterization was to account for the generality of the counterterms, there is no need to apply such general description to the renormalized parameters as well. 

\n To better understand the claim we are making, consider for example the phase $\zeta_a$. At tree-level, we had the freedom to rephase the fields $G^+$ and $H^+$ to absorb $\zeta_a$. Had we used that freedom, though, we would have run into problems; for when we were to consider the theory up to one-loop level, we would have no counterterm $\delta \zeta_a$ (as there would have been no bare parameter $\zeta_{a(0)}$ to start with). It turns out that such counterterm is necessary for the absorption of the divergences of the theory, so that it cannot be discarded.
%
%
That is to say, we could not use the freedom to rephase $\zeta_a$ away at tree-level.
But once in the up-to-one-loop theory, and given that $\delta \zeta_a$ was already generated, there is no longer any reason not to use the rephasing freedom. We can thus rephase the renormalized parameter $\zeta_a$ away in the up-to-one-loop theory; and when we do so, the counterterm $\delta \zeta_a$ (that was meanwhile generated because we did not rephase $\zeta_a$ away at tree-level) will not vanish.

\n To see this explicitly, we may consider a simplified version of eq. \ref{eq:charged-param-original}, where we take $\zeta_b=\chi=0$ (which are irrelevant for the argument):
\be
\begin{pmatrix}
\phi_1^+ \\
\phi_2^+
\end{pmatrix}
=
\begin{pmatrix}
e^{i \zeta_a} & 0\\
0 & e^{-i \zeta_a} 
\end{pmatrix}
\begin{pmatrix}
G^+ \\
H^+
\end{pmatrix}.
\label{eq:ex:tree}
\ee
It is obvious that, in this tree-level relation, one is free to rephase $\zeta_a$ away through $G^+ \to e^{-i \zeta_a} G^+$, $H^+ \to e^{i \zeta_a} H^+$. Yet, as we want to generate a counterterm $\delta \zeta_a$, we do not use that rephasing freedom. So, when considering the theory up to one-loop level, we take the bare version of eq. \ref{eq:ex:tree};
then, expanding the charged scalar states in the mass basis and $\zeta_{a(0)}$ into renormalized quantities and counterterms, according to eq. \ref{eq:charged-diag-bare-exp} and
\be
\zeta_{a(0)} = \zeta_a + \delta \zeta_a,
\label{eq:310}
\ee
we obtain, to first order,
\be
\begin{pmatrix}
\phi_{1(0)}^+ \\
\phi_{2(0)}^+
\end{pmatrix}
=
\begin{pmatrix}
e^{i \zeta_a} \left( i \delta \zeta_a + \delta Z_{G^{+}G^{+}} + 1 \right) &
e^{i \zeta_a} \delta Z_{G^{+}H^{+}} \\
e^{-i \zeta_a} \delta Z_{H^{+}G^{+}} &
e^{-i \zeta_a} \left( -i \delta \zeta_a + \delta Z_{H^{+}H^{+}} + 1 \right)
\end{pmatrix}
\begin{pmatrix}
G^{+} \\ H^{+}
\end{pmatrix}.
\label{eq:ex:first}
\ee
Since $\delta \zeta_a$ was already generated, the renormalized parameter $\zeta_a$ can be rephased away through
$G^+ \to e^{-i \zeta_a} G^+$, 
$H^+ \to e^{i \zeta_a} H^+$,
$Z_{G^{+}H^{+}} = e^{2 i \zeta_a} Z_{G^{+}H^{+}}^{\prime}$,
$Z_{H^{+}G^{+}} = e^{-2 i \zeta_a} Z_{H^{+}G^{+}}^{\prime}$,
yielding:
\be
\begin{pmatrix}
\phi_{1(0)}^+ \\
\phi_{2(0)}^+
\end{pmatrix}
=
\begin{pmatrix}
i \delta \zeta_a + \delta Z_{G^{+}G^{+}} + 1 &
\delta Z_{G^{+}H^{+}}^{\prime} \\
\delta Z_{H^{+}G^{+}}^{\prime} &
-i \delta \zeta_a + \delta Z_{H^{+}H^{+}} + 1
\end{pmatrix}
\begin{pmatrix}
G^{+} \\ H^{+}
\end{pmatrix}.
\label{eq:ex:second}
\ee
This simple example illustrates several relevant points.
First, it shows that one must start with a general description including the tree-level parameter $\zeta_a$, for otherwise the counterterm $\delta \zeta_a$ will not be generated. 
Second, after $\delta \zeta_a$ was generated, the renormalized parameter $\zeta_a$ can be rephased away, and in such a way that  $\delta \zeta_a$ does not vanish.
Third, instead of starting by splitting $\zeta_{a(0)}$ into a non-zero $\zeta_a$ plus a counterterm (as in eq. \ref{eq:310}), and afterwards rephasing $\zeta_a$ away, we can assume beforehand that we will work in a basis for the renormalized parameters where $\zeta_a=0$, which lets us equate $\zeta_{a(0)}$ immediately with the counterterm only. This is what we did in eq. \ref{eq:my-dza} above.
%
%
%
%
%

\n Finally, what we obtained for the renormalized parameters looks just the same as that which we would have obtained at tree-level should we not have been aiming at renormalizing the theory.
As a matter of fact, if we were not concerned with the generation of counterterms, but rather in a mere tree-level description, we would not have needed $\zeta_a$, so that we could have rephased it away in eq. \ref{eq:ex:tree}. In that case, we would have obtained the same which we obtained in eq. \ref{eq:ex:second} for the renormalized terms (i.e. excluding counterterms).%
\fn{Care should be taken with the notion `the same', because in one case what is at stake are renormalized quantities in the context of the up-to-one-theory, whereas in the other case there are only tree-level quantities in a tree-level theory. Hence, they are formally different. Nonetheless, the physical description in both cases is equivalent.}
This also holds for the other parameters that were not rephased away for the sake of renormalization only. Therefore, the basis for the renormalized parameters (i.e. the basis that we can and shall use for the renormalized parameters) is considerably simpler than the general one used in section \ref{section:pot}, and is just the same as the one usually employed when considering the theory solely at tree-level---as in ref.~\cite{Fontes:2017zfn}. Comparing this reference with the general parameterization used in section \ref{section:pot}, we conclude that, for the renormalized parameters in the up-to-one-loop theory, all the fermion masses are real, and%
\fn{The reason for identifying the renormalized phases $\zeta_{1\mathrm{R}}$ and $\zeta_{2\mathrm{R}}$ with the index R is given in note \ref{note:thezetas}.}
\be
\zeta_{1\mathrm{R}} =
\zeta_{2\mathrm{R}} =
\zeta_a =
\alpha_4 =
\alpha_5 =
0,
\qquad
\alpha_0 = \beta.
\label{eq:key}
\ee
As a consequence, the relations one obtains for the renormalized parameters in the context of the up-to-one-loop theory are precisely those that were derived when considering the theory solely at tree-level. We checked this explicitly for different expressions: the minimum equations \cite{Fontes:2017zfn}, the relations for the $\lambda_i$ \cite{Fontes:2014xva}, as well as the relation for $m_{3\mathrm{R}}^2$ \cite{Fontes:2017zfn}. The latter reads:
\be
m_{3\mathrm{R}}^2 = \frac{m_1^2\, R_{13} (R_{12} \tan{\beta} - R_{11})
+ m_2^2\ R_{23} (R_{22} \tan{\beta} - R_{21})}{R_{33} (R_{31} - R_{32} \tan{\beta})}.
\label{m3_derived}
\ee
Here, the matrix $R$ is given by:
\be
R =
\left(
\begin{array}{ccc}
c_1 c_2 & s_1 c_2 & s_2\\
-(c_1 s_2 s_3 + s_1 c_3) & c_1 c_3 - s_1 s_2 s_3  & c_2 s_3\\
- c_1 s_2 c_3 + s_1 s_3 & -(c_1 s_3 + s_1 s_2 c_3) & c_2 c_3
\end{array}
\right),
\label{matrixR}
\ee
and is related to the matrix $Q$ in a simple way. Indeed, given eq. \ref{eq:key}, $Q$ becomes:
\be
Q
=
Q_{321} \, Q_{\beta},
\label{eq:Qsimple1}
\ee
with
\be
Q_{321}
=
Q_3 \, Q_2 \, Q_1
=
\begin{pmatrix}
& & &0\\
& R & &0\\
& & & 0\\
0&0&0& 1
\end{pmatrix},
\qquad
Q_{\beta}
=
\begin{pmatrix}
1&0&0&0\\
0&1&0&0\\
0&0&- s_{\beta} & c_{\beta}\\
0&0&c_{\beta} & s_{\beta}
\end{pmatrix}.
\label{eq:Qsimple2}
\ee
The original parameterization of $Q$ in eqs. \ref{eq:Q1} and \ref{eq:Q2} thus allows a simple connection between the matrix $Q$ for the renormalized parameters and the matrix $R$.
The matrix $X$ becomes orthogonal, and is simply:
\be
X
=
\begin{pmatrix}
c_{\beta} & s_{\beta} \\
- s_{\beta} & c_{\beta}
\end{pmatrix}.
\label{eq:X-reno}
\ee

\n Finally, the total independent renormalized parameters become:
\bs
\label{eq:reno-params}
\bea
\{p^{\text{Y}}_{\text{reno.}}\}
&=&
\{ 
m_{f}, V
\}.
\\
\label{eq:reno-Higgs}
\{p^{\mathrm{H}}_{\text{reno.}}\}
&=&
\{
e, \, m_{\mathrm{W}}, \, m_{\mathrm{Z}},
\, \alpha_1, \, \alpha_2, \, \alpha_3,
\, \beta, \, m_1, \, m_2, \, m_{\mathrm{H}^{+}}, \, \mu^2
\}.
\eea
\es
These parameters are precisely those which are usually taken as independent in the theory considered solely at tree-level \cite{Fontes:2017zfn}.
Moreover, the set of renormalized parameters of the Higgs sector, $\{p^{\mathrm{H}}_{\text{reno.}}\}$, is the same whichever the set of bare parameters $\{p^{\mathrm{H}}_{C_i(0)}\}$ taken as the independent (recall eqs. \ref{eq:bare-params-H}).
Actually, not only it is the same, but it contains less parameters than each of the sets $\{p^{\mathrm{H}}_{C_i(0)}\}$, since the renormalized versions of the last three independent bare parameters of each set were rephased away. 
On the other hand, for each bare independent parameter, there is an independent counterterm. 
Thus, whichever the combination of independent bare parameters chosen, there will be more independent counterterms than independent renormalized parameters.
In fact, identifying $\{\delta p^{\mathrm{H}}_{C_i}\}$ with the set of independent counterterms of the Higgs sector in the combination $C_i$, we have:
\be
\label{eq:CT-params-H}
\hs{-2mm}
\begin{blockarray}{c}
\\[-1.2mm]
\{\delta p^{\mathrm{H}}_{C_1}\}\\[1.5mm]
\{\delta p^{\mathrm{H}}_{C_2}\}\\[1.5mm]
\{\delta p^{\mathrm{H}}_{C_3}\}\\[1.5mm]
\{\delta p^{\mathrm{H}}_{C_4}\}
\end{blockarray}
\hspace{-3mm}
=
\hspace{-1mm}
\{\delta e, \delta m_{\mathrm{W}}, \delta m_{\mathrm{Z}},
 \delta \alpha_{1}, \delta \alpha_{2}, \delta \alpha_{3},
 \delta \beta, \delta m_{1}, \delta m_{2}, \delta m_{\mathrm{H}^{+}}, \delta \mu^2, \delta \zeta_{2}, \delta \zeta_{a},
\begin{blockarray}{c}
\\[-1.2mm]
\delta \alpha_{5}\},\\[1.5mm]
\delta \alpha_{4}\},\\[1.5mm]
\delta \alpha_{0}\},\\[1.5mm]
\delta \zeta_{1}\}.
\end{blockarray}
\ee
Something similar also happens in the Yukawa sector, as the phases of the renormalized fermion masses were rephased away. This means that, while the counterterms $\delta m_f$ are complex, the renormalized masses $m_f$ are real.
In summary, the presence of more independent counterterms than independent renormalized parameters is a consequence of CP violation, which forces the introduction of several parameters; these are required for the one-loop renormalization of the model, but their renormalized versions can be rephased away.

\section{Selection of the true vev}
\label{sec:FJTS-C2HDM}

\n When considering the theory up to one-loop level, it is convenient to assure that scalar fields have zero vacuum expectation value. Such assurance corresponds to the \textit{selection of the true vev}, and is discussed in detail in ref.~\cite{Fontes:PhD}.
%
%
In the present work, we perform that selection using the Fleischer-Jegerlehner tadpole scheme (FJTS)~\cite{Fleischer:1980ub,Denner:2016etu,Denner:2018opp}.%
\fn{Other relevant studies concerning the vev can be found e.g. in refs. \cite{Sperling:2013eva,Sperling:2013xqa}.}

\n We start by considering the bare version of eq. \ref{eq:doublets-first}:
\be
\Phi_{1(0)} =
\left(
\begin{array}{c}
\phi_{1(0)}^+\\
\tfrac{1}{\sqrt{2}} (\bar{v}_1 \, e^{i \bar{\zeta}_1}+ \rho_{1(0)} + i \eta_{1(0)})
\end{array}
\right),
\quad
\Phi_{2(0)} =
\left(
\begin{array}{c}
\phi_{2(0)}^+\\
\tfrac{1}{\sqrt{2}} (\bar{v}_2 \, e^{i \bar{\zeta}_2}+ \rho_{2(0)} + i \eta_{2(0)})
\end{array}
\right).
\label{eq:doublets-bare-first}
\ee
The quantities with a bar correspond to \textit{true} quantities, in the sense that they guarantee that no \textit{proper tadpoles} (i.e. 1-point GFs) show up in the up-to-one-loop theory. In the FJTS, they are split as follows:
\be
\bar{v}_1 = v_{1(0)} + \Delta v_1, 
\qquad
\bar{v}_2 = v_{2(0)} + \Delta v_2,
\qquad
\bar{\zeta}_1 = \zeta_{1(0)} + \Delta \zeta_1,
\qquad
\bar{\zeta}_2 = \zeta_{2(0)} + \Delta \zeta_2.
\label{eq:mybars}
\ee
Thus, each quantity with a bar is split into two terms: the bare one and an extra quantity. These extra quantities are identified in what follows as the $\Delta$ quantities and are responsible for the elimination of proper tadpoles (they are introduced with this purpose). To see this, it is convenient to write eq. \ref{eq:doublets-bare-first} in an alternative formulation:
\be
\Phi_{1(0)} =
\left(
\begin{array}{c}
\phi_{1(0)}^+\\
\tfrac{1}{\sqrt{2}} (\bar{v}_{\rho_1} + \rho_{1(0)} + i \bar{v}_{\eta_1} + i \eta_{1(0)})
\end{array}
\right),
\quad
\Phi_{2(0)} =
\left(
\begin{array}{c}
\phi_{2(0)}^+\\
\tfrac{1}{\sqrt{2}} (\bar{v}_{\rho_2} + \rho_{2(0)} + i \bar{v}_{\eta_2} + i \eta_{2(0)})
\end{array}
\right).
\label{eq:doublets-bare-second}
\ee
with $\bar{v}_{\rho_1}$, $\bar{v}_{\rho_2}$, $\bar{v}_{\eta_1}$, $\bar{v}_{\eta_2}$ real parameters, such that:
\be
\label{eq:mybars-new}
\begin{split}
& \bar{v}_{\rho_1} = v_{\rho_1(0)} + \Delta v_{\rho_1}, 
\qquad
\bar{v}_{\eta_1} = v_{\eta_1(0)} + \Delta v_{\eta_1}, 
\\
& \bar{v}_{\rho_2} = v_{\rho_2(0)} + \Delta v_{\rho_2}, 
\qquad
\bar{v}_{\eta_2} = v_{\eta_2(0)} + \Delta v_{\eta_2}.
\end{split}
\ee
Comparing with eqs. \ref{eq:doublets-bare-first} and \ref{eq:mybars}, we obtain:
\be
\begin{split}
&\Delta v_{\rho_1} = \Delta v_1 \, \cos \zeta_{1(0)} - v_{1(0)} \, \Delta \zeta_1 \sin \zeta_{1(0)},
\hspace{7mm}
\Delta v_{\eta_1} = \Delta v_1 \, \sin \zeta_{1(0)} + v_{1(0)} \, \Delta \zeta_1 \, \cos \zeta_{1(0)},
\\
&\Delta v_{\rho_2} = \Delta v_2 \, \cos \zeta_{2(0)} - v_{2(0)} \, \Delta \zeta_2 \sin \zeta_{2(0)},
\hspace{7mm}
\Delta v_{\eta_2} = \Delta v_2 \, \sin \zeta_{2(0)} + v_{2(0)} \, \Delta \zeta_2 \, \cos \zeta_{2(0)}.
\end{split}
\ee
We neglected terms of order $\Delta^2$, since the $\Delta$ quantities are of one-loop order (as we are about to show). This also means that the bare quantities can be replaced by their renormalized versions.%
\fn{We shall use this argument in the equations that follow.}
Hence, using the fact that we choose a basis for the renormalized parameters such that the renormalized phases $\zeta_{1\mathrm{R}}$ and $\zeta_{2\mathrm{R}}$ vanish (eq. \ref{eq:key}), we get:%
\fn{\label{note:thezetas}%
The phases $\zeta_{1\mathrm{R}}$ and $\zeta_{2\mathrm{R}}$ are the renormalized phases, such that $\zeta_{1(0)} = \zeta_{1\mathrm{R}} + \delta \zeta_1$
and
$\zeta_{2(0)} = \zeta_{2\mathrm{R}} + \delta \zeta_2$.
We use the index $\mathrm{R}$ to distinguish them from the true phases, since the true quantities are usually represented with no index \cite{Denner:2018opp,Denner:2019vbn}. Note that, in the case described above, the quantities with bars ($\bar{v}_1, \bar{v}_2, \bar{\zeta}_1,  \bar{\zeta}_2$) are set equal to the true quantities; for more details, see ref. \cite{Fontes:PhD}.
Finally, note that we keep the bare vevs for convenience in eq. \ref{eq:auxFJTS}.}
\be
\begin{pmatrix}
\Delta v_{\rho_1} \\
\Delta v_{\rho_2} \\
\Delta v_{\eta_1} \\
\Delta v_{\eta_2}
\end{pmatrix}
=
\begin{pmatrix}
\Delta v_1 \\
\Delta v_2 \\
v_{1(0)} \, \Delta \zeta_1 \\
v_{2(0)} \, \Delta \zeta_2
\end{pmatrix}.
\label{eq:auxFJTS}
\ee
%

\n Now, when the Lagrangian is expanded with eq. \ref{eq:doublets-bare-second}, the $\Delta v_{\phi_{n}}$ of eq. \ref{eq:mybars-new} must assure that proper tadpoles vanish. In order for this to happen, the terms with $\Delta v_{\phi_{n,j}}$ that contribute to the 1-point function of a certain neutral scalar field $\phi_{n,i}$ must precisely cancel the one-loop 1-point function in that field. That is, if $T_{\phi_{n,i}}$ represents the one-loop tadpole for $\phi_{n,i}$ (i.e. the one-loop contribution to the 1-point function in $\phi_{n,i}$), we must have:
\be
T_{\phi_{n,i}} = - \sum_j
\dfrac{\partial^2 \mathcal{L}}{\partial \phi_{n,i} \, \partial \Delta v_{\phi_{n,j}}} \bigg|_{\phi=0, \, \Delta v = 0}
\Delta v_{\phi_{n,j}}
.
\label{eq:sacred}
\ee
Using the fact that every $\Delta v_{\phi_{n,j}}$ always shows up together with the corresponding field $\phi_{n,j}$ (cf. eqs. \ref{eq:doublets-bare-second} and \ref{eq:mybars-new}), we can write:
\be
T_{\phi_{n,i}} 
=
\sum_j \dfrac{\partial^2 V}{\partial \phi_{n,i} \, \partial \phi_{n,j}} \bigg|_{\phi=0, \, \Delta v = 0} \Delta v_{\phi_{n,j}} 
=
\sum_j ({\cal M}_n^{2})_{ij} \Delta v_{\phi_{n,j}} ,
\ee
where the mass matrix ${\cal M}_n^{2}$ obeys eq. \ref{eq:neutral-mass-diag}.
Then, going to the diagonal basis,
\be
\begin{pmatrix}
T_{h_1}\\
T_{h_2}\\
T_{h_3}\\
T_{G_0}
\end{pmatrix}
=
Q
\begin{pmatrix}
T_{\rho_1}\\
T_{\rho_2}\\
T_{\eta_1}\\
T_{\eta_2}
\end{pmatrix}
=
Q \,
{\cal M}_n^{2} \,
Q^T \,
Q
\begin{pmatrix}
\Delta v_{\rho_1}\\
\Delta v_{\rho_2}\\
\Delta v_{\eta_1}\\
\Delta v_{\eta_2}
\end{pmatrix}
=
{\cal D}_n^2
\begin{pmatrix}
\Delta v_{h_1}\\
\Delta v_{h_2}\\
\Delta v_{h_3}\\
\Delta v_{G_0}
\end{pmatrix},
\ee
which implies:
\be
\Delta v_{h_1} = \dfrac{T_{h_1}}{m_1^2},
\qquad
\Delta v_{h_2} = \dfrac{T_{h_2}}{m_2^2},
\qquad
\Delta v_{h_3} = \dfrac{T_{h_3}}{m_{3\mathrm{R}}^2},
\qquad
\Delta v_{G_0} = 0,
\label{eq:DelandTrel}
\ee
where we used the fact that $T_{G_0} = 0$, which is a consequence of the Goldstone theorem. Some comments are in order.

\n First, by obeying eq. \ref{eq:sacred}, the $\Delta$ quantities---in the physical basis, $\Delta v_{h_1}$, $\Delta v_{h_2}$, $\Delta v_{h_3}$, $\Delta v_{G_0}$---assure that no proper tadpoles (1-point GFs) show up in the theory. In other words, they guarantee that the true vev up to one-loop level is selected. But as they are contained inside the scalar doublets, they will contribute to other GFs of theory. For example, although no 1-point GF for $h_2$ remains in the theory (that is, no proper tadpole for $h_2$ remains), $\Delta v_{h_2}$ will contribute to some 2-point and 3-point functions; and given eq. \ref{eq:DelandTrel}, the one-loop tadpole $T_{h_2}$ will contribute to those GFs; we dub these one-loop tadpole structures contributing to GFs other than the 1-point GFs \textit{broad tadpoles}.
In this way, $\Delta v_{h_2}$ assures that no proper tadpole for $h_2$ exists, but introduces broad tadpoles in the theory. Broad tadpoles are thus a consequence of the selection of the true vev and must be considered for consistency.

\n Second, all broad tadpoles can be accounted for by considering all possible one-loop tadpole insertions in all possible GFs (as usual in the FJTS \cite{Denner:2016etu}). Indeed,
as suggested above, each renormalized scalar field will have the corresponding $\Delta$ quantity added to it: $\Delta v_{h_1} + h_1$, $\Delta v_{h_2} + h_2$, and so on. This means that, for every term in the Lagrangian with (say) $\Delta v_{h_2}$, there will be a term with $h_2$ instead. As a consequence, for every $n$-point function with a $\Delta v_{h_2}$ contribution, there is a $(n+1)$-point function with a $h_2$ field contribution instead; and with that field, a reducible diagram with a one-loop tadpole can be formed. But it is easy to see that such reducible diagram precisely equals the $n$-point function with $\Delta v_{h_2}$: one just needs to consider the definition of the one-loop tadpole for $h_2$,
\be
\begin{minipage}[h]{.40\textwidth}
\vspace{5mm}
\begin{picture}(0,42)
\begin{fmffile}{1616} 
\begin{fmfgraph*}(50,70) 
\fmfset{arrow_len}{3mm} 
\fmfset{arrow_ang}{20} 
\fmfleft{nJ1} 
\fmfright{nJ2}
\fmflabel{$h_2$}{nJ1}
\fmf{dashes,tension=1}{nJ1,nJ2} 
\fmfv{decor.shape=circle,decor.filled=hatched,decor.size=9thick}{nJ2}
\end{fmfgraph*} 
\end{fmffile} 
\end{picture}
\vspace{-5mm}
\end{minipage}
\hspace{-30mm}
i \, T_{h_2},
\label{eq:loop-tad}
\ee
and the zero-momentum $h_2$ propagator,
\be
\begin{minipage}[h]{.40\textwidth}
\vspace{5mm}
\begin{picture}(0,42)
\begin{fmffile}{1179}
\begin{fmfgraph*}(70,70)
\fmfset{arrow_len}{3mm}
\fmfset{arrow_ang}{20}
\fmfleft{nJ1}
\fmfright{nJ2}
\fmflabel{$h_2$}{nJ1}
\fmflabel{$h_2$}{nJ2}
\fmf{dashes,label=\small $\hspace{12mm} p=0$,label.side=left,tension=3,label.dist=2thick}{nJ1,nJ1nJ2}
\fmf{dashes,tension=3}{nJ1nJ2,nJ2}
\end{fmfgraph*}
\end{fmffile}
\end{picture}
\vspace{-5mm}
\end{minipage}
\hspace{-22mm}
 i \frac{1}{- m_2^2},
\label{eq:0mom}
\ee
and compare with eq. \ref{eq:DelandTrel}.%
\fn{We use the hatched circle (as in eq. \ref{eq:loop-tad}) to represent a generic one-particle irreducible non-renormalized one-loop GF.}
The same argument obviously applies to the remaining broad tadpoles. In sum, although broad tadpoles must be considered for consistency, in the FJTS they can be accounted for by including all possible one-loop tadpole insertions in all possible GFs (cf. also refs. \cite{Krause:2017mal,Krause:2016oke}).

\n Third, the alternative just proposed (calculating reducible diagrams with one-loop tadpoles instead of tracing down the terms with $\Delta$ showing up in the Lagrangian) will be the one adopted in the following. The reason is simply that, with softwares such as \FM, the calculation of reducible diagrams with one-loop tadpoles is trivial;%
\fn{For details, cf. appendix \ref{app:FM}.}
in contrast, the task of expanding the Lagrangian and tracing down all the terms with $\Delta$ is quite cumbersome; and since the two methods are equivalent, we opt for the former. It follows that all GFs that \textit{may} receive an insertion of a one-loop tadpole \textit{must} receive such insertion.
%
In general, then, non-renormalized 2-point functions will get contributions from two different types of terms:
\vspace{4mm}
\be
\begin{minipage}[h]{.40\textwidth}
\vspace{13mm}
\begin{picture}(0,42)
\begin{fmffile}{777} 
\begin{fmfgraph*}(60,80) 
\fmfset{arrow_len}{3mm}
\fmfset{arrow_ang}{20}
\fmfleft{nJ1}
\fmfright{nJ2}
\fmf{plain,tension=3}{nJ1,nJ1nJ2}
\fmf{plain,tension=3}{nJ1nJ2,nJ2}
\fmfv{decor.shape=circle,decor.filled=hatched,decor.size=9thick}{nJ1nJ2}
\end{fmfgraph*} 
\end{fmffile} 
\end{picture}
\end{minipage}
\hspace{-31mm}
+
\hspace{4mm}
\begin{minipage}[h]{.40\textwidth}
\vspace{-3mm}
\begin{picture}(0,100)
\begin{fmffile}{888}
\begin{fmfgraph*}(60,90)
\fmfset{arrow_len}{3mm}
\fmfset{arrow_ang}{20}
\fmfleft{nJ1}
\fmfright{nJ2}
\fmftop{nJ3}
\fmf{plain,tension=3}{nJ1,nJ1nJ2}
\fmf{dashes,tension=0.1}{nJ1nJ2,x}
\fmf{dashes,label=$h_i$,label.dist=4,label.side=right,tension=0.1}{x,y}
\fmf{dashes,tension=0.06}{y,nJ3}
\fmf{plain,tension=3}{nJ1nJ2,nJ2}
\fmfv{decor.shape=circle,decor.filled=70,decor.size=1thick}{x}
\fmfv{decor.shape=circle,decor.filled=70,decor.size=1thick}{y}
\fmfv{decor.shape=circle,decor.filled=hatched,decor.size=9thick}{nJ3}
\end{fmfgraph*}
\end{fmffile}
\end{picture}
\end{minipage}
\hspace{-35mm}
,
\vspace{-10mm}
\label{eq:FJTS-2p}
\ee
where the full lines represent any type of particle and the dashed lines represent $h_1$, $h_2$, $h_3$.%
\fn{We draw small black circles on the dashed lines to highlight the presence of a tree-level propagator.}
The same will happen for non-renormalized 3-point functions, which then include:
\vspace{3mm}
\be
\begin{minipage}[h]{.40\textwidth}
\vspace{9mm}
\begin{picture}(0,42)
\begin{fmffile}{777b} 
\begin{fmfgraph*}(70,70) 
\fmfset{arrow_len}{3mm}
\fmfset{arrow_ang}{20}
\fmfleft{nJ1}
\fmfright{nJ2,nJ4}
\fmf{plain,tension=3}{nJ1,nJ1nJ2}
\fmf{plain,tension=3}{nJ1nJ2,nJ2}
\fmf{plain,tension=3}{nJ1nJ2,nJ4}
\fmfv{decor.shape=circle,decor.filled=hatched,decor.size=9thick}{nJ1nJ2}
\end{fmfgraph*} 
\end{fmffile} 
\end{picture}
\end{minipage}
\hspace{-30mm}
+
\hspace{5mm}
\begin{minipage}[h]{.40\textwidth}
\vspace{-10mm}
\begin{picture}(0,100)
\begin{fmffile}{889}
\begin{fmfgraph*}(70,70)
\fmfset{arrow_len}{3mm}
\fmfset{arrow_ang}{20}
\fmfleft{nJ1} 
\fmfright{nJ2,nJ4} 
\fmftop{nJ3}
\fmf{plain,tension=25}{nJ1,nJ1nJ2nJ4J4} 
\fmf{plain,tension=25}{nJ2,nJ1nJ2nJ4J4} 
\fmf{plain,tension=25}{nJ4,nJ1nJ2nJ4J4}
\fmf{dashes,tension=0.1}{nJ1nJ2nJ4J4,x}
\fmf{dashes,tension=0.1,label=$h_i$,label.dist=3,label.side=left}{x,y}
\fmf{dashes,tension=0.08}{y,nJ3}
\fmfv{decor.shape=circle,decor.filled=70,decor.size=1thick}{x}
\fmfv{decor.shape=circle,decor.filled=70,decor.size=1thick}{y}
\fmfv{decor.shape=circle,decor.filled=hatched,decor.size=9thick}{nJ3}
\end{fmfgraph*}
\end{fmffile}
\end{picture}
\end{minipage}
\hspace{-35mm}
.
\ee
\vspace{-2mm}

\section{Counterterms}
\label{sec:calculation-CTs}

\n As mentioned in section \ref{sec:UTOL}, the renormalization of the theory is completed by fixing the counterterms. 
Independent counterterms are \textit{a priori} not fixed, and must be fixed through independent renormalization conditions.
%
%
%
%
Now, counterterms in general contain a divergent part and a finite part; whichever the renormalization condition used to fix a certain counterterm, its divergent part will always be the same. But the same is not true for the finite part, which in general depends on the prescription used. The different prescriptions that can be used to define the finite part of counterterms are called subtraction schemes.
In what follows, most counterterms will be fixed either through the \textit{modified minimal subtraction} ($\overline{\text{MS}}$)  scheme or through the \textit{on-shell subtraction} (OSS) scheme.
%
The former is simpler: one selects a process that depends on the counterterm to be calculated; then, the renormalization condition simply states that the terms of the renormalized one-loop amplitude proportional to $\Delta_{\epsilon}$ are zero. Here, $\Delta_{\epsilon}$ is 
\be
\Delta_{\epsilon} = \dfrac{2}{\epsilon} - \gamma_{\text{E}} + \ln 4 \pi,
\ee
where $\epsilon = 4-d$ ($d$ being the dimension) and $\gamma_{\text{E}}$ is the Euler-Mascheroni constant. The counterterm at stake will then be proportional to $\Delta_{\epsilon}$ only. When $\overline{\text{MS}}$ is used to calculate a parameter counterterm,
the corresponding renormalized parameter will have no clear physical meaning.
In contrast, OSS is defined precisely so that all renormalized parameters have a direct physical meaning \cite{Denner:2019vbn}; the renormalization conditions in this case are less direct, though, and are discussed in detail in appendix \ref{app:OSS}.
In what follows, we adopt the conventions therein defined.

\n In section \ref{sec:calc}, we fix all of the counterterms of the theory, proceeding by sectors.
Several counterterms are fixed through usual renormalization conditions (cf. e.g. ref. \cite{Denner:2019vbn}); in those cases, we omit the derivation, and present just the results.%
\fn{A detailed derivation of all the counterterms can be found in ref. \cite{Fontes:PhD}.}
The remaining cases will be considered in detail.
Finally, in section \ref{sec:gauge-dep}, we discuss the gauge dependence of the counterterms.

\subsection{Calculation of counterterms}
\label{sec:calc}

\subsubsection{Gauge masses and fields}
\label{sec:CT-gauge}

\n The counterterms for the gauge sector are determined through OSS. The results are as follows: for the charged sector \cite{Fontes:PhD,Denner:2019vbn},
\be
\delta m_{\mathrm{W}}^{2} = \widetilde{\operatorname{Re}} \, \Sigma_{\mathrm{T}}^{W^+W^-}\left(m_{\mathrm{W}}^{2}\right), 
\qquad
\delta Z_{W} = -\left.\widetilde{\operatorname{Re}} \frac{\partial \Sigma_{\mathrm{T}}^{W^+W^-}\left(k^{2}\right)}{\partial k^{2}}\right|_{k^{2}=m_{\mathrm{W}}^{2}},
\ee
while for the neutral sector \cite{Fontes:PhD,Denner:2019vbn},
\bs
\bea
\delta m_{\mathrm{Z}}^{2} &=& \widetilde{\operatorname{Re}} \, \Sigma_{\mathrm{T}}^{Z Z}\left(m_{\mathrm{Z}}^{2}\right),\\
\begin{pmatrix}
\delta Z_{A A} & \delta Z_{A Z}\\
\delta Z_{Z A} & \delta Z_{Z Z}
\end{pmatrix}
&=&
\begin{pmatrix}
-\left.\widetilde{\operatorname{Re}} \frac{\partial \Sigma_{\mathrm{T}}^{A A}\left(k^{2}\right)}{\partial k^{2}}\right|_{k^{2}=0}
&
-2 \widetilde{\operatorname{Re}} \frac{\Sigma_{\mathrm{T}}^{A Z}\left(m_{\mathrm{Z}}^{2}\right)}{m_{\mathrm{Z}}^{2}} \\
2 \frac{\Sigma_{\mathrm{T}}^{A Z}(0)}{m_{\mathrm{Z}}^{2}}
&
-\left.\widetilde{\operatorname{Re}} \frac{\partial \Sigma_{\mathrm{T}}^{Z Z}\left(k^{2}\right)}{\partial k^{2}}\right|_{k^{2}=m_{\mathrm{Z}}^{2}}
\end{pmatrix}.
\label{eq:my50b}
\eea
\es

\subsubsection{Fermion masses and fields}

\n The counterterms for fermion masses and fields are also fixed through OSS conditions---which, in this case, leave some freedom. We show in detail in appendix \ref{sec:apD} how to derive the counterterms. The results are:
\bs
\label{eq:reno-ferm}
\begin{flalign}
&\delta m_{f,i}^{\mathrm{L}}
=
\dfrac{1}{2} \widetilde{\operatorname{Re}} 
\left[
m_{f,i} \, \Sigma_{ii}^{f, \mathrm{L}}(m_{f,i}^2)
+ m_{f,i} \, \Sigma_{ii}^{f, \mathrm{R}}(m_{f,i}^2)
+ 2 \, \Sigma_{ii}^{f, \mathrm{l}}(m_{f,i}^2)
\right], \\[3mm]
&\delta m_{f,i}^{\mathrm{R}}
=
\dfrac{1}{2} \widetilde{\operatorname{Re}} 
\left[
m_{f,i} \, \Sigma_{ii}^{f, \mathrm{L}}(m_{f,i}^2)
+ m_{f,i} \, \Sigma_{ii}^{f, \mathrm{R}}(m_{f,i}^2)
+ 2 \, \Sigma_{ii}^{f, \mathrm{r}}(m_{f,i}^2)
\right], \\[5mm]
&\delta Z_{i i}^{f, \mathrm{L}}
=
-\widetilde{\operatorname{Re}} \, \Sigma_{i i}^{f, \mathrm{L}}\left(m_{f, i}^{2}\right) - m_{f, i} \frac{\partial}{\partial p^{2}} \widetilde{\operatorname{Re}}\Big[m_{f, i}\left(\Sigma_{i i}^{f, \mathrm{L}}(p^2)+\Sigma_{i i}^{f, \mathrm{R}}(p^2)\right) \nonumber \\[-2mm]
&\hs{80mm}+ \Sigma_{i i}^{f, \mathrm{l}}(p^2)+\Sigma_{i i}^{f, \mathrm{r}}(p^2)\Big]\Big|_{p^{2}=m_{f, i}^{2}}, \\
&\delta Z_{i i}^{f, \mathrm{R}}
=
-\widetilde{\operatorname{Re}} \, \Sigma_{i i}^{f, \mathrm{R}}\left(m_{f, i}^{2}\right) - m_{f, i} \frac{\partial}{\partial p^{2}} \widetilde{\operatorname{Re}}\Big[m_{f, i}\left(\Sigma_{i i}^{f, \mathrm{L}}(p^2)+\Sigma_{i i}^{f, \mathrm{R}}(p^2)\right) \nonumber \\[-2mm]
&\hs{80mm}+\Sigma_{i i}^{f, \mathrm{l}}(p^2)+\Sigma_{i i}^{f, \mathrm{r}}(p^2)\Big]\Big|_{p^{2}=m_{f, i}^{2}}, \\
&\delta Z_{i j}^{f, \mathrm{L}}
\stackrel{i \neq j}{=}
\frac{2}{m_{f, i}^{2}-m_{f, j}^{2}} \widetilde{\operatorname{Re}}\Big[m_{f, j}^{2} \Sigma_{i j}^{f, \mathrm{L}}\left(m_{f, j}^{2}\right)+m_{f, i} m_{f, j} \Sigma_{i j}^{f, \mathrm{R}}\left(m_{f, j}^{2}\right) \nonumber \\[-2mm]
&\hs{60mm}+m_{f, i} \Sigma_{i j}^{f, \mathrm{l}}\left(m_{f, j}^{2}\right)+m_{f, j} \Sigma_{i j}^{f, \mathrm{r}}\left(m_{f, j}^{2}\right)\Big],
\label{eq:reno-fermB-pre}
\\
&\delta Z_{i j}^{f, \mathrm{R}}
\stackrel{i \neq j}{=}
\frac{2}{m_{f, i}^{2}-m_{f, j}^{2}} \widetilde{\operatorname{Re}}\Big[m_{f, j}^{2} \Sigma_{i j}^{f, \mathrm{R}}\left(m_{f, j}^{2}\right)+m_{f, i} m_{f, j} \Sigma_{i j}^{f, \mathrm{L}}\left(m_{f, j}^{2}\right)
\nonumber \\[-2mm]
&\hs{60mm}+m_{f, j} \Sigma_{i j}^{f, \mathrm{l}}\left(m_{f, j}^{2}\right)+m_{f, i} \Sigma_{i j}^{f, \mathrm{r}}\left(m_{f, j}^{2}\right)\Big].
\label{eq:reno-fermB}
\end{flalign}
\es

\subsubsection{Scalar masses and fields}
\label{sec:CT-scalar}

\n The counterterms relative to the charged scalar sector are easily defined in the OSS scheme; the results are \cite{Fontes:PhD,Krause:2017mal}:
\bs
\bea
\hs{-5mm}
\delta m_{\mathrm{H}^{+}}^2 &=& \widetilde{\operatorname{Re}} \, \Sigma^{H^+H^-}(m_{\mathrm{H}^{+}}^2), \\
\hs{-5mm}
\begin{pmatrix}
\delta Z_{G^{+}G^{+}} & \delta Z_{G^{+}H^{+}}\\
\delta Z_{H^{+}G^{+}} & \delta Z_{H^{+}H^{+}}
\end{pmatrix}
&=&
\begin{pmatrix}
-\left.\widetilde{\operatorname{Re}} \, \frac{\partial \Sigma^{G^{+}G^{-}}\left(k^{2}\right)}{\partial k^{2}}\right|_{k^{2}=0}
&
- 2 \,\widetilde{\operatorname{Re}}  \frac{\Sigma^{H^{+}G^{-}}\left(m_{\mathrm{H}^{+}}^2\right)}{m_{\mathrm{H}^{+}}^2}
\\[4mm]
2 \, \widetilde{\operatorname{Re}}  \frac{\Sigma^{G^{+}H^{-}}\left(0\right)}{m_{\mathrm{H}^{+}}^2}
&
-\left.\widetilde{\operatorname{Re}} \, \frac{\partial \Sigma^{H^{+}H^{-}}\left(k^{2}\right)}{\partial k^{2}}\right|_{k^{2}=m_{\mathrm{H}^{+}}^2}
\end{pmatrix}.
\eea
\es
The case of the neutral fields is more subtle, and is considered in detail in appendix \ref{app:OSS}. The independent counterterms are:%
\fn{In the case $S_n = h_3$, $m_{S_n}^2$ represents $m_{3\mathrm{R}}^2$.}
\bs
\begin{gather}
\delta m_1^2 = \widetilde{\operatorname{Re}} \, \Sigma^{h_1h_1}(m_1^2), \qquad
\delta m_2^2 = \widetilde{\operatorname{Re}} \, \Sigma^{h_2h_2}(m_2^2),
\\[5mm]
\delta Z_{S_n S_n} = -\left.\widetilde{\operatorname{Re}} \, \frac{\partial \Sigma^{S_nS_n}\left(k^{2}\right)}{\partial k^{2}}\right|_{k^{2}=m_{S_n}^{2}},
\qquad
\delta Z_{S_n^{\prime} S_n} \stackrel{S_n^{\prime} \neq S_n}{=}  2 \, \frac{\widetilde{\operatorname{Re}}  \, \Sigma^{S_n S_n^{\prime}}\left(m_{S_n}^{2}\right)}{m_{S_n^{\prime}}^2 - m_{S_n}^2},
\label{eq:CT-scalar-neutral}
\end{gather}
\es
and the pole squared mass $m_{3\mathrm{P}}^2$ is given by:
\be
m_{3\mathrm{P}}^2 = m_{\mathrm{3R}}^2 - \widetilde{\operatorname{Re}} \, \Sigma^{h_3h_3}(m_{3\mathrm{R}}^2) + \delta m_3^2.
\ee
%


\subsubsection{Electric charge}
\label{sec:charge}

\n The counterterm for the electric charge is also fixed through OSS; the result is \cite{Fontes:PhD,Krause:2017mal}:
\be
\delta Z_{e} = -\frac{1}{2} \delta Z_{A A} + \frac{s_{\text{w}}}{c_{\text{w}}} \frac{1}{2} \delta Z_{Z A}=\left.\frac{1}{2} \frac{\partial \Sigma_{\mathrm{T}}^{A A}\left(k^{2}\right)}{\partial k^{2}}\right|_{k^{2}=0} + \frac{s_{\text{w}}}{c_{\text{w}}} \frac{\Sigma_{\mathrm{T}}^{A Z}(0)}{m_{\mathrm{Z}}^{2}}.
\label{eq:deltaZe}
\ee
%

\subsubsection{CKM matrix}
\label{sec:CKM}

\n For the CKM matrix, we follow the prescription originally presented in ref.~\cite{Denner:1990yz}. We fix the expression in the Feynman gauge, which will be justified in section \ref{sec:gauge-dep} below. We then have:
\be
\delta V_{i j}=\frac{1}{4} \left.\bigg[\left(\delta Z_{i k}^{u, \mathrm{L}}-\delta Z_{i k}^{u, \mathrm{L} \dagger}\right) V_{k j}-V_{i k}\left(\delta Z_{k j}^{d, \mathrm{L}}-\delta Z_{k j}^{d, \mathrm{L} \dagger}\right)\bigg]\right|_{\xi=1},
\ee
where $\xi=1$ represents the Feynman gauge.

\subsubsection{Mixing parameters}
\label{sec:RenoMixingParameters}

\n The renormalization of mixing parameters has been recurrently discussed in recent years \cite{Kanemura:2004mg, Kanemura:2015mxa, Krause:2016oke, Altenkamp:2017ldc, Denner:2016etu, Denner:2017vms, Denner:2018opp, Krause:2017mal}, especially in ref.~\cite{Denner:2018opp}.%
\fn{Mixing parameters usually reduce to mixing \textit{angles}. Yet, in the present model (and due to CP violation), there are not only mixing angles (like $\beta$), but also mixing phases (like $\zeta_a$). Hence, we promote the designation `mixing angles' to `mixing parameters'.}
In the latter, several methods for fixing counterterms for mixing parameters are proposed and analysed.
In the present work, we consider one of those methods, which involves fixing the independent counterterms for mixing parameters using symmetry relations. This method is simple and leads to well-behaved results, as shall be discussed below.
A description of the procedure, as well as the derivation of the expressions for the counterterms for mixing parameters of the C2HDM, can be found in appendix \ref{app:Sym}.
In this section, we present the results therein derived. For the charged sector,
\bs
\label{eq:rela-charged-real}
\bea
{\delta \beta}
&=& \dfrac{1}{4}
\operatorname{Re}
\left.\Big[\delta Z_{G^+H^+} - \delta Z_{H^+G^+}\Big]
\right|_{\xi=1},
\\
\delta \zeta_a
&=&
-\dfrac{1}{2} \cot(2 \beta) \operatorname{Im} \left[\delta Z_{G^+H^+} \right]\big|_{\xi=1}.
\eea
\es
For the neutral sector, the set of independent counterterms for mixing parameters depends on the combination $C_i$ (recall eq. \ref{eq:CT-params-H}). There are three counterterms which are independent in all the combinations:
\bs
\begin{flalign}
&
\delta \alpha_1
=
\dfrac{1}{4} \sec(\alpha_2) \left. \Big[  \cos(\alpha_3) \left( \delta Z_{h_1h_2} - \delta Z_{h_2h_1} \right)  +  \sin(\alpha_3) \left( \delta Z_{h_3h_1} -\delta Z_{h_1h_3} \right)  \Big] \right|_{\xi=1},
\\[3mm]
&
\delta \alpha_2
= 
\left.
\dfrac{1}{4}  \sin(\alpha_3) \Big[ \delta Z_{h_1h_2} - \delta Z_{h_2h_1} + \cot(\alpha_3) \left( \delta Z_{h_1h_3} - \delta Z_{h_3h_1} \right)  \Big] \right|_{\xi=1},
\\[3mm]
&
\delta \alpha_3
=
\dfrac{1}{4} \bigg[\delta Z_{h_2h_3} - \delta Z_{h_3h_2} -   \cos(\alpha_3) \tan(\alpha_2)  \left( \delta Z_{h_1h_2} - \delta Z_{h_2h_1} \right) \nonumber\\[-3mm]
& \hs{50mm} + \sin(\alpha_3) \tan(\alpha_2) \left( \delta Z_{h_1h_3} - \delta Z_{h_3h_1} \right)  \bigg] \bigg|_{\xi=1}.
\end{flalign}
\es
Then, the combinations $C_1$, $C_2$ and $C_3$ have $\delta \alpha_5$, $\delta \alpha_4$ and $\delta \alpha_0$ as independent, respectively, given by:%
\fn{The notation $\stackrel{C_i}{=}$ clarifies which combination $C_i$ the expression is applicable to. For example, eq. \ref{eq:indep-dalfa5} is only valid if $C_1$ is chosen; in other combinations, $\delta \alpha_5$ will be a dependent counterterm, so that eq. \ref{eq:indep-dalfa5} is not valid.
Finally, note that $\delta \alpha_5$, $\delta \alpha_4$ and $\delta \alpha_0$ are all dependent counterterms in the combination $C_4$.}
\bs
\label{eq:indep-dalfas}
\begin{flalign}
\label{eq:indep-dalfa5}
&
\delta \alpha_5
\stackrel{C_1}{=}
\dfrac{1}{4} \bigg[\delta Z_{h_2G_0} -\delta Z_{G_0h_2} + \tan(\alpha_3) \left( \delta Z_{G_0h_3} - \delta Z_{h_3G_0} \right)
\bigg]
\bigg|_{\xi=1}
,
\\[3mm]
&
\label{eq:indep-dalfa4}
\delta \alpha_4
\stackrel{C_2}{=}
\dfrac{1}{4} \bigg[\delta Z_{h_1G_0} -\delta Z_{G_0h_1} + \sec(\alpha_3) \tan(\alpha_2) \left( \delta Z_{G_0h_3} - \delta Z_{h_3G_0} \right)  \bigg]
\bigg|_{\xi=1},
\\[3mm]
&
\label{eq:indep-dalfa0}
\delta \alpha_0
\stackrel{C_3}{=}
\dfrac{1}{4} \sec(\alpha_2) \sec(\alpha_3)  \big( \delta Z_{G_0h_3} - \delta Z_{h_3G_0} \big)
\big|_{\xi=1}.
\end{flalign}
\es

\subsubsection{Remaining parameters}
\label{sec:remaining}

\n By now, all the field counterterms are determined; as for the independent parameter counterterms, considering eq. \ref{eq:CT-params-H} we see that we still need to fix $\delta \mu^2$ and $\delta \zeta_2$ (both required for all combinations $C_i$) and, in the specific case of the combination $C_4$, also $\delta \zeta_1$.
%
%
We calculate the latter and $\delta \mu^2$ in the $\overline{\text{MS}}$ scheme.
%
%
As described above, this requires selecting a process where the counterterm to be fixed intervenes.
\n For $\delta \mu^2$, we choose $h_3 \to H^+H^-$. Here, the total counterterm is such that:
\be
\vs{3mm}
\begin{minipage}[h]{.35\textwidth}
\begin{picture}(0,70)
\begin{fmffile}{h3HPHMCTb} 
\begin{fmfgraph*}(70,70) 
\fmfset{arrow_len}{3mm} 
\fmfset{arrow_ang}{20} 
\fmfleft{nJ1} 
\fmfright{nJ2,nJ4} 
\fmf{dashes,label=$h_3$,label.side=left,tension=3,label.dist=3thick}{nJ1,nJ2nJ4J2}
\fmf{scalar,label=$H^-$,label.side=right,tension=3}{nJ2nJ4J2,nJ2} 
\fmf{scalar,label=$H^+$,label.side=left,tension=3}{nJ2nJ4J2,nJ4} 
\fmfv{decor.shape=pentagram,decor.filled=full,decor.size=6thick}{nJ2nJ4J2}
\end{fmfgraph*} 
\end{fmffile}
\end{picture}
\end{minipage}
\hs{-30mm}
\ni 
\hs{1mm}
i \dfrac{e}{2 m_{\mathrm{W}} s_{\text{w}} c_{\beta} s_{\beta}}
\left(s_{\beta} R_{31} + c_{\beta} R_{32}\right) \delta \mu^2
.
\vspace{-2mm}
\ee
Then, by requiring the renormalized 3-point function to be such that the terms proportional to $\Delta_{\epsilon}$ are zero,
\vspace{2mm}
\be
i \hat{\Gamma}^{h_3 H^+ H^-}\Big|_{\Delta_{\epsilon}}
=
\hs{2mm}
\left.
\begin{bmatrix}
\begin{minipage}[h]{.35\textwidth}
\begin{picture}(0,70)
\begin{fmffile}{h3HPHM1L} 
\begin{fmfgraph*}(70,70) 
\fmfset{arrow_len}{3mm} 
\fmfset{arrow_ang}{20} 
\fmfleft{nJ1} 
\fmfright{nJ2,nJ4} 
\fmf{dashes,label=$h_3$,label.side=left,tension=3,label.dist=3thick}{nJ1,nJ2nJ4J2}
\fmf{scalar,label=$H^-$,label.dist=3,tension=3}{nJ2nJ4J2,nJ2} 
\fmf{scalar,label=$H^+$,label.dist=3,label.side=right,tension=3}{nJ2nJ4J2,nJ4} 
\fmfv{decor.shape=circle,decor.filled=hatched,decor.size=11thick}{nJ2nJ4J2}
\end{fmfgraph*} 
\end{fmffile}
\end{picture}
\end{minipage}
\hs{-23mm}
+
\begin{minipage}[h]{.35\textwidth}
\begin{picture}(0,70)
\begin{fmffile}{h3HPHM1LB} 
\begin{fmfgraph*}(70,70)
\fmfset{arrow_len}{3mm}
\fmfset{arrow_ang}{20}
\fmfleft{nJ1} 
\fmfright{nJ2,nJ4} 
\fmftop{nJ3}
\fmf{dashes,label=$h_3$,label.side=right,tension=35}{nJ1,nJ1nJ2nJ4J4} 
\fmf{scalar,label=$H^-$,label.side=left,label.dist=3,tension=25}{nJ1nJ2nJ4J4,nJ2} 
\fmf{scalar,label=$H^+$,label.side=right,label.dist=3,tension=25}{nJ1nJ2nJ4J4,nJ4}
\fmf{dashes,tension=0.1}{nJ1nJ2nJ4J4,x}
\fmf{dashes,label=$h_i$,label.dist=3,label.side=left,tension=0.1}{x,y}
\fmf{dashes,tension=0.08}{y,nJ3}
\fmfv{decor.shape=circle,decor.filled=70,decor.size=1thick}{x}
\fmfv{decor.shape=circle,decor.filled=70,decor.size=1thick}{y}
\fmfv{decor.shape=circle,decor.filled=hatched,decor.size=9thick}{nJ3}
\end{fmfgraph*}
\end{fmffile}
\end{picture}
\end{minipage}
\hs{-23mm}
+
\hs{1mm}
\begin{minipage}[h]{.35\textwidth}
\begin{picture}(0,70)
\begin{fmffile}{h3HPHMCT} 
\begin{fmfgraph*}(70,70) 
\fmfset{arrow_len}{3mm} 
\fmfset{arrow_ang}{20} 
\fmfleft{nJ1} 
\fmfright{nJ2,nJ4} 
\fmf{dashes,label=$h_3$,label.side=left,tension=4,label.dist=3thick}{nJ1,nJ2nJ4J2}
\fmf{scalar,label=$H^-$,label.side=right,label.dist=3,tension=3}{nJ2nJ4J2,nJ2} 
\fmf{scalar,label=$H^+$,label.side=left,label.dist=3,tension=3}{nJ2nJ4J2,nJ4} 
\fmfv{decor.shape=pentagram,decor.filled=full,decor.size=6thick}{nJ2nJ4J2}
\end{fmfgraph*} 
\end{fmffile}
\end{picture}
\end{minipage}
\hs{-28mm}
\hphantom{.}
\end{bmatrix}
\right|_{\Delta_{\epsilon}}
= 0 ,
\ee
%
%
%
%
we fix $\delta \mu^2$. Concerning $\delta \zeta_1$ in $C_4$, we choose $h_3 \to  \bar{\tau} \tau$. Knowing that:
\be
\vs{3mm}
\begin{minipage}[h]{.35\textwidth}
\begin{picture}(0,70)
\begin{fmffile}{h3tTCTb} 
\begin{fmfgraph*}(70,70) 
\fmfset{arrow_len}{3mm} 
\fmfset{arrow_ang}{20} 
\fmfleft{nJ1} 
\fmfright{nJ2,nJ4} 
\fmf{dashes,label=$h_3$,label.side=left,tension=3,label.dist=3thick}{nJ1,nJ2nJ4J2}
\fmf{fermion,label=$\tau$,label.side=right,tension=3}{nJ2nJ4J2,nJ2} 
\fmf{fermion,label=$\bar{\tau}$,tension=3}{nJ4,nJ2nJ4J2} 
\fmfv{decor.shape=pentagram,decor.filled=full,decor.size=6thick}{nJ2nJ4J2}
\end{fmfgraph*} 
\end{fmffile}
\end{picture}
\end{minipage}
\hs{-30mm}
\ni 
\hs{1mm}
i \, \delta \zeta_1 \left(\dfrac{e \, m_{\tau} \, R_{33} \tan \beta}{2 m_{\mathrm{W}} s_{\text{w}}} + ...\right),
\ee
where the ellipses represent terms proportional to $\gamma_5$, 
we fix $\delta \zeta_1$ in $C_4$ by imposing:
\be
\vs{3mm}
i \hat{\Gamma}^{h_3  \bar{\tau} \tau}\Big|_{\Delta_{\epsilon}}
=
\hs{2mm}
\left.
\begin{bmatrix}
\begin{minipage}[h]{.35\textwidth}
\begin{picture}(0,70)
\begin{fmffile}{h3tT1L} 
\begin{fmfgraph*}(70,70) 
\fmfset{arrow_len}{3mm} 
\fmfset{arrow_ang}{20} 
\fmfleft{nJ1} 
\fmfright{nJ2,nJ4} 
\fmf{dashes,label=$h_3$,label.side=left,tension=3,label.dist=3thick}{nJ1,nJ2nJ4J2}
\fmf{fermion,label=$\tau$,tension=3}{nJ2nJ4J2,nJ2} 
\fmf{fermion,label=$\bar{\tau}$,tension=3}{nJ4,nJ2nJ4J2} 
\fmfv{decor.shape=circle,decor.filled=hatched,decor.size=11thick}{nJ2nJ4J2}
\end{fmfgraph*} 
\end{fmffile}
\end{picture}
\end{minipage}
\hs{-28mm}
+
\hs{5mm}
\begin{minipage}[h]{.35\textwidth}
\begin{picture}(0,70)
\begin{fmffile}{h3tTCT} 
\begin{fmfgraph*}(70,70) 
\fmfset{arrow_len}{3mm} 
\fmfset{arrow_ang}{20} 
\fmfleft{nJ1} 
\fmfright{nJ2,nJ4} 
\fmf{dashes,label=$h_3$,label.side=left,tension=3,label.dist=3thick}{nJ1,nJ2nJ4J2}
\fmf{fermion,label=$\tau$,label.side=right,tension=3}{nJ2nJ4J2,nJ2} 
\fmf{fermion,label=$\bar{\tau}$,tension=3}{nJ4,nJ2nJ4J2} 
\fmfv{decor.shape=pentagram,decor.filled=full,decor.size=6thick}{nJ2nJ4J2}
\end{fmfgraph*} 
\end{fmffile}
\end{picture}
\end{minipage}
\hs{-28mm}
\hphantom{.}
\end{bmatrix}
\right|_{\Delta_{\epsilon}}
\stackrel{C_4}{=} 0 .
\label{eq:delta-zeta1}
\ee

\n As for $\delta \zeta_2$, we checked that its divergent part is related to that of $\delta \zeta_1$ via:%
\fn{The divergent part of $\delta \zeta_2$ can be easily calculated by requiring e.g. the process $h_3 \to \bar{u} u$ to be finite.}
\be
\delta \zeta_2 \big|_{\Delta_{\epsilon}} = -\cot^2(\beta) \,  \delta \zeta_1 \big|_{\Delta_{\epsilon}}.
\label{eq:rel-z1-z2-divs}
\ee
Then, similarly to what can be done with symmetry relations (cf. appendix \ref{app:Sym}), we fix $\delta \zeta_2$ as a whole by requiring that eq. \ref{eq:rel-z1-z2-divs} also holds for the finite parts. That is, we fix $\delta \zeta_2$ by imposing:%
\fn{Eq. \ref{eq:rel-z1-z2-divs} thus provides a simple prescription to fix $\delta \zeta_2$. This simplicity is the reason why we chose this counterterm as independent in all the four combinations discussed above.}
%
%
\be
\delta \zeta_2 = -\cot^2(\beta) \, \delta \zeta_1 \big|_{\xi=1}.
\label{eq:rel-z1-z2-fin}
\ee
Note that this condition applies to all four combinations $C_i$. Accordingly, $\delta \zeta_2$ will have a general finite part in the first three combinations, whereas in $C_4$ it will be proportional to $\Delta_{\epsilon}$ only.

\subsection{Gauge dependence}
\label{sec:gauge-dep}


\n \n Physical observables cannot depend on the gauge chosen, which implies that $S$-matrix elements must be gauge independent.
In a renormalized up-to-one-loop process, both the (non-renormalized) one-loop diagrams and the counterterms are in general gauge dependent.
Now, it can be shown that, if parameter counterterms are gauge independent, the gauge dependences of the one-loop diagrams and those of the field counterterms precisely cancel, thus rendering the $S$-matrix elements gauge independent \cite{Bohm:2001yx}.
Conversely, gauge dependent parameter counterterms in general lead to gauge dependent $S$-matrix elements. 
Motivated by this, several authors have insisted in recent years on the importance on finding gauge independent parameter counterterms \cite{Krause:2016oke,Krause:2017mal,Denner:2016etu,Kanemura:2017wtm}.

\n However, as noted in ref.~\cite{Denner:2018opp}, the circumstance that parameter counterterms are gauge dependent does not by itself ruin the gauge independence of $S$-matrix elements. In fact, all that is required for $S$-matrix elements to be gauge independent is that parameter counterterms are \textit{considered} gauge independent. In order to assure the gauge independence of $S$-matrix elements, indeed, it does not really matter whether the parameter counterterms are truly gauge independent or not; all what is required is that they are considered gauge independent, i.e. that they do not change when the gauge is changed \cite{Denner:2018opp,Bohm:2001yx}.%
\fn{We thank Ansgar Denner for clarifications on this point.
}
But this can be easily implemented by calculating a gauge dependent counterterm in a particular gauge. When calculating observables in other gauges, one must stick to the value of the parameter counterterm in the gauge it was calculated in. Then, the gauge dependence of field counterterms cancels the gauge dependence of the one-loop diagrams, and $S$-matrix elements will be gauge independent.
In face of this, there does not seem to be a clear advantage of true gauge independence in parameter counterterms over considered gauge independence. In both scenarios, $S$-matrix elements end up being gauge-independent, and both lead to correct relations between physical observables.%
\fn{It goes without saying that, when using considered gauge independent counterterms, one has to make sure that no inconsistencies are introduced (i.e. one must make sure not to change the gauge in the gauge dependent parameter counterterms). In this sense, truly gauge independent counterterms may be preferable.
It has been claimed that true gauge independence is a desirable property, as it allows a ``more physical interpretation'' of non-observable parameters \cite{Freitas:2002um}. However, a non-observable parameter is always non-physical anyway; the fact that it is truly gauge independent does not render it any more physical (in whatever physically meaningful way).}

\n Now, this procedure is obviously irrelevant when parameter counterterms are truly gauge independent. In the FJTS, this happens for parameter counterterms fixed in either the OSS or the $\overline{\text{MS}}$ schemes.%
\fn{For details, cf. ref. \cite{Fontes:PhD}.}
However, the method described in the previous paragraph is very convenient for the remaining independent parameter counterterms, which in this model (and according to our choices) consist of those for the CKM matrix elements, those for the independent mixing parameters, and $\delta \zeta_2$ (sections \ref{sec:CKM}, \ref{sec:RenoMixingParameters} and \ref{sec:remaining} above, respectively).
We thus applied the method in those cases, following the suggestion proposed in ref.~\cite{Denner:2018opp} of selecting the Feynman gauge as the gauge in which the counterterms are calculated, which avoids artificially large parameters.


\section{Results}
\label{sec:num-res-main}

\n We now study the behaviour of the different combinations $C_i$ introduced above. This agenda is quite novel in studies of renormalization, due to the mere existence of combinations. In fact, there is usually no such thing as different combinations of independent parameters, as the number of independent renormalized parameters usually equals the number of independent counterterms. The studies of the renormalization of models thus tend to be focused either on the choice of different subtraction schemes for the counterterms, or on the choice of different renormalization scales.%
\fn{Should there be counterterms fixed in a subtraction scheme that introduces a renormalization scale dependence.}

\n In our case, the fact that there are more independent counterterms than independent renormalized parameters means that, for the same set of the latter, there can be different combinations of the former. Our goal, then, is to compare those combinations, ascertaining how the choice of one or the other combination (i.e. the choice of one or the other set of independent counterterms) affects the predictions of the theory at NLO.%
\fn{In this paper, we restrict ourselves to the four combinations $C_i$ implied in eq. \ref{eq:CT-params-H} and, for a certain combination, to the subtraction schemes proposed in section \ref{sec:calculation-CTs}.}
%
To do that, we consider three specific NLO processes: the decays of $h_2$ to $ZZ$, $h_1 Z$ and $h_1 h_1$.
%

\n In the present section, we start by describing both the computational tools and the simulation procedure used to study these decays. Then, we discuss how the decays are influenced by the different combinations $C_i$. Finally, after presenting the expressions for the one-loop decay widths, we show numerical results that allow to compare the different combinations.

\subsection{Computational tools and simulation procedure}

The generation of Feynman rules for both the tree-level and the counterterm interactions, the drawing of the Feynman rules and diagrams, and the calculation of one-loop amplitudes, counterterms and one-loop decay widths were all performed with \FMT. These different tasks are discussed in detail in appendix \ref{app:FM}.
\FMTS was also used for two additional tasks.
The first was a confirmation that the model is renormalized; by considering processes of the different sectors, we numerically checked that all the counterterms calculated in section \ref{sec:calc} lead to finite results.
The second task was a conversion of the results to \t{\ts{Fortran}}, which were then numerically evaluated using \ts{LoopTools}~\cite{Hahn:1998yk}.

\n For the scatter plots of section \ref{sec:num-res} below, we generated points in the parameter space of the C2HDM, assuming that $h_1$ corresponds to the SM-like Higgs boson, such that $m_1 = 125 \, \, \textrm{GeV}$.
We imposed both theoretical and experimental restrictions on the points.
From the theory side, we required them to comply with
boundeness from below \cite{Kanemura:1993hm},
perturbative unitarity \cite{Akeroyd:2000wc,Ginzburg:2003fe},
vacuum globality \cite{Ivanov:2015nea}
and the oblique parameters $S$, $T$ and $U$ \cite{Branco:2011iw}.
As for the experimental restrictions, we required points to have a $2 \sigma$ compatibility with the results coming from different experiments.
One of them is $B \to X_s \gamma$ \cite{Deschamps:2009rh,Mahmoudi:2009zx,Hermann:2012fc,Misiak:2015xwa,Misiak:2017bgg},
which forces $m_{\textrm{H}^\pm} > 580 \mbox{ GeV}$ (for the Type II model); the constraints from $R_b$ are also included \cite{Haber:1999zh,Deschamps:2009rh};
concerning the different signals for the SM-like Higgs boson, we required compatibility with the fits presented in ref.
\cite{Aad:2019mbh}, whereas exclusion bounds from additional Higgs searches are taken into account via {\tt HiggsBounds5} \cite{Bechtle:2020pkv}.
We also considered experiments that restrict the amount of CP violation of the C2HDM; we used the most stringent limit on the electron EDM, $|d_e| < 1.1 \times 10^{-29} e \mbox{ cm}$
at $90\%$ confidence level, provided by the ACME collaboration \cite{Andreev:2018ayy}.
The Higgs boson production cross sections were calculated with {\tt SusHiv1.6.0} \cite{Harlander:2012pb,Harlander:2016hcx}; for the ratio of rates, we followed the procedure described in ref. \cite{Fontes:2017zfn}.
Finally, we allowed the not-fixed input parameters from eq. \ref{eq:reno-params} to vary according to:
\begin{gather}
- \frac{\pi}{2} \le \alpha_{1,2,3} < \frac{\pi}{2},
\qquad
0.8 \le \tan \beta  \le 35,
\qquad
30\mbox{ GeV } \leq m_2 < 1 \mbox{ TeV },
\no
580 \mbox{ GeV } \le m_{H^\pm} < 1 \mbox{ TeV },
\qquad
0 \mbox{ GeV}^2 \le m_{12 \mathrm{R}}^2  < 500 000 \mbox{ GeV}^2.
\end{gather}
%


\subsection{The influence of the different combinations}
\label{sec:influ}

\n The renormalized NLO amplitude for the process $j$ (with $j = \{ h_2 \to ZZ, h_2 \to h_1Z, h_2 \to h_1h_1\}$),
which we represent as $\hat{\mathcal{M}}_j$, can generically be written as:
\be
\hat{\mathcal{M}}_j = \mathcal{M}^{\mathrm{tree}}_j + \hat{\mathcal{M}}^{\mathrm{loop}}_j,
\ee
where $\mathcal{M}^{\mathrm{tree}}_j$ and $\hat{\mathcal{M}}^{\mathrm{loop}}_j$ respectively represent the tree-level amplitude and the renormalized one-loop amplitude, the latter of which can in turn be split according to:
\be
\hat{\mathcal{M}}^{\mathrm{loop}}_j =
 \mathcal{M}^{\mathrm{loop}}_j + \mathcal{M}^{\mathrm{CT}}_j,
\ee
where the terms on the right-hand side represent the non-renormalized one-loop amplitude and the counterterms, respectively.
Here, $\mathcal{M}^{\mathrm{CT}}_j$ includes all the counterterms that, after expanding the bare quantities into renormalized plus counterterm, end up contributing to the process $j$. Now, the set of counterterms that constitute $\mathcal{M}^{\mathrm{CT}}_j$ is just a consequence of such an expansion, and is the same for all the combinations $C_i$. The differences between the combinations start
when the counterterms are to be given an expression,
since the latter in general depend on the combination. For example, suppose that $\delta \zeta_1$ contributes to $\mathcal{M}^{\mathrm{CT}}_j$; recalling eq. \ref{eq:CT-params-H}, it is clear that the expression for this counterterm depends on $C_i$. Indeed, in $C_4$, $\delta \zeta_1$ is an independent counterterm, so that it is determined by eq. \ref{eq:delta-zeta1}. In the remaining combinations, in contrast, $\delta \zeta_1$ is dependent, which means that it is determined by a function (which in general depends on the particular combination at stake) of independent counterterms. As a consequence, $\mathcal{M}^{\mathrm{CT}}_j$ will in general have different values according to the combination $C_i$ chosen. In this way, the different $C_i$ in general lead to different predictions for the renormalized NLO amplitudes.

\n At this point, it is worth noting that, in a renormalized function such that all its counterterms are fixed through OSS or other momentum-dependent subtraction schemes, the dependence on the renormalization scale $\mu_{\mathrm{R}}$ drops out.%
\fn{A simple way to see this is to note that, in a one-loop integral calculated through dimensional regularization, the UV divergent term ($\frac{2}{\epsilon}$) and the renormalization scale dependent term ($\ln\mu_{\mathrm{R}}^2$) always come together ($\frac{2}{\epsilon} + \ln\mu_{\mathrm{R}}^2$). Now, a momentum-dependent subtraction scheme is such that the counterterms are defined as a function of a one-loop integral evaluated at a specific momentum. Thus, since neither $\frac{2}{\epsilon}$ nor $\ln\mu_{\mathrm{R}}^2$ depend on the momentum, the counterterms will contain both. And just as they (i.e. the counterterms) assure that the renormalized function is UV finite (i.e. assure that the $\frac{2}{\epsilon}$ of the counterterms cancel against that of the original one-loop integral), they also assure that it does not depend on the renormalization scale (i.e. the $\ln\mu_{\mathrm{R}}^2$ cancel).
A final note: the renormalization scale $\mu_{\mathrm{R}}$ should not be confused with the parameter $\mu^2$ introduced in eq. \ref{eq:mu}.
}
Conversely, if there is at least a counterterm which is fixed in $\overline{\text{MS}}$, such counterterm will not cancel the dependence of the non-renormalized function on the renormalization scale, so that the renormalized function at stake will depend on $\mu_{\mathrm{R}}$.
In section \ref{sec:calc}, all independent counterterms were fixed through a momentum-dependent subtraction scheme, except for $\delta \mu^2$ and (whenever it is independent) $\delta \zeta_1$.%
\fn{Whenever $\delta \zeta_1$ is fixed in $\overline{\text{MS}}$, $\delta \zeta_2$ will also be (cf. eq. \ref{eq:rel-z1-z2-fin}).} 
These exceptions will be quite relevant for the impact of the different combinations on the three processes $j$ introduced above.

\n In order to better illustrate such impact, we show in table \ref{tab:CTs-per-process} all the independent counterterms contributing to the different $\mathcal{M}^{\mathrm{CT}}_{j}$, for the four combinations $C_i$.%
\fn{These counterterms are fixed according to the prescriptions presented in section \ref{sec:calc}. We should mention that $\delta \zeta_2$ contributes to all processes, in all combinations; and since it is an independent counterterm, it should in principle be included in table \ref{tab:CTs-per-process}. However, we fixed it by reference to $\delta \zeta_1$ (recall eq. \ref{eq:rel-z1-z2-fin}), which is only independent in $C_4$. In the remaining combinations, the counterterms which $\delta \zeta_1$ depends on are already accounted for in the table. Hence, for the sake of clarity, we do not show $\delta \zeta_2$.}
\begin{table}[!h]%
\begin{normalsize}
\normalsize
\begin{center}
\begin{tabular}
{@{\hspace{0.1mm}}
>{\centering\arraybackslash}m{1.8cm} >{\centering\arraybackslash}m{8.7cm}
>{\centering\arraybackslash}m{0.4cm}
>{\centering\arraybackslash}m{0.4cm}
>{\centering\arraybackslash}m{0.4cm}
>{\centering\arraybackslash}m{0.4cm}
@{\hspace{2mm}}}
\hlinewd{1.1pt}
Process $j$ & Common to all the $C_i$ & $C_1$ & $C_2$ & $C_3$ & $C_4$ \\
\hline\\[-4mm]
{\vspace{3mm}
$h_2 \to ZZ$} &
\multirow{2}{9.3cm}{
\centering
$
{\delta Z_e},  \,  {\delta m_{\mathrm{W}}^2},  \,  {\delta m_{\mathrm{Z}}^2},  \,  {\delta \alpha_1},  \,  {\delta \alpha_2}, \, {\delta \alpha_3}, \, {\delta \beta},
$ \\[0.3mm]
$
{\delta Z_{ZZ}},  \,   {\delta Z_{h_1h_2}},  \,  {\delta Z_{h_2h_2}},  \,  {\delta Z_{h_3h_2}} 
$
}
& $\delta \alpha_5$ & $\delta \alpha_4$ & $\delta \alpha_0$ & $\delta \zeta_1$
\\[7mm]
{\vspace{3mm}
$h_2 \to h_1 Z$} &
\multirow{2}{9.3cm}{
\centering
$
{\delta Z_e},  \,  {\delta m_{\mathrm{W}}^2},  \,  {\delta m_{\mathrm{Z}}^2},  \,  {\delta \alpha_1},  \,  {\delta \alpha_2},  \,  {\delta \alpha_3}, \, {\delta \beta}, \, {\delta Z_{ZZ}},$ \\[0.3mm]
${\delta Z_{h_1h_1}},  \,  {\delta Z_{h_2h_2}},  \,  {\delta Z_{h_3h_1}},  \,  {\delta Z_{h_3h_2}},  \,  {\delta Z_{G_0h_1}},  \,  {\delta Z_{G_0h_2}} 
$
} & $\delta \alpha_5$ & $\delta \alpha_4$ & $\delta \alpha_0$
& $\delta \zeta_1$
\\[7mm]
{\vspace{3mm}
$h_2 \to h_1 h_1$} 
&
\multirow{3}{9.3cm}{
\centering
${\delta Z_e},  \,  {\delta m_{\mathrm{W}}^2},  \,  {\delta m_{\mathrm{Z}}^2},  \,  {\delta \alpha_1},  \,  {\delta \alpha_2},  \,  {\delta \alpha_3}, \, {\delta \beta}, \, {\delta m_1^2}, $\\[0.3mm]
$ {\delta m_2^2}, \,  {\delta \mu^2}, \, {\delta Z_{h_1h_1}},  \,  {\delta Z_{h_1h_2}},  \,  {\delta Z_{h_2h_1}}, \, {\delta Z_{h_2h_2}}, $\\[0.3mm]
${\delta Z_{h_3h_1}},  \,  {\delta Z_{h_3h_2}},  \,  {\delta Z_{G_0h_1}},  \,  {\delta Z_{G_0h_2}}$
}
& $\delta \alpha_5$ & $\delta \alpha_4$ & $\delta \alpha_0$ & $\delta \zeta_1$
\\[13mm]
\hlinewd{1.1pt}
\end{tabular}
\end{center}
\vspace{-5mm}
\end{normalsize}
\caption{Total set of independent counterterms contributing to $\mathcal{M}^{\mathrm{CT}}_{j}$ for the combinations $C_i$. For each process $j$, the counterterms are separated in two groups: those that are common to all the four combinations (second column), and those that are specific to each combination (last four columns).}
\label{tab:CTs-per-process}
\end{table}
\normalsize
Some comments are in order.
First, in all three processes, the complete set of independent counterterms always depends on the combination $C_i$ (note that the last four columns of the table reflect eq. \ref{eq:CT-params-H}).
Moreover, processes $h_1 \to ZZ$ and $h_2 \to h_1 Z$ should have no scale dependence in $C_1$, $C_2$ and $C_3$, since they do not involve any counterterm fixed in $\overline{\text{MS}}$ in those combinations. By contrast, the combination $C_4$ in all processes is expected to lead to scale dependent results, as it includes the $\overline{\text{MS}}$-fixed $\delta \zeta_1$.
Finally, process $h_2 \to h_1 h_1$ depends on $\delta \mu^2$, so that we expect scale dependence in all combinations of that process.

\subsection{Decay widths}
\label{sec:decays}

\n Using \FMT, the expressions for the renormalized NLO decay width $\Gamma^{\mathrm{NLO}}_j$ for the process $j$ can be easily calculated from the renormalized NLO amplitude $\hat{\mathcal{M}}_j$. We assume that all external particles are on-shell (OS). The final expressions are considerably clear, as long as they are written in terms of form factors.%
\fn{In what follows, we follow the conventions of \FMT; we omit the Feynman diagrams, as well as the expressions for the form factors.}

\n We start with the decay $h_2 \to ZZ$. Defining the momenta and Lorentz indices such that $h_2(p_1) \to Z^{\nu}(q_1) Z^{\sigma}(q_2)$, we have:
\bs
\label{eq:ff-h2ZZ}
\bea
\mathcal{M}_{h_2 \to ZZ}^{\mathrm{tree}}
= \varepsilon^*_{\nu}(q_1) \, \varepsilon^*_{\sigma}(q_2) && \hs{-1mm} f_3^{\mathrm{tree}} \, g^{\nu\sigma}, \\[2mm]
\hat{\mathcal{M}}_{h_2 \to ZZ}^{\mathrm{loop}}
= \varepsilon^*_{\nu}(q_1) \, \varepsilon^*_{\sigma}(q_2)
\bigg(
\hspace{-2mm}
&&
f_3 \, g^{\nu\sigma} +
f_6 \, p_1^{\nu} \, p_1^{\sigma} +
f_9 \, p_1^{\nu} \, q_1^{\sigma} +
f_{24} \, p_1^{\sigma} \, q_2^{\nu} +
f_{27} \, q_1^{\sigma} \, q_2^{\nu} \nonumber\\[-3mm]
&&
+ f_{33} \, p_1^{\omega} \, q_1^{\upsilon} \, \epsilon^{\nu\sigma\omega\upsilon} +
f_{15} \, p_1^{\sigma} \, q_1^{\nu} +
f_{18} \, q_1^{\nu} \, q_1^{\sigma} +
f_{21} \, q_1^{\nu} \, q_2^{\sigma}
\bigg), \hs{8mm}
\eea
\es
where the different $f_k$
correspond to form factors.%
\fn{We identify tree-level form factors with the superscript `tree'.
The form factor $f_{33}$ corresponds to the CP-violating component predicted in ref. \cite{Huang:2020zde}. It turns out that it does not contribute to $\Gamma^{\mathrm{NLO}}_{h_2 \to ZZ}$, as can be seen in eq. \ref{eq:GNLOh2ZZ}.}
These in general contain contributions from the counterterms, so that they will in general have different values in the different combinations.
The form factors $f_{15}$, $f_{18}$ and $f_{21}$ do not contribute, since the fact that the $Z$ bosons are OS implies $\varepsilon^*_{\nu}(q_1) \, q_1^{\nu} = \varepsilon^*_{\sigma}(q_2) \, q_2^{\sigma} = 0$.
The decay width is:
\bea
\label{eq:GNLOh2ZZ}
\Gamma^{\mathrm{NLO}}_{h_2 \to ZZ}
&=&
\dfrac{\sqrt{m_2^4 - 4 \, m_2^2 \, m_{\mathrm{Z}}^2}}{128\, \pi \, m_2^3 \, m_{\mathrm{Z}}^4} f_3^{\mathrm{tree}}
\bigg\{ \left( m_2^4 - 4 \, m_2^2 \, m_{\mathrm{Z}}^2 + 12 \, m_{\mathrm{Z}}^4 \right) \left( f_3^{\mathrm{tree}}  + 2 \, \mathrm{Re} [f_3] \right) \nonumber\\
&& \hs{17mm}
+ m_2^2 \,  \left( m_2^4 - 6 \, m_2^2 \, m_{\mathrm{Z}}^2 + 8 \, m_{\mathrm{Z}}^4 \right)  \, \operatorname{Re}[f_6 + f_9 + f_{24} + f_{27}] \bigg\}.
\eea
As for the decay $h_2 \to h_1 Z$, we define the momenta and Lorentz indices such that $h_2(p_1) \to h_1(q_1) Z^{\sigma}(q_2)$, so that
\bs
\label{eq:ff-h2h1Z}
\bea
\mathcal{M}_{h_2 \to h_1 Z}^{\mathrm{tree}}
&=&
\varepsilon^*_{\sigma}(q_2)
\Big(
f_{10}^{\mathrm{tree}} \, q_1^{\sigma} + f_{12}^{\mathrm{tree}} \, p_1^{\sigma}
\Big), \\
\hat{\mathcal{M}}_{h_2 \to h_1 Z}^{\mathrm{loop}}
&=& \varepsilon^*_{\sigma}(q_2)
\Big(
f_{10} \, q_1^{\sigma} +
f_{11} \, q_2^{\sigma} +
f_{12} \, p_1^{\sigma}
\Big),
\eea
\es
but, as the $Z$ boson is OS, $f_{11}$ does not contribute. The decay width is:
\bea
\Gamma^{\mathrm{NLO}}_{h_2 \to h_1 Z}
&=&
\dfrac{1}{64 \, \pi \, m_2^3 \, m_{\textrm{Z}}^2}
\Big(
m_1^4 + m_2^4 + m_{\textrm{Z}}^4 - 2 \, m_1^2 \, m_2^2 - 2 \, m_1^2 \, m_{\textrm{Z}}^2 - 2 \, m_2^2 \, m_{\textrm{Z}}^2
\Big)^{3/2} \nonumber\\[-2mm]
&& \hs{20mm} \times 
\bigg\{ 
\left|f_{10}^{\mathrm{tree}} + f_{12}^{\mathrm{tree}}\right|^2
+
2 \, \mathrm{Re}
\Big[
\left(f_{10}^{\mathrm{tree}} + f_{12}^{\mathrm{tree}}\right)
\left(f_{10}^* + f_{12}^*\right)
\Big]
\bigg\}.
\eea
Finally, the decay $h_2 \to h_1 h_1$ is such that the renormalized NLO amplitude is simply given by:
\be
\hat{\mathcal{M}}_{h_2 \to h_1 h_1}
=
\mathcal{M}_{h_2 \to h_1 h_1}^{\mathrm{tree}}
+ 
\hat{\mathcal{M}}_{h_2 \to h_1 h_1}^{\mathrm{loop}}
=
f_1^{\mathrm{tree}} + f_1,
\label{eq:ff-h2h1h1}
\ee
and the decay width is:
\be
\Gamma^{\mathrm{NLO}}_{h_2 \to h_1 h_1}
=
\dfrac{\sqrt{m_2^4 -4 \, m_1^2 \, m_2^2}}{32 \, \pi \, m_2^3} \, f_1^{\mathrm{tree}} (f_1^{\mathrm{tree}} + 2 \, \mathrm{Re}[f_1]).
\ee

\subsection{Numerical results}
\label{sec:num-res}

\n At last, we investigate the behaviour of the different combinations on the decay widths using numerical results.
First and foremost, we checked that all combinations in all processes lead to gauge independent decay widths.
This is in agreement with the discussion presented in section \ref{sec:gauge-dep}, and validates the simple prescription for the renormalization of mixing parameters proposed in section \ref{sec:RenoMixingParameters}.

\n We compare the different combinations by ascertaining how their results affect the numerical stability of the perturbative expansion. If perturbation theory is to be trusted, higher orders should contribute with smaller corrections. In particular, the NLO results should generate small corrections when compared to the LO ones. This leads us to define the quantity \cite{Krause:2016xku,Krause:2017mal}:
\be
\Delta \Gamma_j \equiv \dfrac{\Gamma^{\mathrm{NLO}}_j - \Gamma^{\mathrm{LO}}_j}{\Gamma^{\mathrm{LO}}_j},
\label{eq:def-Gamma}
\ee
where $\Gamma^{\mathrm{LO}}_j$ represents the decay width at LO for the process $j$.

\begin{figure}[htb]
\centering
\includegraphics[width=0.55\textwidth]{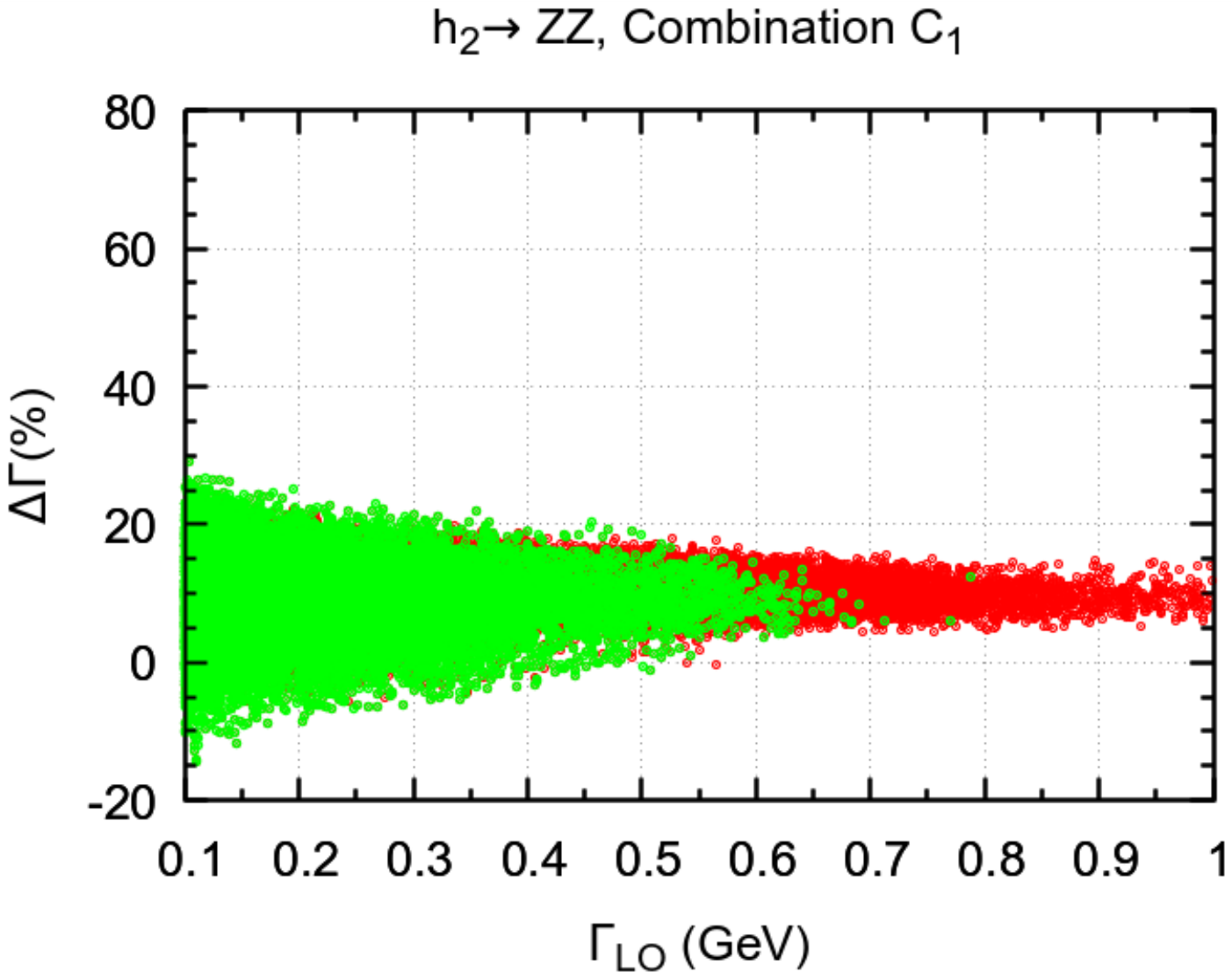}
\caption{
$\Delta \Gamma_{h_2 \to ZZ}$ in percentage as a function of $\Gamma^{\mathrm{LO}}_{h_2 \to ZZ}$, for the combination $C_1$.
Only the interval $ 0.1 \, \, \textrm{GeV} < \Gamma^{\mathrm{LO}} < 1 \, \, \textrm{GeV}$ is shown. 
In red/dark-gray, points passing all constraints except \texttt{HiggBounds-5}; in green/gray, points passing all constraints.}
\label{fig:h2ZZ-C1}
\end{figure}

\n In figure \ref{fig:h2ZZ-C1}, we show $\Delta \Gamma_{h_2 \to ZZ}$ in percentage as a function of $\Gamma^{\mathrm{LO}}_{h_2 \to ZZ}$ for the combination $C_1$.
Several aspects should be stressed here.
First if all, we checked numerically that the results are invariant under the renormalization scale, as expected from the discussion of section \ref{sec:influ}. 
Besides, the plot shows that \texttt{HiggBounds-5} hinders points with large values of $\Gamma^{\mathrm{LO}}$;
this is consistent with the fact that, had $h_2 \to ZZ$ a large value of $\Gamma^{\mathrm{LO}}$, $h_2$ would probably already have been discovered by now.
The majority of points passing all constraints thus obey $\Gamma^{\mathrm{LO}} < 0.7 \, \, \textrm{GeV}$; points with $\Gamma^{\mathrm{LO}}$ almost up to $0.8 \, \, \textrm{GeV}$ are still allowed, albeit very scarcely (a stringent fine tuning is required).
All in all, the points describe a smooth and well-behaved pattern; in particular, there is no region of parameter space generating singularities in $\Delta \Gamma$.
Finally, although there are points with small $\Delta \Gamma$ for all allowed values of $\Gamma^{\mathrm{LO}}$---indicating that, for those points, perturbation theory is valid---, larger values of $\Delta \Gamma$ are also possible for smaller values of  $\Gamma^{\mathrm{LO}}$.%
\fn{From eq. \ref{eq:def-Gamma}, larger values of $\Delta \Gamma$ are expected for smaller values of $\Gamma^{\mathrm{LO}}$.}
For those points, one would in principle need to calculate the following order in perturbation theory to check whether or not a perturbative description is possible.

\begin{figure}[htb]
\centering
\includegraphics[width=0.48\textwidth]{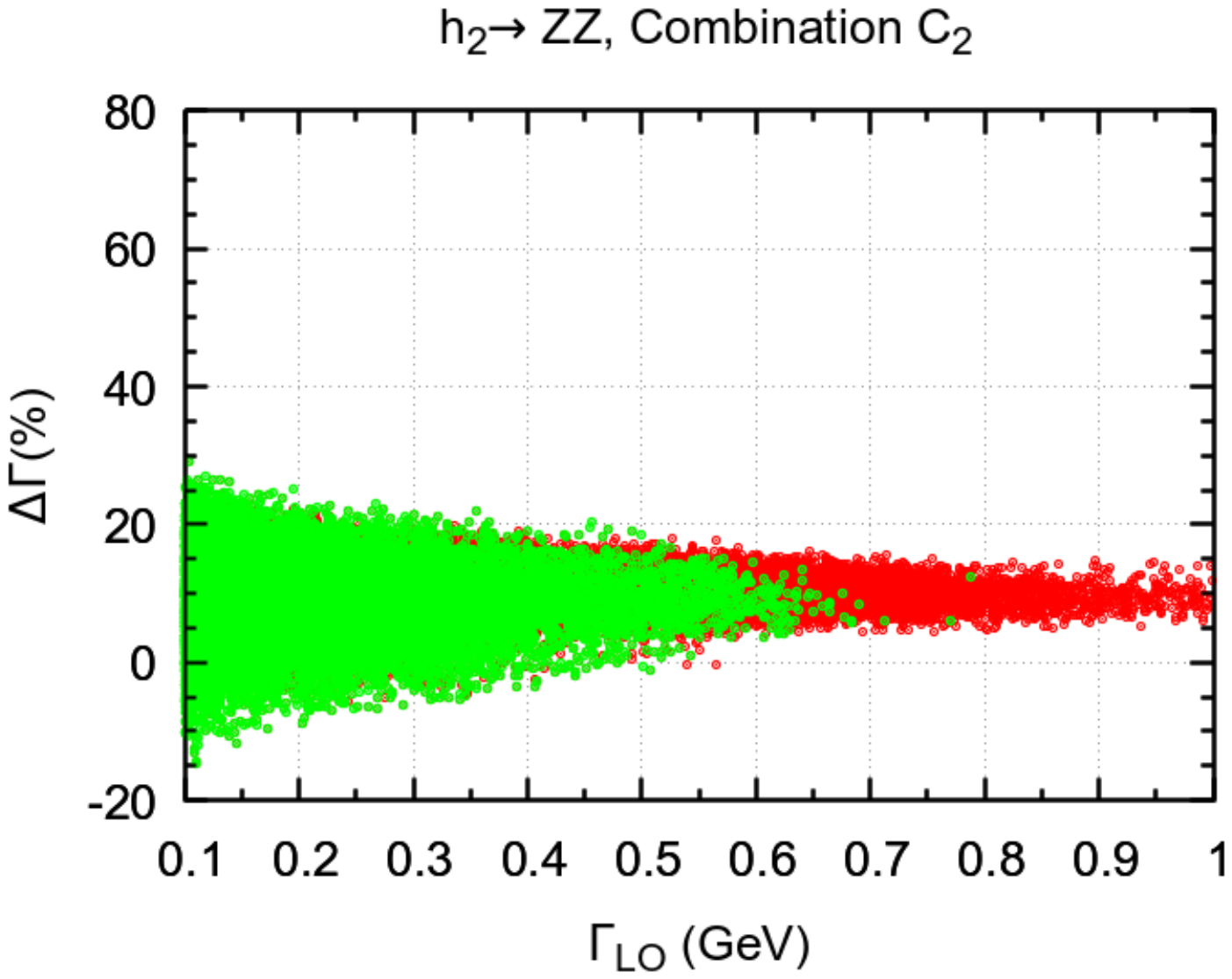}
\hfill
\includegraphics[width=0.48\textwidth]{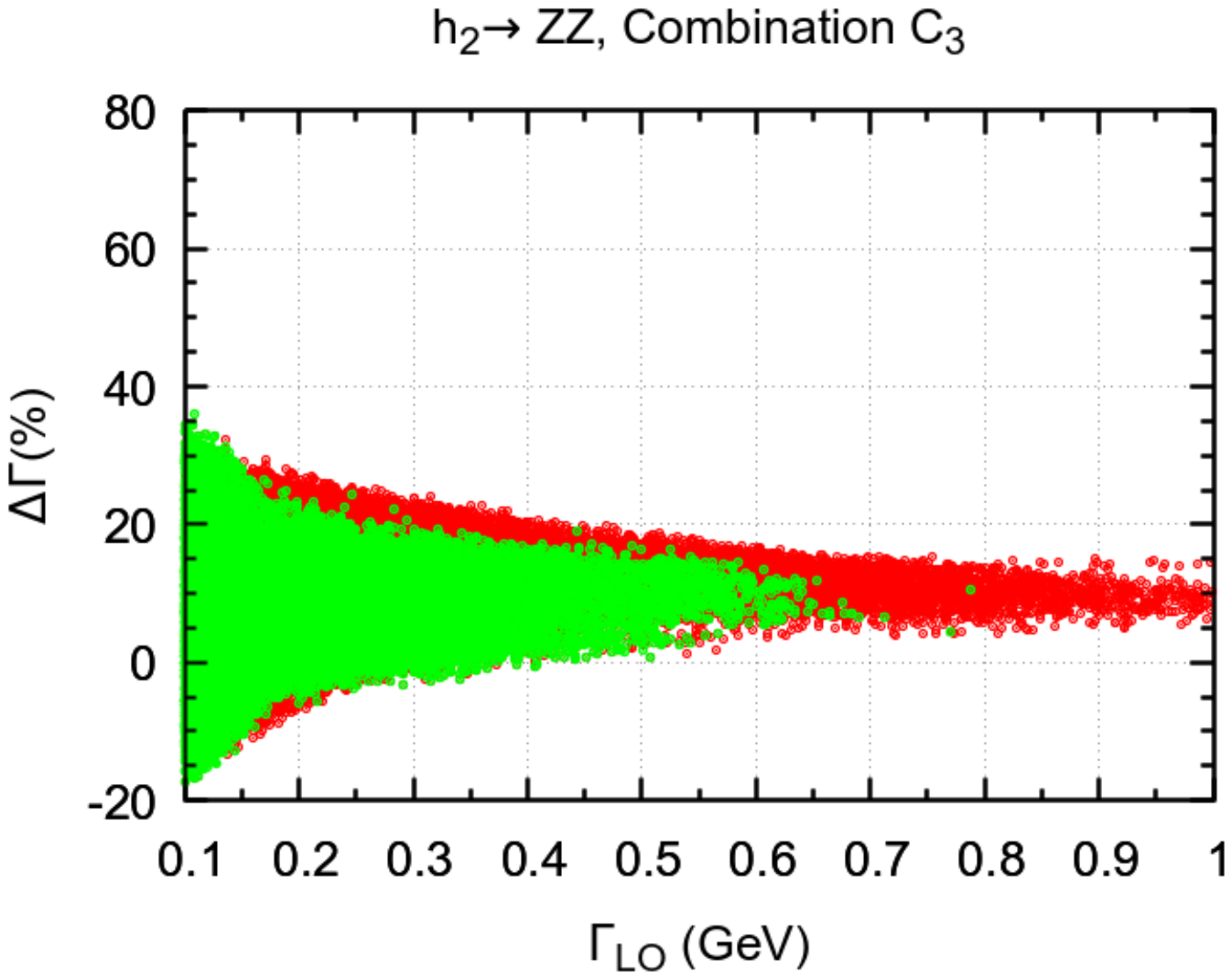}
\caption{
$\Delta \Gamma_{h_2 \to ZZ}$ in percentage as a function of $\Gamma^{\mathrm{LO}}_{h_2 \to ZZ}$, for the combinations $C_2$ (left) and $C_3$ (right).
Only the interval $ 0.1 \, \, \textrm{GeV} < \Gamma^{\mathrm{LO}} < 1 \, \, \textrm{GeV}$ is shown. 
The color conventions are those of figure \ref{fig:h2ZZ-C1}.}
\label{fig:h2ZZ-C2-C3}
\end{figure}

\n Similar to figure \ref{fig:h2ZZ-C1} are the two panels of figure \ref{fig:h2ZZ-C2-C3}, where we show equivalent results for the combinations $C_2$ and $C_3$.
We checked that, as before, there is no scale dependence.
The values of $\Delta \Gamma_{h_2 \to ZZ}$ are essentially the same as in figure \ref{fig:h2ZZ-C1}; moreover, we find the same smooth pattern in the points. This allows us to attest the quality of our prescription for the determination of the counterterms for the mixing parameters, proposed in section \ref{sec:RenoMixingParameters}.%
\fn{Except for $\delta \zeta_a$, which we do not investigate in this paper.}
Indeed, from table \ref{tab:CTs-per-process}, the combinations $C_1$, $C_2$ and $C_3$ for the process $h_2 \to ZZ$, besides depending all on $\delta \beta$, $\delta \alpha_1$, $\delta \alpha_2$ and $\delta \alpha_3$, depend individually on $\delta \alpha_5$, $\delta \alpha_4$ and $\delta \alpha_0$, respectively. Then, by considering figures \ref{fig:h2ZZ-C1} and \ref{fig:h2ZZ-C2-C3}, we conclude that all these counterterms lead to well-behaved results. Actually, not only are the results all well-behaved, but they are also very similar. 
This is consistent with the circumstance that $\delta \alpha_5$, $\delta \alpha_4$ and $\delta \alpha_0$ were all determined via symmetry relations, so that one would expect similar behaviours from the different lines of eqs. \ref{eq:indep-dalfas}.
%
All in all, then,
the combinations $C_1$, $C_2$ and $C_3$ are essentially equivalent for the process $h_2 \to ZZ$.

\begin{figure}[htb]
\centering
\includegraphics[width=0.55\textwidth]{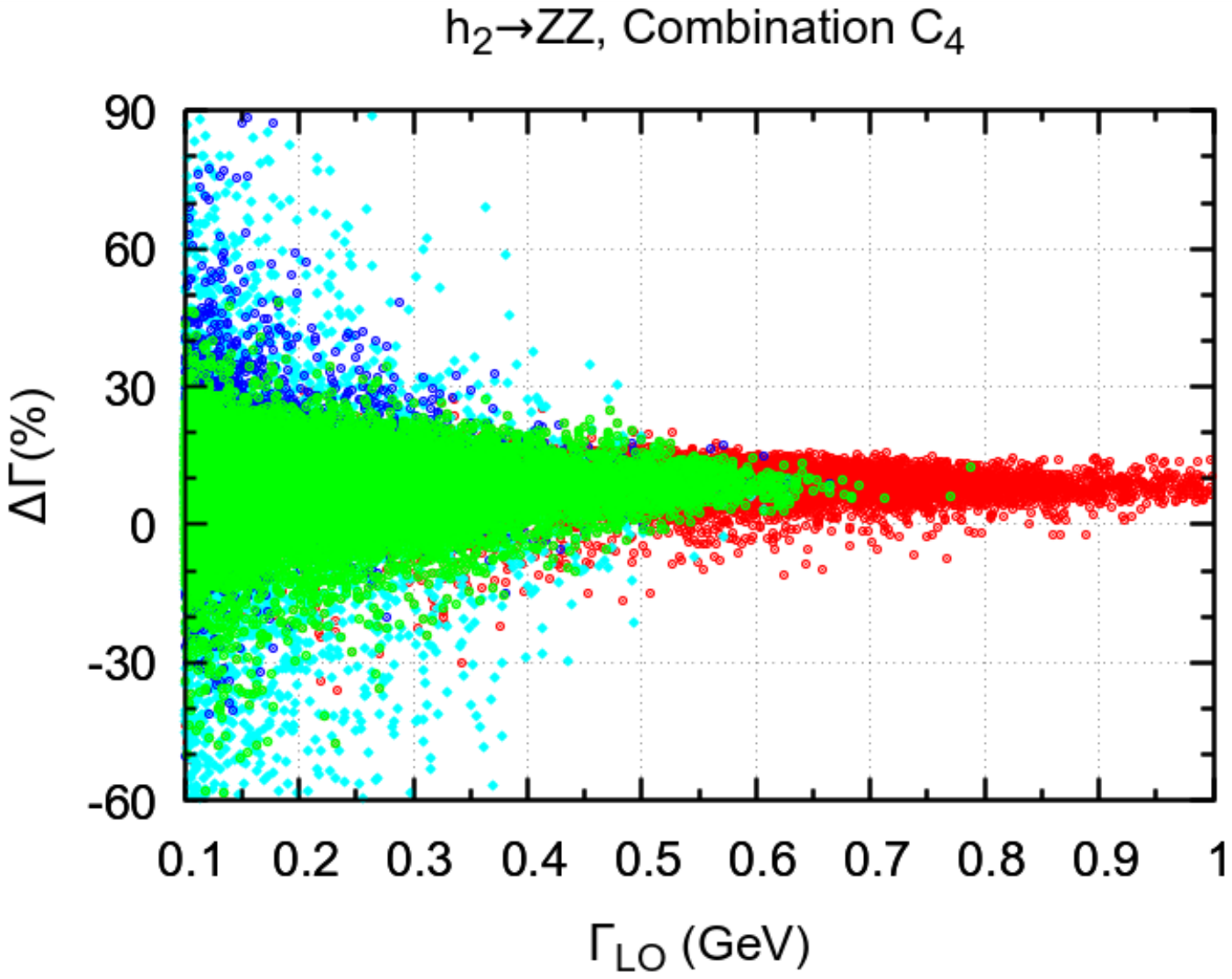}
\caption{
$\Delta \Gamma_{h_2 \to ZZ}$ in percentage as a function of $\Gamma^{\mathrm{LO}}_{h_2 \to ZZ}$, for the combination $C_4$.
Only the interval $ 0.1 \, \, \textrm{GeV} < \Gamma^{\mathrm{LO}} < 1 \, \, \textrm{GeV}$ is shown. 
In red/dark-gray, points with $\mu_{\mathrm{R}}= 350$ GeV passing all constraints except \texttt{HiggBounds-5}. The remaining colors represent different values of the renormalization scale for points passing all the constraints: in cyan/light-gray, $\mu_{\mathrm{R}} = 125$ GeV; in green/gray, $\mu_{\mathrm{R}}= 350$ GeV; in blue/black, $\mu_{\mathrm{R}}= 700$ GeV.}
\label{fig:h2ZZ-C4}
\end{figure}

\n The results for $C_4$ are shown in figure \ref{fig:h2ZZ-C4}.
Here, as expected from the discussion of section \ref{sec:influ}, the results depend on the renormalization scale $\mu_{\mathrm{R}}$. We plot three different scales, for points passing all constraints. It is clear that the lighter scale leads to a big dispersion in the values of $\Delta \Gamma$, and that the intermediate scale leads to the most stable results. 
For $\Gamma^{\mathrm{LO}}_{h_2 \to ZZ} \gtrsim 0.4$ GeV, these turn out to be quite similar to those of the points passing all constraints in the remaining combinations: the ranges of values of $\Delta \Gamma$ are essentially the same, and the same well-behaved pattern is observed. This could hardly be expected from the expressions for the different counterterms; not only is the specific counterterm in $C_4$ ($\delta \zeta_1$) fixed through a method completely different from the one used to fix the specific counterterms of the combinations $C_1$, $C_2$ and $C_3$ ($\delta \alpha_5$, $\delta \alpha_4$ and $\delta \alpha_0$, respectively), but the role of counterterms is also different: $\delta \zeta_1$ is a counterterm for a phase of the potential, whereas $\delta \alpha_5$, $\delta \alpha_4$ and $\delta \alpha_0$ are counterterms for mixing parameters. For smaller values of $\Gamma^{\mathrm{LO}}_{h_2 \to ZZ}$, however, and whichever the scale chosen, $C_4$ leads to more unstable results than the remaining combinations.

\begin{figure}[htbp]
\centering
\subfloat{\includegraphics[width=0.46\linewidth]{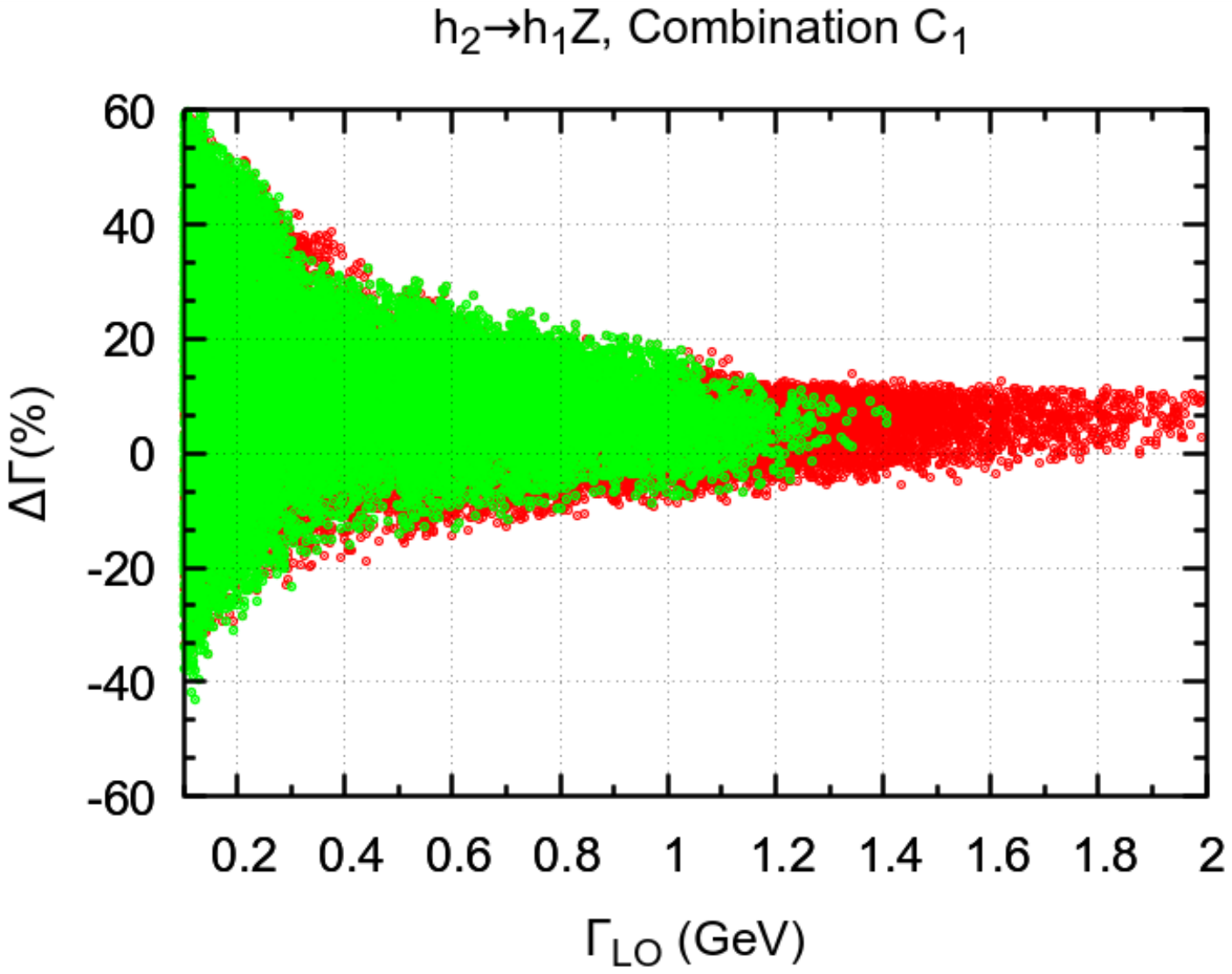}}\qquad
\subfloat{\includegraphics[width=0.46\linewidth]{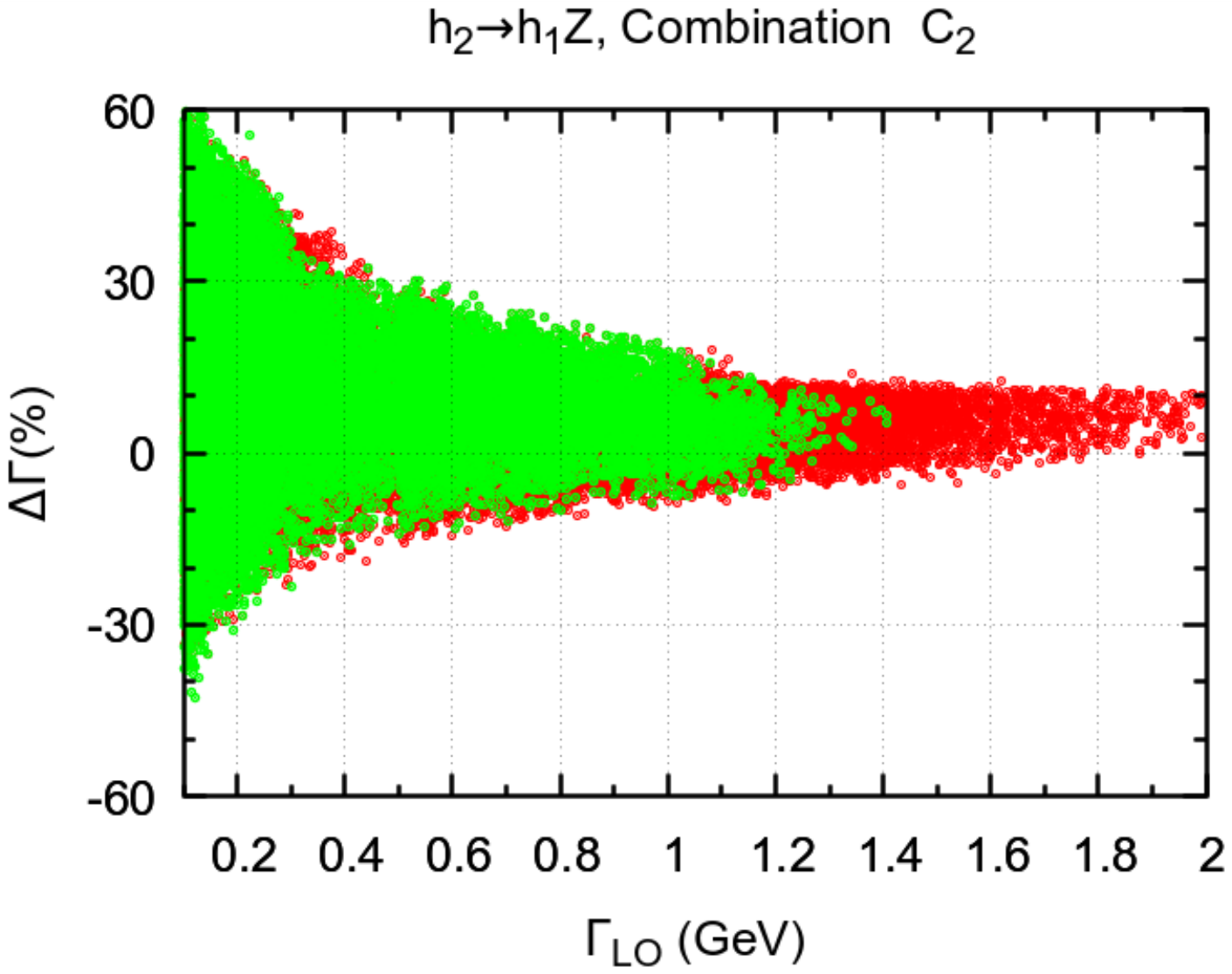}}\\
\subfloat{\includegraphics[width=0.46\textwidth]{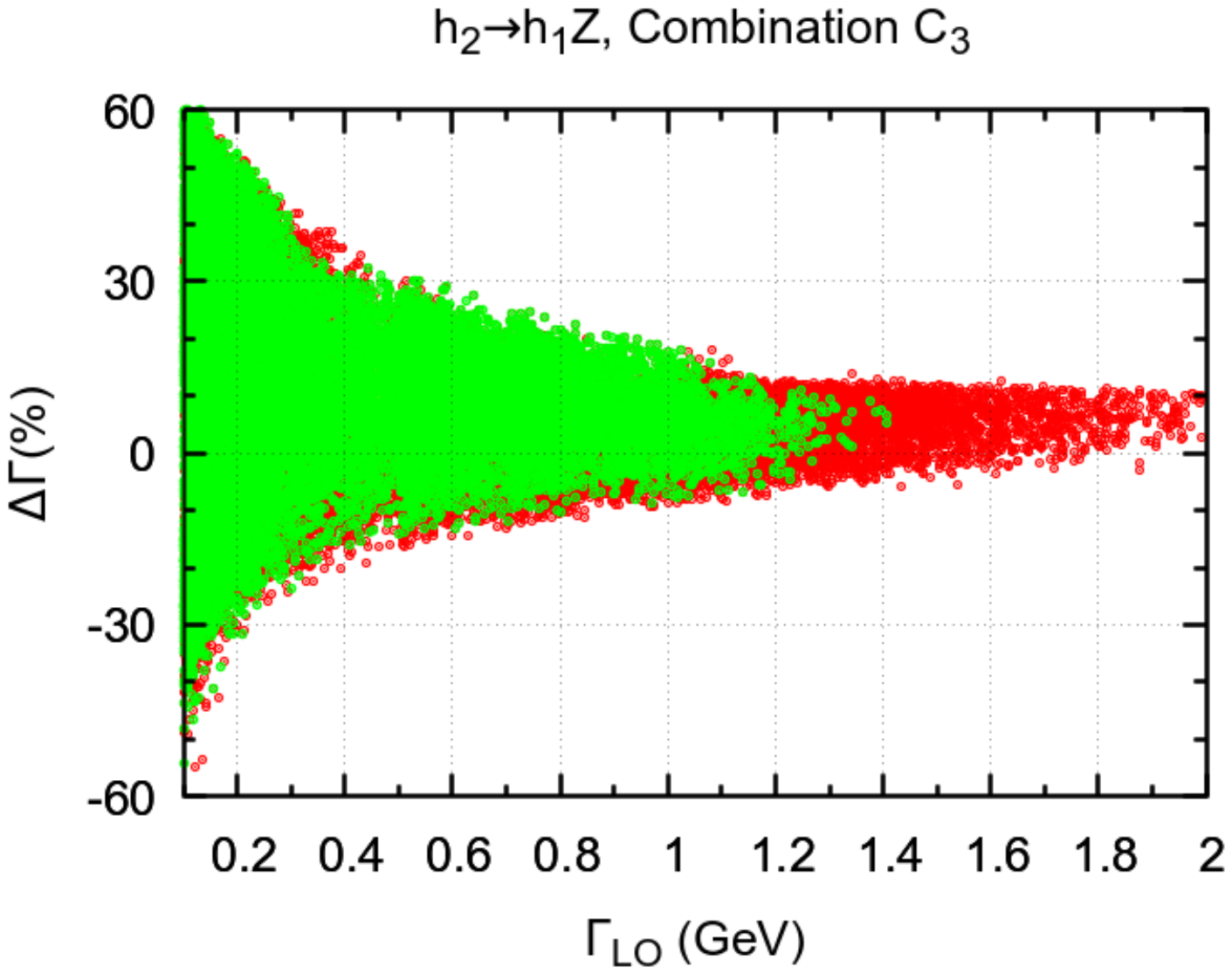}}\qquad%
\subfloat{\includegraphics[width=0.46\textwidth]{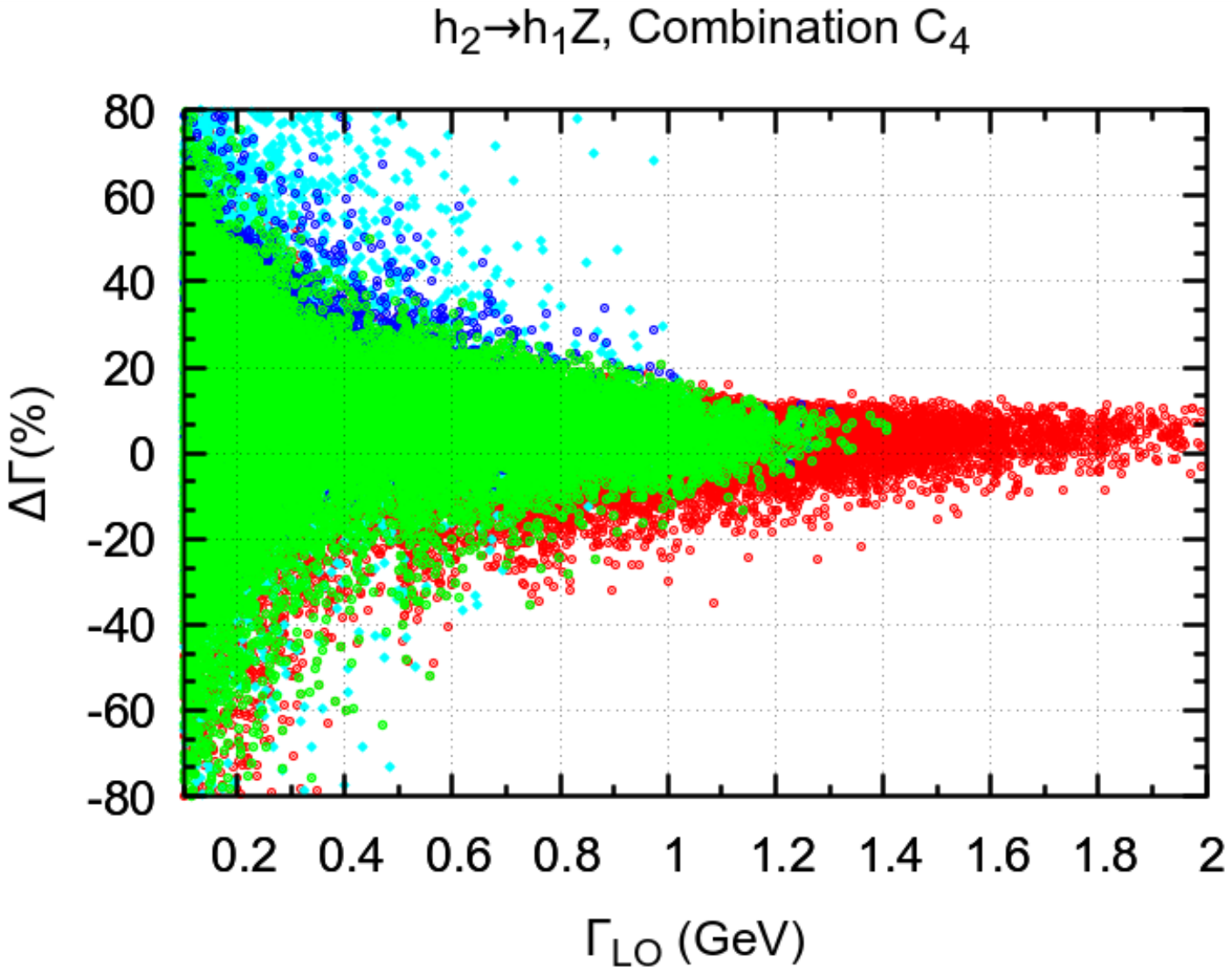}}%
\caption{
$\Delta \Gamma_{h_2 \to h_1Z}$ in percentage as a function of $\Gamma^{\mathrm{LO}}_{h_2 \to h_1Z}$, for the combinations $C_1$ (top left), $C_2$ (top right), $C_3$ (down left) and $C_4$ (down right).
Only the interval $ 0.1 \, \, \textrm{GeV} < \Gamma^{\mathrm{LO}} < 2 \, \, \textrm{GeV}$ is shown.
The color conventions for $C_1$, $C_2$ and $C_3$ are those of figure \ref{fig:h2ZZ-C1}, whereas for $C_4$ are those of figure \ref{fig:h2ZZ-C4}.
}
\label{fig:h2h1Z}
\end{figure}

\n In figure \ref{fig:h2h1Z}, we show the results for the decay $h_2 \to h_1Z$, for the four combinations.
The conclusions are similar to the ones for $h_2 \to ZZ$, which we just discussed.
In particular, the combinations $C_1$, $C_2$ and $C_3$ are scale independent, whereas $C_4$ is scale dependent (in which case the intermediate scale is again the most stable one); \texttt{HiggBounds-5} hampers points with large values of $\Gamma^{\mathrm{LO}}$, and all points passing all constraints in the four combinations show a similar and well-behaved pattern. Broadly speaking, these points are similar to the equivalent ones in $h_2 \to ZZ$ (except that, for small values of $\Gamma^{\mathrm{LO}}_{h_2 \to h_1Z}$, there are valid points with larger values---in modulus---of $\Delta \Gamma$, especially for $C_4$).
This similarity is expected from table \ref{tab:CTs-per-process}, which shows that the only difference in the set of independent counterterms contributing to the two decays concerns field counterterms; but since all field counterterms are fixed through the same method (namely, OSS), it is not surprising that the two processes turn out to lead to similar NLO corrections.

\begin{figure}[htbp]
\centering
\subfloat{\includegraphics[width=0.46\linewidth]{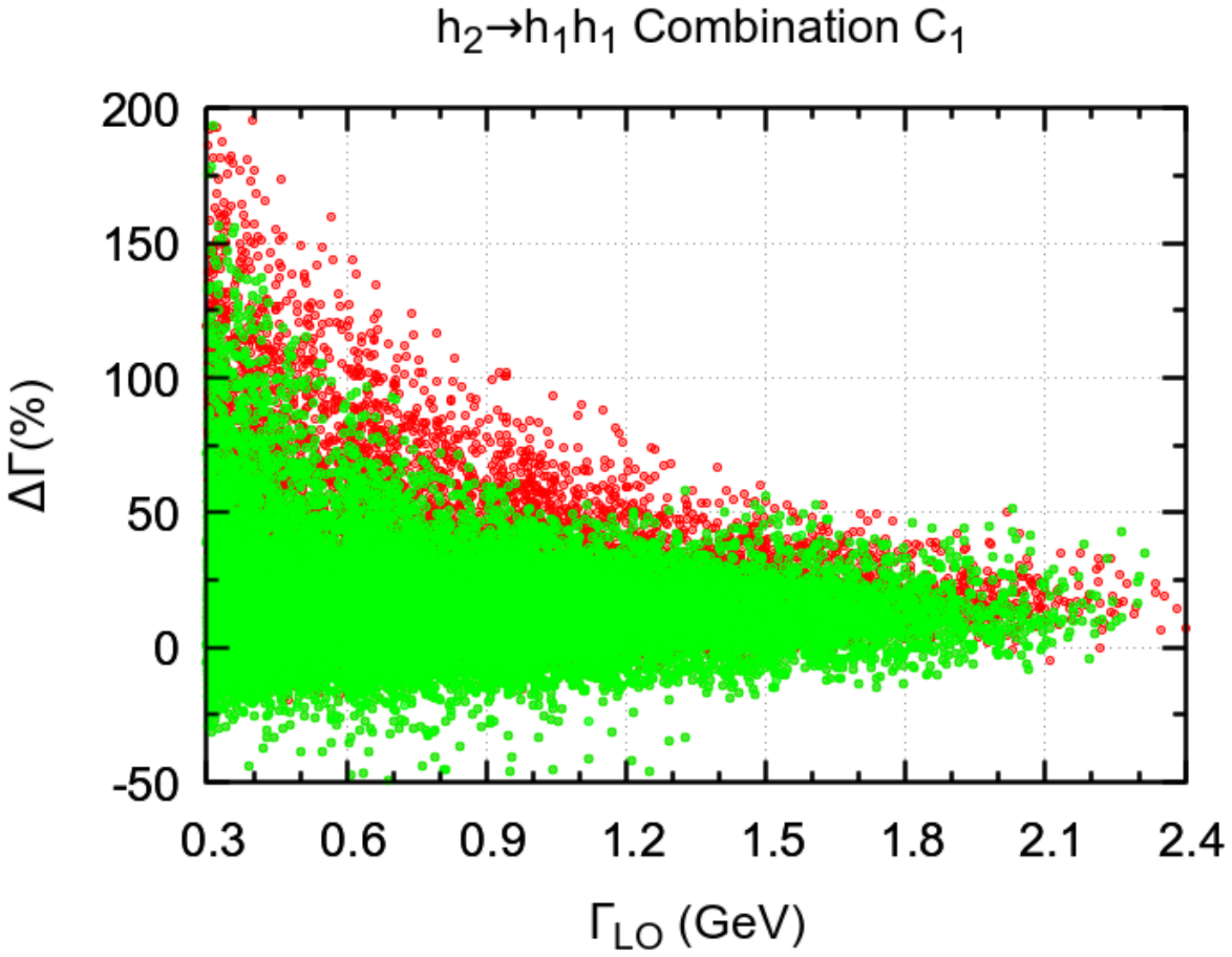}}\qquad
\subfloat{\includegraphics[width=0.46\linewidth]{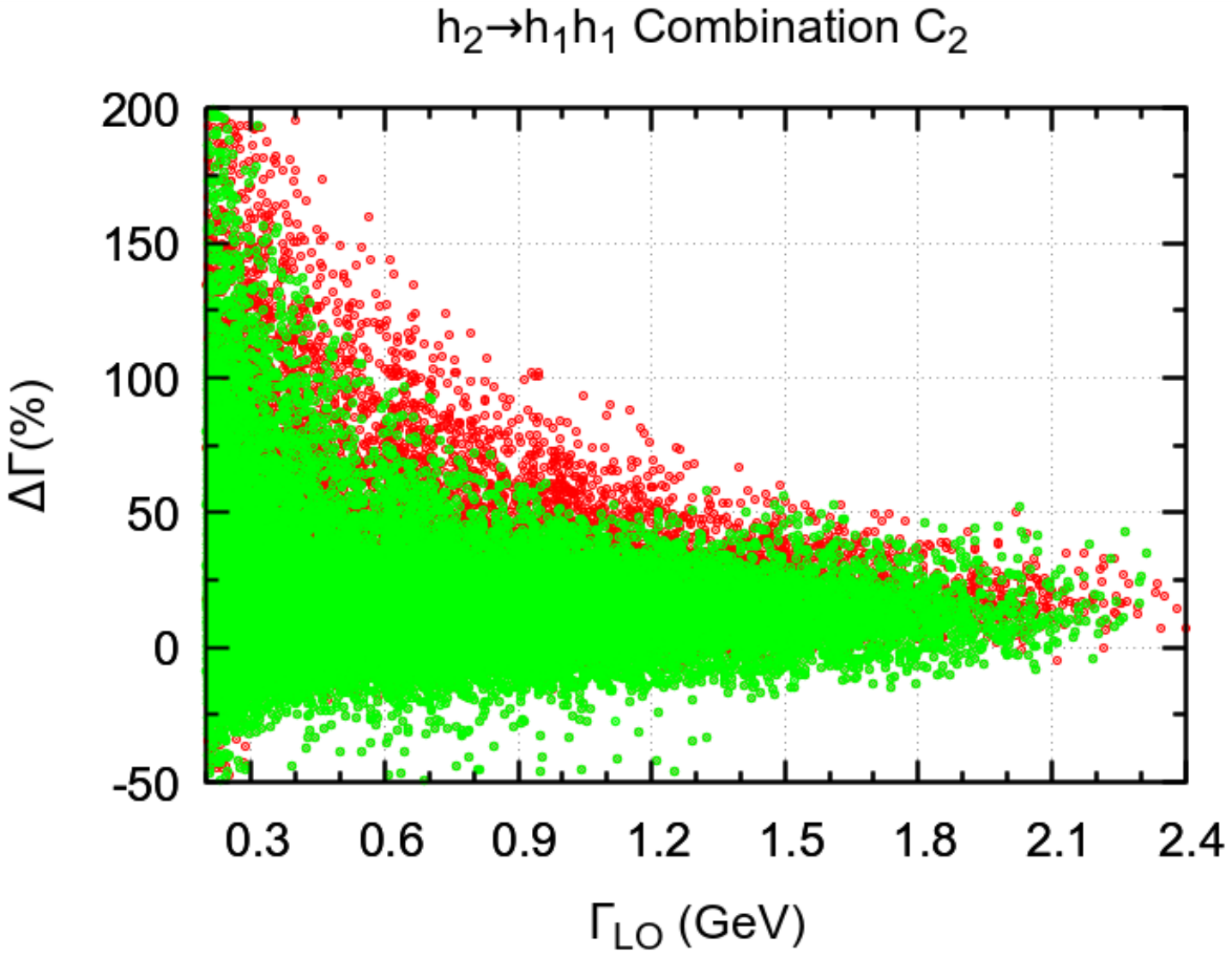}}\\
\subfloat{\includegraphics[width=0.46\textwidth]{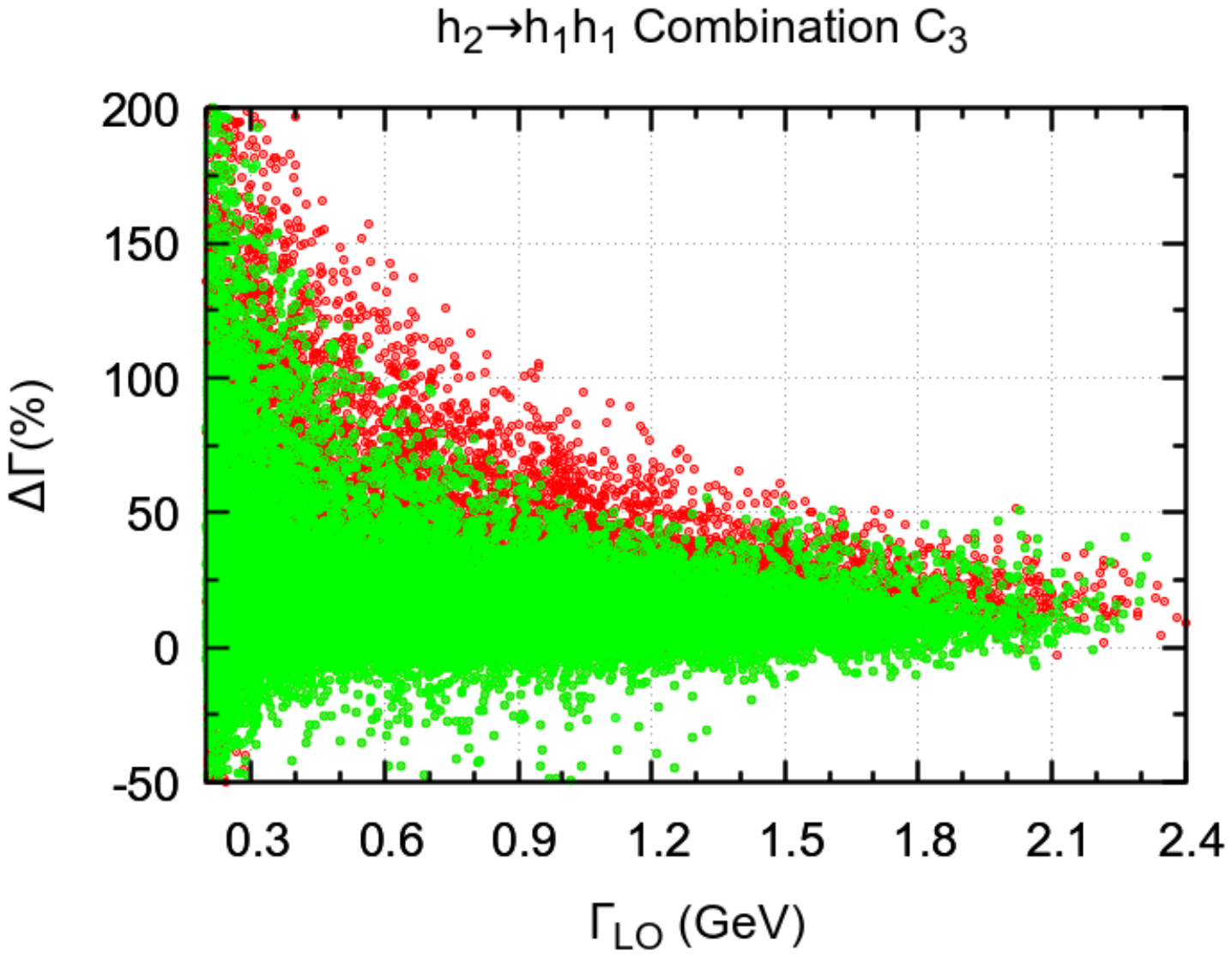}}\qquad%
\subfloat{\includegraphics[width=0.46\textwidth]{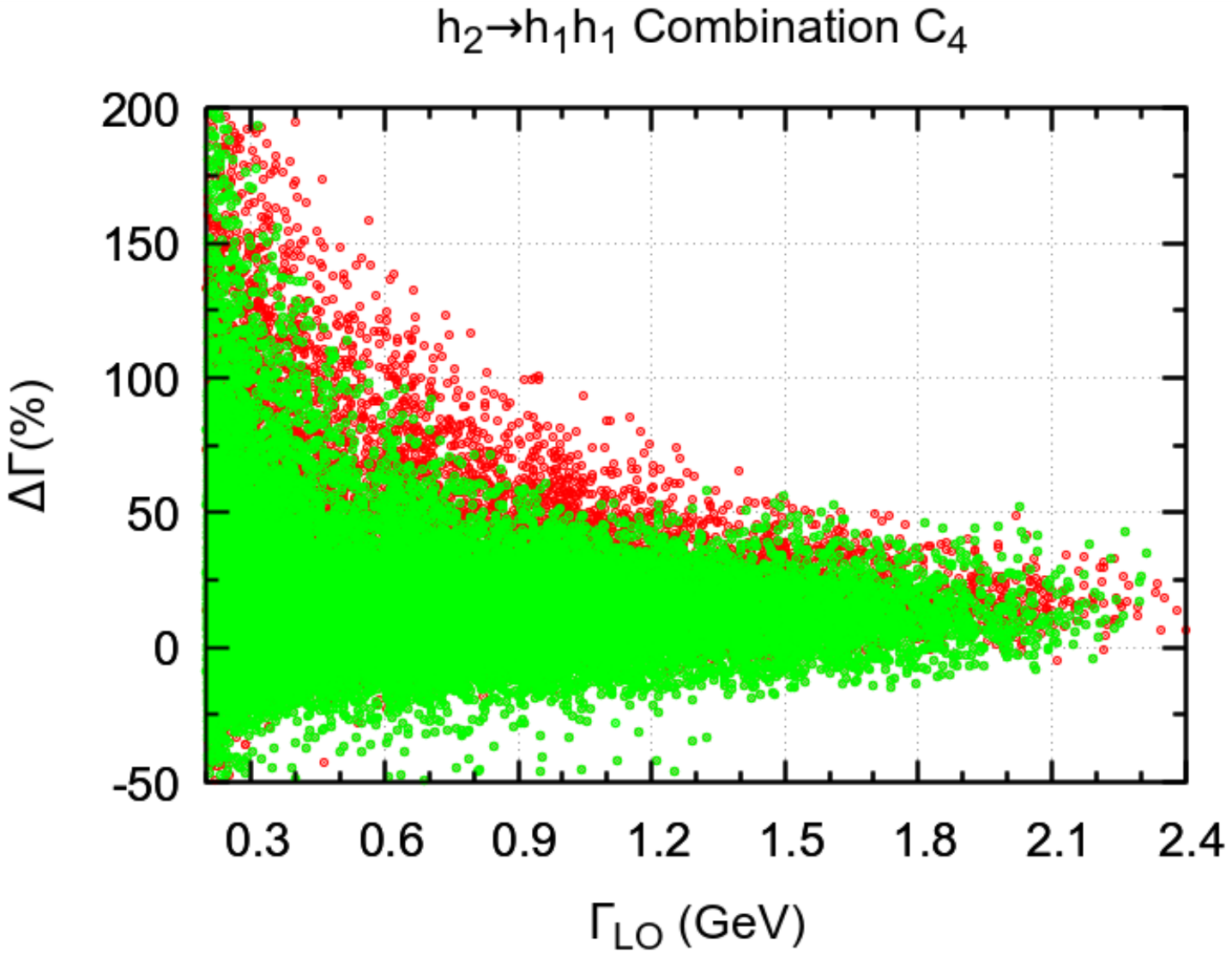}}%
\caption{
$\Delta \Gamma_{h_2 \to h_1h_1}$ in percentage as a function of $\Gamma^{\mathrm{LO}}_{h_2 \to h_1h_1}$, for the combinations $C_1$ (top left), $C_2$ (top right), $C_3$ (down left) and $C_4$ (down right).
Only the interval $ 0.1 \, \, \textrm{GeV} < \Gamma^{\mathrm{LO}} < 2.4 \, \, \textrm{GeV}$ is shown.
All points correspond to $\mu_{\mathrm{R}}= 350$ GeV; the color conventions are those of figure \ref{fig:h2ZZ-C1}.
}
\label{fig:h2h1h1}
\end{figure}

\begin{figure}[htb]
\centering
\includegraphics[width=0.55\textwidth]{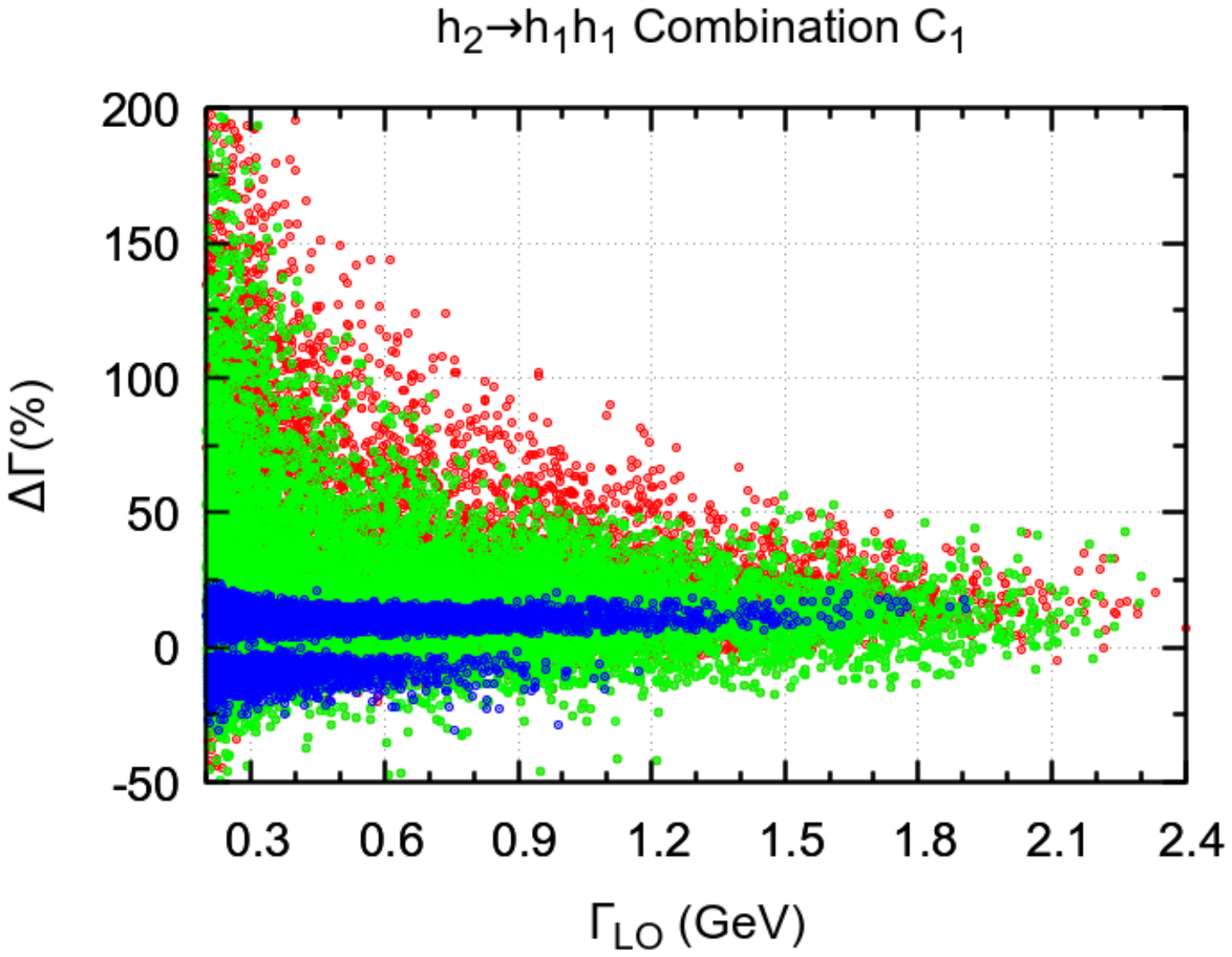}
\caption{
Identical to figure \ref{fig:h2h1h1} top left, except that blue/black points are included, satisfying all constraints and, additionally, $|\lambda_i|<2$ (see text for details).}
\label{fig:h2h1h1-ll}
\end{figure}

\n In figure \ref{fig:h2h1h1}, we show the results for $h_2 \to h_1 h_1$, for the four combinations.
Here, as predicted in section \ref{sec:influ} above, all combinations are scale dependent, due to the contribution of the counterterm $\delta \mu^2$; we plot the intermediate scale only, which we checked is again the most stable one.
The four combinations lead to similar results; a well-behaved pattern is once again to be observed in all of them. In contrast to what happened previously, \texttt{HiggBounds-5} does not preclude points with large values of $\Gamma^{\mathrm{LO}}_{h_2 \to h_1h_1}$; we checked that these points lead to small values of $\Gamma^{\mathrm{LO}}_{h_2 \to ZZ}$ and  $\Gamma^{\mathrm{LO}}_{h_2 \to h_1Z}$, and that only large values of these would be excluded by the experimental results.
Moreover, values of $\Delta \Gamma$ as high as $40\%$ are allowed for all the region of allowed points in all combinations. We found that more stable results can be obtained by requiring smaller values for the different $\lambda_i$ (parameters of the potential), as can be seen in figure \ref{fig:h2h1h1-ll} for the combination $C_1$, where we constrained $|\lambda_i| < 2$ for all $\lambda_i$.
This can be explained by the circumstance that smaller values for the $\lambda_i$ correspond to a more stringent constraint regarding the tree-level perturbative unitarity. Hence, it is not surprising that such values lead to a more stable perturbative expansion \cite{Krause:2016xku,Denner:2016etu}.

\section{Conclusions}
\label{sec:conclu}

\n We presented the renormalization of the C2HDM, which is one of the simplest models containing an extended scalar sector with CP violation. We showed in detail how the presence of CP violation in the scalar sector leads to a rather peculiar program of renormalization, due to two main reasons.
First, several non-physical parameters must be introduced in order to assure the generation of the necessary counterterms. This happens in such a way that, once the counterterms are obtained, the corresponding renormalized parameters are not needed and can be rephased away.
%
Second, for the same set of independent physical parameters, multiple combinations $C_i$ of independent counterterms can be chosen. It thus becomes relevant to study the impact of different combinations on observables.

\n We considered four combinations and applied them to three specific one-loop decays of $h_2$; the first three combinations choose as independent a counterterm for a mixing parameter, and they fix it through symmetry relations, whereas $C_4$ takes a parameter of the potential as independent, and fixes it through $\overline{\text{MS}}$---thus yielding a dependence on the renormalization scale $\mu_{\mathrm{R}}$.
We checked that all combinations in all processes lead to gauge-independent and well-behaved results, which validates the prescriptions used to fix the counterterms.
For all processes, we found that the combination $C_4$ leads (at least for some values of $\mu_{\mathrm{R}}$ and of the LO decay width) to results very similar to those of the remaining combinations.
Among the points passing all constraints in all processes, one finds regions in the four combinations that lead to a numerical stability of the perturbative expansion; points leading to large NLO corrections are also allowed, in which case the results of higher orders should be investigated.
 
\n We stressed several aspects related to an adequate treatment of the C2HDM when considered up to one-loop level. 
We used the Fleischner-Jegerlehner tadpole scheme to assure the selection of the true vev; this was the first time that this scheme was applied to a model with CP violation in scalar sector (and thus to a model with generally complex vevs).
We discussed the renormalization of a field whose mass is a dependent parameter.
We showed that the OSS conditions in the fermion sector leave some freedom, which can be used to define complex counterterms for the fermion masses.
We proposed a simple and easily generalizable prescription to calculate counterterms for mixing parameters, which assures that observables are gauge independent.
Finally, \FMTS was shown to be an ideal tool for renormalizing models such as the C2HDM and study them at NLO, since it calculates a multiplicity of crucial elements---Feynman rules, counterterms, one-loop amplitudes, decay widths, etc.---in a simultaneously automatic and flexible way. Both \FMTS and the model file for the C2HDM can be downloaded at:
\begin{center}
\url{https://porthos.tecnico.ulisboa.pt/FeynMaster/}
\end{center}

\n Given its pioneering character, this work opens the door to an exploration of precise predictions in the C2HDM, as well as to the renormalization of models with CP violation in the scalar sector. One interesting direction of future investigation concerns processes in the C2HDM with external $h_3$ fields, whose pole mass is in general different from the renormalized mass.

\section*{Acknowledgments}

\n D.F. is especially grateful to Ansgar Denner for many valuable discussions. Both authors also thank Augusto Barroso, Luís Lavoura, Rui Santos, Vladyslav Shtabovenko and João P. Silva for discussions.
This work is supported by projects
CFTP-FCT Unit 777 (UID/FIS/00777/2013 and UID/FIS/00777/2019), and PTDC/FIS-PAR/ 29436/2017, which are partially funded through POCTI (FEDER), COMPETE, QREN and EU. D.F. is also supported by the Portuguese \textit{Funda\c{c}\~ao para a Ci\^encia e Tecnologia} under the project SFRH/BD/135698/2018.

\appendix

\section{OSS and dependent masses}
\label{app:OSS}

\n The on-shell subtraction (OSS) scheme, originally proposed in ref.~\cite{Ross:1973fp}, has become one of the most common subtraction schemes in the renormalization of gauge theories, having been recurrently described and used by many authors \cite{Aoki:1982ed,Hollik:1988ii,Bohm:1986rj,Denner:1991kt,Schwartz:2013pla,Grimus:2016hmw,Denner:2019vbn}.
In this appendix, we present a clear and simple description of the one-loop renormalization conditions that characterize the OSS scheme, explaining how they allow to derive the mass and field counterterms. We consider both the scenario with no mixing of fields, as well as that with mixing; we also study the scenarios where a dependent mass prevents the application of the OSS conditions, and we show how the counterterms are determined in those cases.
This appendix can thus be seen as an introduction to sections \ref{sec:CT-gauge} to \ref{sec:CT-scalar} above.%

\n We will assume scalars particles, for simplicity; in the case of fermions and gauge bosons, the 2-point functions must be projected onto (i.e. multiplied by) physical states: the spinor $u(p)$ for fermions and the polarization vector $\varepsilon_{\mu}(k)$ for gauge bosons.

\subsection{No mixing, independent mass}

\n We start with the simplest case: a generic single scalar neutral field of a generic theory.
Suppose a real scalar field $\phi$ with mass $m$, such that:
\be
\mathcal{L} \ni - \dfrac{1}{2} \phi \Box \phi - \dfrac{1}{2} m^2 \phi^2.
\label{eq:LagToy}
\ee
We assume that $m$ is an independent parameter, i.e. that it is not written in terms of other parameters.
When considering the theory up to one-loop level, we follow the usual procedure, namely: we identify the tree-level quantities as bare ones, and split them into renormalized quantities and counterterms,
\be
m_{(0)}^2 = m_{\mathrm{R}}^2 + \delta m^2,
\quad
\phi_{(0)} = \phi + \dfrac{1}{2} \delta Z_{\phi} \phi.
\label{eq:ExpanToy}
\ee
Note that we explicitely represent the renormalized mass $m_{\mathrm{R}}$ with an index $\mathrm{R}$. Moreover, no mixed fields counterterms are introduced, since we are assuming that $\phi$ mixes with no other field at one-loop.
Identifying the up-to-one-loop 2-point function with $\Gamma$ and the one-loop 2-point function with $\Sigma$, we have:
\be
\hat{\Gamma}(p^2) = p^2 - m_{\mathrm{R}}^2 + \hat{\Sigma}(p^2),
\label{eq:OSaux1}
\ee
where the caret means `renormalized'. Then, expanding the bare version of eq. \ref{eq:LagToy} with eq. \ref{eq:ExpanToy}, we find:%
\fn{Recall that, as seen in eq. \ref{eq:FJTS-2p}, the non-renormalized one-loop 2-point function $\Sigma(p^2)$ in general includes diagrams with one-loop tadpoles if the FJTS is adopted.}
\be
\hat{\Sigma}(p^2) = \Sigma(p^2) + (p^2 - m_{\mathrm{R}}^2) \delta Z_{\phi} - \delta m^2.
\label{eq:OSaux2}
\ee
Now, the  pole mass $m_{\mathrm{P}}$ is defined as the value of the momentum for which $\hat{\Gamma}(p^2)$ is zero, that is:%
\fn{We are using the definition of \textit{real} pole mass (also known as Breit-Wigner mass), where $m_{\mathrm{P}}$ is real as a consequence of the fact that we are considering the real part of $\hat{\Sigma}(m_{\mathrm{P}}^2)$. This is what is done in the traditional OSS scheme \cite{Aoki:1982ed, Denner:2019vbn}.
We are using the operator $\widetilde{\operatorname{Re}}$ (and not $\mathrm{Re}$), which neglects the imaginary---also called absorptive---parts of loop integrals, while keeping the imaginary parts of complex parameters. The use of $\widetilde{\operatorname{Re}}$ in the renormalization conditions is also common practice in the traditional OSS scheme.
%
}
\be
\widetilde{\operatorname{Re}} \, \hat{\Gamma}(m_{\mathrm{P}}^2) = m_{\mathrm{P}}^2 - m_{\mathrm{R}}^2 + \widetilde{\operatorname{Re}} \,  \hat{\Sigma}(m_{\mathrm{P}}^2) = 0.
\label{eq:def:polemass}
\ee
This allows us to see how the OSS conditions fix the counterterms $\delta m^2$ and $\delta Z_{\phi}$. In the case we are considering (a single field), there are only two such conditions. The first one simply states that the renormalized mass is equal to the pole mass,
\be
m_{\mathrm{R}} \stackrel{\mathrm{OSS}}{=} m_{\mathrm{P}},
\label{eq:my-first-OSS}
\ee
thus giving a physical meaning to the renormalized mass.
In that case, eq. \ref{eq:def:polemass} implies:
\be
\widetilde{\operatorname{Re}} \, \hat{\Gamma}(m_{\mathrm{R}}^2) \stackrel{\mathrm{OSS}}{=} 0,
\label{eq:OSSfirst}
\ee
which, together with eqs. \ref{eq:OSaux1} and \ref{eq:OSaux2}, fixes $\delta m^2$ to:
\be
\delta m^2 \stackrel{\mathrm{OSS}}{=} \widetilde{\operatorname{Re}} \, \Sigma(m_{\mathrm{R}}^2).
\label{eq:dm-OSS}
\ee
One should not confuse eqs. \ref{eq:def:polemass} and \ref{eq:OSSfirst}: the former is the definition of the pole mass, valid in every subtraction scheme.
In contrast, eq. \ref{eq:OSSfirst} is specific of OSS, and results from the application of the OSS condition \ref{eq:my-first-OSS} to eq. \ref{eq:def:polemass}.%
%

\n The second OSS condition takes the (renormalized, real) residue of the renormalized propagator at the pole mass,
\be
\hat{R}
\equiv
\lim_{p^2 \to m_{\mathrm{P}}^2} (-i) (p^2 - m_{\mathrm{P}}^2) \, \widetilde{\operatorname{Re}} \, \hat{G}(p^2)
=
\lim_{p^2 \to m_{\mathrm{P}}^2} (-i) (p^2 - m_{\mathrm{P}}^2) \, \widetilde{\operatorname{Re}} \left[ i \hat{\Gamma}^{-1}(p^2) \right],
\label{eq:resi}
\ee
and sets it to one. Actually, one usually sets the inverse of the residue equal to one, which means:%
\fn{In this equation, $m_{\mathrm{P}}$ and $m_{\mathrm{R}}$ can be used indifferently, as they are chosen to be the same in OSS.}
\be
\lim_{p^{2} \to m_{\mathrm{R}}^2} \frac{1}{p^2 - m_{\mathrm{R}}^{2}} \widetilde{\operatorname{Re}} \, \hat{\Gamma}(p^2) \stackrel{\mathrm{OSS}}{=} 1.
\ee
Then, using eqs. \ref{eq:OSaux1} and \ref{eq:OSaux2}, as well as L'Hôpital's rule, the expression for $\delta Z_{\phi}$ follows:
\be
\delta Z_{\phi} \stackrel{\mathrm{OSS}}{=} -\left.\widetilde{\operatorname{Re}} \, \frac{\partial \Sigma(p^2)}{\partial p^{2}}\right|_{p^{2}=m_{\mathrm{R}}^{2}}.
\label{eq:dZ-OSS}
\ee

\subsection{No mixing, dependent mass}
\label{sec:dep-mass}

\n If we suppose instead that the bare mass $m_{(0)}$ is a dependent parameter, then both the renormalized mass $m_{\mathrm{R}}$ and the counterterm $\delta m^2$ are also dependent, i.e. fixed.%
\fn{The case of dependent masses is common in the minimal supersymmetric extension of the SM (MSSM); see e.g. ref. \cite{Diaz:1991ki}.}
Now, since $m_{\mathrm{R}}$ is fixed, it cannot be set to be equal to the pole mass $m_{\mathrm{P}}$, so that the first OSS condition cannot be used. In that case, then, the two masses will in general be different, and eq. \ref{eq:def:polemass} (which is always valid) assures that their difference is of one-loop order. But this means that, to first order in perturbation theory, $m_{\mathrm{P}}^2$ can be replaced by $m_{\mathrm{R}}^2$ when $m_{\mathrm{P}}^2$ is an argument of a one-loop function. In particular, the relation
\be
\widetilde{\operatorname{Re}} \, \hat{\Sigma}(m_{\mathrm{P}}^2) = \widetilde{\operatorname{Re}} \, \hat{\Sigma}(m_{\mathrm{R}}^2)
\ee
in valid to first order, so that eqs. \ref{eq:def:polemass} and \ref{eq:OSaux2} imply, to that order,
\be
m_{\mathrm{P}}^2 = m_{\mathrm{R}}^2 - \widetilde{\operatorname{Re}} \, \Sigma(m_{\mathrm{R}}^2) + \delta m^2.
\label{eq:def:polemass2}
\ee
Hence, the pole mass is completely determined.%
\fn{Note that $m_{\mathrm{P}}^2$ is finite, as it should be. 
A simple way to see this is to recall that the divergent parts of counterterms must be the same whichever the subtraction scheme chosen. Then, the divergent part of the left-hand side of eq. \ref{eq:dm-OSS} is equal to that of the right-hand side, for all subtraction schemes. Therefore, the divergent parts of the right-hand side of eq. \ref{eq:def:polemass2} cancel.}

\n Just as in the case of the independent mass, $\delta Z_{\phi}$ can be fixed by setting the residue equal to one. Only, one must be careful with the fact that $m_{\mathrm{P}}^2$ is different from $m_{\mathrm{R}}^2$. Inserting eqs. \ref{eq:OSaux1} and \ref{eq:OSaux2} inside eq. \ref{eq:resi}, equating the latter to one, using L'Hôpital's rule and the relation
\be
\left.\widetilde{\operatorname{Re}} \, \frac{\partial \hat{\Sigma}(p^2)}{\partial p^{2}}\right|_{p^{2}=m_{\mathrm{P}}^{2}}
=
\left.\widetilde{\operatorname{Re}} \, \frac{\partial \hat{\Sigma}(p^2)}{\partial p^{2}}\right|_{p^{2}=m_{\mathrm{R}}^{2}},
\ee
which holds to first order, we find:
\be
\delta Z_{\phi}
=
-\left.\widetilde{\operatorname{Re}} \, \frac{\partial \Sigma(p^2)}{\partial p^{2}}\right|_{p^{2}=m_{\mathrm{R}}^{2}},
\ee
which is precisely the same as the field counterterm fixed through OSS in the case of the independent mass, eq. \ref{eq:dZ-OSS}.

\subsection{Mixing, all masses independent}

\n We now consider a generalization of eqs. \ref{eq:LagToy} to \ref{eq:OSaux2} for two fields:
\begin{gather}
\mathcal{L} \ni
- \dfrac{1}{2} \phi_1 \Box \phi_1 - \dfrac{1}{2} m_1^2 \phi_1^2
- \dfrac{1}{2} \phi_2 \Box \phi_2 - \dfrac{1}{2} m_2^2 \phi_2^2,
\label{eq:LagToy2}
\\
m_{1(0)}^2 = m_{1\mathrm{R}}^2 + \delta m_1^2,
\quad
m_{2(0)}^2 = m_{2\mathrm{R}}^2 + \delta m_2^2, \\
\begin{pmatrix}
\phi_{1(0)} \\ \phi_{2(0)}
\end{pmatrix}
=
\begin{pmatrix}
1 + \dfrac{1}{2} \delta Z_{11} & \dfrac{1}{2} \delta Z_{12} \\
\dfrac{1}{2} \delta Z_{21} & 1 + \dfrac{1}{2} \delta Z_{22}
\end{pmatrix}
\begin{pmatrix}
\phi_1 \\ \phi_2
\end{pmatrix},
\label{eq:ExpanToy2}
\\
\hat{\Gamma}_{ij} (p^2) =
\begin{pmatrix}
p^2 - m_{1\mathrm{R}}^2 + \hat{\Sigma}_{11}(p^2) &
\hat{\Sigma}_{12}(p^2) \\
\hat{\Sigma}_{21}(p^2) & p^2 - m_{2\mathrm{R}}^2 + \hat{\Sigma}_{22}(p^2)
\end{pmatrix},
\label{eq:OSaux3}
\\
\hat{\Sigma}_{ij}(p^2) = \Sigma_{ij}(p^2) + \frac{1}{2}\left(p^{2}-m_{j\mathrm{R}}^{2}\right) \delta Z_{ji}
+ \frac{1}{2}\left(p^{2}-m_{i\mathrm{R}}^{2}\right) \delta Z_{ij}
-\delta_{ij} \, \delta m_{i}^{2},
\label{eq:OSaux4}
\end{gather}
where $i,j = \{1,2\}$, and where both $m_{1(0)}$ and $m_{2(0)}$ are assumed to be independent parameters.
In OSS, the counterterms are once again calculated by reference to the pole masses and the residues.

\n But, now that there is mixing, what is the definition of the pole masses and the residues?
Because $\hat{\Gamma}(p^2)$ is now a matrix (eq. \ref{eq:OSaux3}), the pole masses are determined as the values of the momentum for which its eigenvalues are zero.
And while that leads to complicated expressions, these can be simplified by fixing the off-diagonal field counterterms in an appropriate way. More specifically, if we choose those counterterms to be such that:
\be
\widetilde{\operatorname{Re}} \, \hat{\Gamma}_{ij} (m_{i\mathrm{P}}^2)
\stackrel{j \neq i}{=}
0,
\qquad
\widetilde{\operatorname{Re}} \, \hat{\Gamma}_{ji} (m_{i\mathrm{P}}^2)
\stackrel{j \neq i}{=}
0,
\label{eq:aux3OSS}
\ee
then the definition of the pole masses $m_{i\mathrm{P}}$ is a trivial generalization of eq. \ref{eq:def:polemass}, namely,
\be
\widetilde{\operatorname{Re}} \, \hat{\Gamma}_{ii}(m_{i\mathrm{P}}^2) = m_{i\mathrm{P}}^2 - m_{i\mathrm{R}}^2 + \widetilde{\operatorname{Re}} \,  \hat{\Sigma}_{ii}(m_{i\mathrm{P}}^2) = 0,
\label{eq:def:polemass-i}
\ee
in which case the residues become:
\be
\hat{R}_i
=
\lim_{p^2 \to m_{i\mathrm{P}}^2} (p^2 - m_{i\mathrm{P}}^2) \, \widetilde{\operatorname{Re}} \, \hat{\Gamma}_{ii}^{-1}(p^2).
\label{eq:def:residue-i}
\ee
Eq. \ref{eq:aux3OSS} assures that OS particles do not mix with each other, and allows to calculate the off-diagonal field counterterms ($\delta Z_{12}$ and $\delta Z_{21}$) through eqs. \ref{eq:OSaux3} and \ref{eq:OSaux4}. 
Now, OSS assumes not only eq. \ref{eq:aux3OSS}, but also that the renormalized masses are equal to the pole masses, $m_{i\mathrm{R}} \stackrel{\mathrm{OSS}}{=} m_{i\mathrm{P}}$ (which we can always do, since the renormalized masses $m_{i\mathrm{R}}$ are assumed to be free parameters). In that case, eq. \ref{eq:aux3OSS} becomes:
\be
\widetilde{\operatorname{Re}} \, \hat{\Gamma}_{ij} (m_{i\mathrm{R}}^2)
\stackrel{j \neq i, \, \mathrm{OSS}}{=}
0,
\qquad
\widetilde{\operatorname{Re}} \, \hat{\Gamma}_{ji} (m_{i\mathrm{R}}^2)
\stackrel{j \neq i, \, \mathrm{OSS}}{=}
0,
\label{eq:aux3OSS-OSS}
\ee
while eq. \ref{eq:def:polemass-i} becomes:
\be
\widetilde{\operatorname{Re}} \, \hat{\Gamma}_{ii}(m_{i\mathrm{R}}^2) \stackrel{\mathrm{OSS}}{=} 0,
\label{eq:OS-mixing-2}
\ee
which allows to determine the mass counterterms. Besides, and also in a trivial generalization of the case with no mixing, OSS fixes the diagonal field counterterms by requiring the different inverse residues to be equal to one:
\be
\lim_{p^{2} \to m_{i\mathrm{R}}^2} \frac{1}{p^2 - m_{i\mathrm{R}}^{2}} \widetilde{\operatorname{Re}} \, \hat{\Gamma}_{ii}(p^2) \stackrel{\mathrm{OSS}}{=} 1.
\label{eq:OS-mixing-3}
\ee
The expressions for the complete set of mass and field counterterms are, thus,
\be
\delta m_i^2 \stackrel{\mathrm{OSS}}{=} \widetilde{\operatorname{Re}} \, \Sigma_{ii}(m_{i\mathrm{R}}^2),
\quad
\delta Z_{ij} \stackrel{j \neq i, \, \mathrm{OSS}}{=} 2 \, \frac{\widetilde{\operatorname{Re}}  \, \Sigma_{ij}(m_{j\mathrm{R}}^{2})}{m_{i\mathrm{R}}^2 - m_{j\mathrm{R}}^2},
\quad
\delta Z_{ii} \stackrel{\mathrm{OSS}}{=} -\left.\widetilde{\operatorname{Re}} \, \frac{\partial \Sigma_{ii}(p^2)}{\partial p^{2}}\right|_{p^{2}=m_{i\mathrm{R}}^{2}}.
\label{eq:total-set-OSS}
\ee

\subsection{Mixing, one dependent mass}

\n Finally, we consider the situation where $m_{2(0)}$ is a dependent parameter (while $m_{1(0)}$ is still independent). It follows that $m_{2\mathrm{R}}^2$ and $\delta m_2^2$ are dependent. Nonetheless, we can always impose eq. \ref{eq:aux3OSS}, which not only fixes the counterterms $\delta Z_{12}$ and $\delta Z_{21}$, but also implies that eqs. \ref{eq:def:polemass-i} and \ref{eq:def:residue-i} are valid. Now, while we have the freedom to choose $m_{1\mathrm{R}} \stackrel{\mathrm{OSS}}{=} m_{1\mathrm{P}}$ (so that eqs. \ref{eq:OS-mixing-2} and \ref{eq:OS-mixing-3} hold for $i=1$, and these respectively allow to calculate $\delta m_1^2$ and $\delta Z_{11}$), we do not have the freedom to fix $m_{2\mathrm{R}}^2$.
In this case, though, we have a trivial generalization of section \ref{sec:dep-mass}: the expression for $m_{2\mathrm{P}}^2$ follows from eq. \ref{eq:def:polemass-i}, 
while $\delta Z_{22}$ can be determined by fixing the residue of the OS propagator of $\phi_2$ to one.%
\fn{As a consequence, the LSZ factors become trivial. For details, cf. ref. \cite{Fontes:PhD}.}
The expressions then become:
\begin{gather}
m_{2\mathrm{P}}^2 = m_{\mathrm{2R}}^2 - \widetilde{\operatorname{Re}} \, \Sigma_{22}(m_{2\mathrm{R}}^2) + \delta m_2^2, \\
\delta m_1^2 = \widetilde{\operatorname{Re}} \, \Sigma_{11}(m_{1\mathrm{R}}^2),
\quad
\delta Z_{ij} \stackrel{j \neq i}{=} 2 \, \frac{\widetilde{\operatorname{Re}}  \, \Sigma_{ij}(m_{j\mathrm{R}}^{2})}{m_{i\mathrm{R}}^2 - m_{j\mathrm{R}}^2},
\quad
\delta Z_{ii} = -\left.\widetilde{\operatorname{Re}} \, \frac{\partial \Sigma_{ii}(p^2)}{\partial p^{2}}\right|_{p^{2}=m_{i\mathrm{R}}^{2}},
\label{eq:total-set}
\end{gather}
where we neglected higher order terms.
%
%
Hence, 
eq. \ref{eq:total-set} only differs from eq. \ref{eq:total-set-OSS} in the fact that $\delta m_2^2$ is \textit{a priori} fixed.

\n The generalization to the case with four fields with one dependent mass (as in section \ref{sec:CT-scalar} above) is straightforward.

\section{CP violation in fermionic 2-point functions}
\label{sec:apD}

\n In this appendix, we study CP violation in fermionic 2-point functions. As we will show, this is particularly relevant for renormalization purposes---and hence for the renormalization of the C2HDM. However, we do not restrict ourselves to a particular theory; in fact, the appendix can be applied to every model with Dirac fermions.
We organize it as follows:
in a first part, we show how a complex mass shows up in the Dirac representation of the Lorentz group;
then, we consider CP violation in fermionic 2-point functions and prove that a complex mass is intrinsically related with it;
after that, we investigate the consequences of a CP-violating one-loop 2-point function on counterterms;
finally, we ascertain how to handle CP violation using the OSS scheme.

\subsection{Complex mass}
\label{sec:D1}

\n Let us consider two fields: a left-handed Weyl spinor $f^{\text{w}}_{\mathrm{L}}$---transforming in the $\left(\frac{1}{2}, 0\right)$ representation of the Lorentz group---and a right-handed Weyl spinor $f^{\text{w}}_{\mathrm{R}}$---transforming in the $\left(0, \frac{1}{2}\right)$ representation of the Lorentz group.%
\fn{We consider a single flavour for simplicity.}
The Lagrangian for the mass term is such that:
\be
-\mathcal{L}_{\text{mass},f} = m_f \, f_{\mathrm{L}}^{\text{w}\dagger} f_{\mathrm{R}}^{\text{w}} + \text{h.c.}
=
m_f \, f_{\mathrm{L}}^{\text{w}\dagger} f_{\mathrm{R}}^{\text{w}} + m_f^* \, f_{\mathrm{R}}^{\text{w}\dagger} f_{\mathrm{L}}^{\text{w}}.
\label{eq:apD:mass-base}
\ee
Note that the parameter $m_f$ is in general complex.
Moreover, $f_{\mathrm{L}}^{\text{w}}$ and $f_{\mathrm{R}}^{\text{w}}$ are two separate and unrelated fields, which implies in particular that they can be rotated in independent ways. Later on, they (or some rotated versions of them) will be embedded in a Dirac spinor as its left-handed and right-handed components. But until then, $f_{\mathrm{L}}^{\text{w}}$ and $f_{\mathrm{R}}^{\text{w}}$ are not connected with each other in any way. Now, the fact that we can redefine them differently means that the phase of $m_f$ can be absorbed in the fields. Indeed, defining $m_f = |m_f| e^{i \theta_m}$, we can for example rephase $f_{\mathrm{R}}^{\text{w}}$ as
\be
f_{\mathrm{R}}^{\text{w}} \rightarrow e^{-i \theta_m} f_{\mathrm{R}}^{\text{w}}
\label{eq:apD:7}
\ee
to render $m_f$ real.
Such rephasing (or another one equivalent to it) is usually performed, since $m_f$ is usually identified with the (real) pole mass.
Yet, it is illustrative to keep $m_f$ complex. In particular, we want to ascertain how a complex mass shows up in the Dirac representation $\left(\frac{1}{2}, 0\right) \oplus \left(0, \frac{1}{2}\right)$ of the Lorentz group. Thus, we now embed the Weyl spinors in a Dirac spinor $f$, such that:
\be
f = \begin{pmatrix}
f_{\mathrm{L}}^{\text{w}} \\
f_{\mathrm{R}}^{\text{w}}
\end{pmatrix} = f_{\mathrm{L}} + f_{\mathrm{R}},
\qquad
\text{with}
\quad
f_{\mathrm{L}} = \gamma_{\mathrm{L}} \, f
\equiv \dfrac{1-\gamma_5}{2} f,
\quad 
f_{\mathrm{R}} = \gamma_{\mathrm{R}} \, f
\equiv \dfrac{1+\gamma_5}{2} f.
\ee
%
In the Weyl representation, 
\be
f_L
=
\begin{pmatrix}
f_{\mathrm{L}}^{\text{w}} \\
0
\end{pmatrix},
\quad
f_R =
\begin{pmatrix}
0 \\
f_{\mathrm{R}}^{\text{w}}
\end{pmatrix},
\ee
which means that eq. \ref{eq:apD:mass-base} can be rewritten as:
\be
-\mathcal{L}_{\text{mass},f} = m_f \, \bar{f}_{\mathrm{L}} f_{\mathrm{R}} + m_f^* \, \bar{f}_{\mathrm{R}} f_{\mathrm{L}}.
\label{eq:apD:my11}
\ee
Separating $m_f$ into its real and imaginary parts, we find:
\be
\begin{split}
-\mathcal{L}_{\text{mass},f} 
&=  \Big(\mathrm{Re} [m_f] + i \,  \mathrm{Im}[m_f]\Big) \bar{f}_{\mathrm{L}} f_{\mathrm{R}} + \Big(\mathrm{Re} [m_f] - i \,  \mathrm{Im}[m_f]\Big) \bar{f}_{\mathrm{R}} f_{\mathrm{L}} \\
&= \mathrm{Re} [m_f] \Big(\bar{f}_{\mathrm{L}} f_{\mathrm{R}} + \bar{f}_{\mathrm{R}} f_{\mathrm{L}}\Big) + i \, \mathrm{Im} [m_f] \Big(\bar{f}_{\mathrm{L}} f_{\mathrm{R}} -  \bar{f}_{\mathrm{R}} f_{\mathrm{L}}\Big).
\label{eq:apD:mass-2}
\end{split}
\ee
Using the properties $\gamma_5^{\dagger}=\gamma_5$ and $\{\gamma_5, \gamma_{\mu}\}=0$, it is straightforward to conclude that
\be
\bar{f}_{\mathrm{L}} f_{\mathrm{R}} + \bar{f}_{\mathrm{R}} f_{\mathrm{L}} = \bar{f} f,
\qquad
\bar{f}_{\mathrm{L}} f_{\mathrm{R}} - \bar{f}_{\mathrm{R}} f_{\mathrm{L}} = \bar{f} \gamma_5 f,
\ee
so that eq. \ref{eq:apD:mass-2} becomes:
\be
-\mathcal{L}_{\text{mass},f} = \mathrm{Re} [m_f] \, \bar{f} f + i \, \mathrm{Im} [m_f] \, \bar{f} \gamma_5 f.
\label{eq:apD:13}
\ee
In the Dirac representation, then, a mass $m_f$ with non-null imaginary part is equivalent to the presence of a term of the form $\bar{f} \gamma_5 f$.%
\fn{This term has no physical meaning whatsoever, as it depends on the basis chosen: given eq. \ref{eq:apD:7}, that term will not show up if, before we embed the Weyl spinors $f_{\mathrm{L}}^{\text{w}}$ and $f_{\mathrm{R}}^{\text{w}}$ in the Dirac spinor $f$, we rotate them to render $m_f$ real.}
We now show that such term violates CP.

\subsection{CP violation in 2-point functions}

\n We start by considering the 2-point function for $f$. We can parameterize it as \cite{Denner:2019vbn}:
\be
\Gamma^{\bar{f} f}(p) = \slashed{p} \frac{1-\gamma_{5}}{2} \Gamma^{f,\mathrm{L}}(p^2)+ \slashed{p} \frac{1+\gamma_{5}}{2} \Gamma^{f,\mathrm{R}}(p^2)+\frac{1-\gamma_{5}}{2} \Gamma^{f,\mathrm{l}}(p^2)+\frac{1+\gamma_{5}}{2} \Gamma^{f,\mathrm{r}}(p^2),
\label{eq:apD:17}
\ee
or equivalently as
\be
\Gamma^{\bar{f} f} (p)
=
\slashed{p} \, \Gamma^{f,{\mathrm{V}}}(p^2)
+ \slashed{p} \gamma_5 \, \Gamma^{f,{\mathrm{A}}}(p^2)
+ \Gamma^{f,{\mathrm{S}}}(p^2)
+ \gamma_5 \, \Gamma^{f,{\mathrm{P}}}(p^2),
\label{eq:apD:1}
\ee
with the correspondence
\be
\Gamma^{f,{\mathrm{V}}} = \dfrac{\Gamma^{f,\mathrm{L}}+\Gamma^{f,\mathrm{R}}}{2},
\quad 
\Gamma^{f,{\mathrm{A}}} = \dfrac{\Gamma^{f,\mathrm{R}}-\Gamma^{f,\mathrm{L}}}{2},
\quad 
\Gamma^{f,{\mathrm{S}}} = \dfrac{\Gamma^{f,\mathrm{l}}+\Gamma^{f,\mathrm{r}}}{2},
\quad 
\Gamma^{f,{\mathrm{P}}} = \dfrac{\Gamma^{f,\mathrm{r}}-\Gamma^{f,\mathrm{l}}}{2}.
\label{eq:apD:conv}
\ee
Now, eq. \ref{eq:apD:1} corresponds to the effective Lagrangian
\be
\mathcal{L}_{\text{eff}}^{\bar{f} f} =  i \, \bar{f} \, \slashed{\partial} f \, \Gamma^{f,{\mathrm{V}}}(\partial^2)  + i \, \bar{f} \, \slashed{\partial} \gamma_5 f \, \Gamma^{f,{\mathrm{A}}}(\partial^2) + \bar{f} f \, \Gamma^{f,{\mathrm{S}}} (\partial^2)+ \bar{f} \gamma_5 f \, \Gamma^{f,{\mathrm{P}}}(\partial^2).
\label{eq:apD:2}
\ee
In Table \ref{tab:apD:CPtable},
\newcolumntype{?}{!{\vrule width 0.005mm}}
\begin{table}[!h]%
\begin{normalsize}
\normalsize
\begin{center}
\begin{tabular}
{@{\hspace{3mm}}
>{\centering\arraybackslash}p{1.5cm}
?
>{\centering\arraybackslash}p{2.0cm}
>{\centering\arraybackslash}p{2.0cm}
>{\centering\arraybackslash}p{2.0cm}
@{\hspace{3mm}}}
\hlinewd{1.1pt}
& C & P & CP\\
\hline\\[-3mm]
$\bar{\psi}_i \, \psi_j$ & $\bar{\psi}_j \, \psi_i$ & $\bar{\psi}_i \, \psi_j$ & $\bar{\psi}_j \, \psi_i$\\[1mm]
$\bar{\psi}_i \, \gamma_5 \, \psi_j$ & $\bar{\psi}_j \, \gamma_5 \, \psi_i$ & $- \bar{\psi}_i \, \gamma_5 \, \psi_j$ & $- \bar{\psi}_j \, \gamma_5 \, \psi_i$ \\[1mm]
$\bar{\psi}_i \, \slashed{\partial} \, \psi_j$ & $\bar{\psi}_j \, \slashed{\partial} \, \psi_i$ & $\bar{\psi}_i \, \slashed{\partial} \, \psi_j$ & $\bar{\psi}_j \, \slashed{\partial} \, \psi_i$ \\[1mm]
$\bar{\psi}_i \, \slashed{\partial} \, \gamma_5 \psi_j$ & $- \bar{\psi}_j \, \slashed{\partial} \, \gamma_5 \, \psi_i$ & $- \bar{\psi}_i \, \slashed{\partial} \, \gamma_5 \, \psi_j$ & $\bar{\psi}_j \, \slashed{\partial} \, \gamma_5 \, \psi_i$ \\[1mm]
\hlinewd{1.1pt}
\end{tabular}
\end{center}
\vspace{-5mm}
\end{normalsize}
\caption{Results of the application of the symmetries C, P and CP to the terms in the left column.}
\label{tab:apD:CPtable}
\end{table}
\normalsize
we show the action of the operators C, P and CP on the four bilinear structures of eq. \ref{eq:apD:2}, but for general fields $\psi_i$ and $\psi_j$. In the particular case $\psi_i=\psi_j=f$, there is CP violation in eq. \ref{eq:apD:2} if and only if $\Gamma^{f,{\mathrm{P}}} \neq 0$. Using eq. \ref{eq:apD:conv}, we thus conclude:
\be 
\text{CP violation in} \,  \mathcal{L}_{\text{eff}}^{\bar{f} f}
\quad \Leftrightarrow
\quad 
\Gamma^{f,{\mathrm{P}}} \neq 0
\quad \Leftrightarrow
\quad 
\Gamma^{f,\mathrm{l}} \neq \Gamma^{f,\mathrm{r}}.
\label{eq:apD:box}
\ee
Comparing eqs.  \ref{eq:apD:13} and \ref{eq:apD:2}, it is clear that the term in eq.  \ref{eq:apD:13} proportional to $\mathrm{Im} [m_f]$ violates CP.
One might wonder why this conclusion is relevant; after all, and as we showed, one can always avoid such term by a rephasing of the Weyl spinors. But in that case, what happens at loop level? In particular, if we perform such rephasing and if CP violation contributes to fermionic 2-point functions at one-loop, how are we assured that there will be a counterterm for such contribution?
It is to these questions that we now turn.

\subsection{Fermionic counterterms and CP violation}

\n When renormalizing the theory, the tree-level parameters are identified as usual with bare parameters. Since the tree-level parameter $m_f$ was in general complex,
the bare parameter $m_{f(0)}$ is also in general complex. Writing eq. \ref{eq:apD:my11} in terms of bare quantities, we have:
\be
-\mathcal{L}_{\text{mass},f}
= m_{f(0)} \, \bar{f}_{\mathrm{L}(0)} \, f_{\mathrm{R}(0)} + m_{f(0)}^* \, \bar{f}_{\mathrm{R}(0)} \, f_{\mathrm{L}(0)},
\label{eq:bare-mass-Lag}
\ee
such that
\be
\label{eq:m-expa}
m_{f(0)} = m_f + \delta m_f,
\qquad
f_{\mathrm{L}(0)} = f_{\mathrm{L}} + \frac{1}{2} \delta Z^{f,\mathrm{L}} f_{\mathrm{L}},
\qquad
f_{\mathrm{R}(0)} = f_{\mathrm{R}} + \frac{1}{2} \delta Z^{f,\mathrm{R}} f_{\mathrm{R}}.
\ee
Here, $m_f$ and $\delta m_f$ are in general complex.
However, based on what we showed one could do at tree-level, we can take the freedom in the basis choice of the (renormalized) chiral spinors $f_{\mathrm{L}}$ and $f_{\mathrm{R}}$ to render $m_f$ real. After that, $\delta m_f$ is still in general complex.%
\fn{Cf. section \ref{sec:simple} for an explicit example of a case where a counterterm remains complex even after the corresponding renormalized parameter is rendered real.}
%
%
%
Separating the renormalized terms and the counterterms, eq. \ref{eq:bare-mass-Lag} then becomes:
\be
-\mathcal{L}_{\text{mass},f}
=  
\mathcal{L}^{\text{reno.}}_{\text{mass},f}
+ \mathcal{L}^{\text{CT}}_{\text{mass},f},
\ee
where
\bs
\label{eq:reno-and-CT}
\begin{flalign}
\label{eq:f-reno}
&\mathcal{L}^{\text{reno.}}_{\text{mass},f}
=
m_f \left( \bar{f}_{\mathrm{L}} f_{\mathrm{R}} + \bar{f}_{\mathrm{R}} f_{\mathrm{L}}\right) 
= m_f \bar{f} f,
\\
\label{eq:f-CT}
&\mathcal{L}^{\text{CT}}_{\text{mass},f}
=
\delta M_f \, \bar{f}_{\mathrm{L}} f_{\mathrm{R}}
+ \delta M_f^* \, \bar{f}_{\mathrm{R}} f_{\mathrm{L}}
=
\mathrm{Re} [\delta M_f] \, \bar{f} f + i \, \mathrm{Im} [\delta M_f] \, \bar{f} \gamma_5 f,
\end{flalign}
\es
with
\be
\delta M_f \equiv \left(\delta m_f + \frac{1}{2} m_f \, \delta Z^{f,L *} + \frac{1}{2} m_f \, \delta Z^{f,R}\right).
\label{eq:delta-Mf}
\ee
Eqs. \ref{eq:reno-and-CT} clarify the importance of the discussion of the previous section: we see that, due to our basis choice for the chiral spinors, there is no CP violation in the renormalized terms, eq. \ref{eq:f-reno}. But it is also clear that, even after that basis choice, there \textit{is} CP violation in the counterterms, eq. \ref{eq:f-CT}, as long as $\mathrm{Im} [\delta M_f] \neq 0$ (recall eqs. \ref{eq:apD:2} and \ref{eq:apD:box}). Therefore, if CP-violating effects show up at one-loop level, there will in general be counterterms to absorb their divergences.

\subsection{CP violation in the OSS scheme}

\n We now focus on the consequences of the OSS renormalization conditions on the different terms of $\delta M_f$.
We start by noting that, motivated by the fact that
\be
\delta m_f \, \bar{f}_{\mathrm{L}} f_{\mathrm{R}}
+\delta m_f^*  \, \bar{f}_{\mathrm{R}} f_{\mathrm{L}} =
\delta m_f \, \bar{f} \gamma_{\mathrm{R}} f
+\delta m_f^* \, \bar{f} \gamma_{\mathrm{L}} f,
\ee
we make the identification:
\be
\delta m_f^{\mathrm{R}} \equiv \delta m_f,
\qquad
\delta m_f^{\mathrm{L}} \equiv \delta m_f^*,
\ee
which implies, in particular,
\be
\delta m_f^{\mathrm{R}} = \delta m_f^{\mathrm{L} *},
\qquad
\mathrm{Im} \left[\delta m_f\right] = \dfrac{\delta m_f^{\mathrm{R}} - \delta m_f^{\mathrm{L}}}{2 i}.
\label{eq:apD:reldeltams}
\ee

\n Now, the fermionic renormalized 2-point function reads:
\be
\hat{\Gamma}^{\bar{f} f}_{i j} (p) = \delta_{i j} \left(\slashed{p} - m_{f,i}\right) + \hat{\Sigma}^{\bar{f} f}_{i j} (p),
\label{eq:GammaRenFermion}
\ee
and we can use the parameterization of eq. \ref{eq:apD:17} for renormalized quantities:
\be
\hat{\Gamma}_{i j}^{\bar{f} f}(p) = \slashed{p} \frac{1-\gamma_{5}}{2} \hat{\Gamma}_{i j}^{f, \mathrm{L}}(p^2)+ \slashed{p} \frac{1+\gamma_{5}}{2} \hat{\Gamma}_{i j}^{f, \mathrm{R}}(p^2)+\frac{1-\gamma_{5}}{2} \hat{\Gamma}_{i j}^{f, \mathrm{l}}(p^2)+\frac{1+\gamma_{5}}{2} \hat{\Gamma}_{i j}^{f, \mathrm{r}}(p^2),
\label{eq:GammaRenoFermion}
\ee
which, in the particular case of the renormalized one-loop GFs, reads:
\be
\hat{\Sigma}_{i j}^{\bar{f} f}(p) = \slashed{p} \frac{1-\gamma_{5}}{2} \hat{\Sigma}_{i j}^{f, \mathrm{L}}(p^2)+ \slashed{p} \frac{1+\gamma_{5}}{2} \hat{\Sigma}_{i j}^{f, \mathrm{R}}(p^2)+\frac{1-\gamma_{5}}{2} \hat{\Sigma}_{i j}^{f, \mathrm{l}}(p^2)+\frac{1+\gamma_{5}}{2} \hat{\Sigma}_{i j}^{f, \mathrm{r}}(p^2).
\label{eq:SigmaRenoFermion}
\ee

\n Expanding the Lagrangian using eqs. \ref{eq:mass-fermion-expa} and \ref{eq:field-fermion-expa}, we find%
\bs
\label{eq:SigmaRenoDecompFermion}
\bea
\hat{\Sigma}_{i j}^{f, \mathrm{L}}(p^2) &=& \Sigma_{i j}^{f, \mathrm{L}}(p^2)+\frac{1}{2}\left(\delta Z_{i j}^{f, \mathrm{L}}+\delta Z_{i j}^{f, \mathrm{L}^{\dagger}}\right), \\
\hat{\Sigma}_{i j}^{f, \mathrm{R}}(p^2) &=& \Sigma_{i j}^{f, \mathrm{R}}(p^2)+\frac{1}{2}\left(\delta Z_{i j}^{f, \mathrm{R}}+\delta Z_{i j}^{f, \mathrm{R}^{\dagger}}\right), \\
\hat{\Sigma}_{i j}^{f, \mathrm{l}}(p^2) &=& \Sigma_{i j}^{f, \mathrm{l}}(p^2)-\frac{1}{2}\left(m_{f, i} \delta Z_{i j}^{f, \mathrm{L}}+m_{f, j} \delta Z_{i j}^{f, \mathrm{R}^{\dagger}}\right)-\delta_{i j} \, \delta m_{f, i}^{\mathrm{L}}, \\
\hat{\Sigma}_{i j}^{f, \mathrm{r}}(p^2) &=& \Sigma_{i j}^{f, \mathrm{r}}(p^2)-\frac{1}{2}\left(m_{f, i} \delta Z_{i j}^{f, \mathrm{R}}+m_{f, j} \delta Z_{i j}^{f, \mathrm{L} \dagger}\right)-\delta_{i j} \, \delta m_{f, i}^{\mathrm{R}},
\eea
\es
so that, using eqs. \ref{eq:GammaRenFermion}, \ref{eq:SigmaRenoFermion} and \ref{eq:SigmaRenoDecompFermion}, we obtain:
\bs
\label{eq:GammaRenoDecompFermion}
\bea
\hat{\Gamma}_{i j}^{f, \mathrm{L}}(p^2) &=& \delta_{ij} + \Sigma_{i j}^{f, \mathrm{L}}(p^2)+\frac{1}{2}\left(\delta Z_{i j}^{f, \mathrm{L}}+\delta Z_{i j}^{f, \mathrm{L}^{\dagger}}\right), \\
\hat{\Gamma}_{i j}^{f, \mathrm{R}}(p^2) &=& \delta_{ij} + \Sigma_{i j}^{f, \mathrm{R}}(p^2)+\frac{1}{2}\left(\delta Z_{i j}^{f, \mathrm{R}}+\delta Z_{i j}^{f, \mathrm{R}^{\dagger}}\right), \\
\hat{\Gamma}_{i j}^{f, \mathrm{l}}(p^2) &=& - \delta_{ij} m_{f,i} + \Sigma_{i j}^{f, \mathrm{l}}(p^2)-\frac{1}{2}\left(m_{f, i} \delta Z_{i j}^{f, \mathrm{L}}+m_{f, j} \delta Z_{i j}^{f, \mathrm{R}^{\dagger}}\right)-\delta_{i j} \, \delta m_{f, i}^{\mathrm{L}}, \\
\hat{\Gamma}_{i j}^{f, \mathrm{r}}(p^2) &=& - \delta_{ij} m_{f,i} + \Sigma_{i j}^{f, \mathrm{r}}(p^2)-\frac{1}{2}\left(m_{f, i} \delta Z_{i j}^{f, \mathrm{R}}+m_{f, j} \delta Z_{i j}^{f, \mathrm{L} \dagger}\right)-\delta_{i j} \, \delta m_{f, i}^{\mathrm{R}}.
\eea
\es

\n The OSS renormalization conditions are (no summation in the indices is implicit):
\bs
\bea
\left[
\widetilde{\operatorname{Re}} \left\{\hat{\Gamma}_{i j}^{\bar{f} f}(p)\right\} u_{j}(p)
\right]
\Big|_{\slashed{p} \, u_{j}(p) = m_{f, j} u_{j}(p)} &=& 0, \\
\lim_{\slashed{p} \, u_{i}(p) \to m_{f, i} u_{i}(p)}
\,
\frac{\slashed{p}+m_{f,i}}{p^{2}-m_{f, i}^{2}}\widetilde{\operatorname{Re}} \left[\hat{\Gamma}_{i i}^{\bar{f} f}(p)\right] u_{i}(p) &=& u_{i}(p).
\eea
\es

\n Inserting eq. \ref{eq:GammaRenoFermion}, we have respectively:
\bs
\label{eq:AuxFermions}
\begin{flalign}
&
m_{f, j} \, \widetilde{\operatorname{Re}} \, \hat{\Gamma}_{i j}^{f, \mathrm{L}}\left(m_{f, j}^{2}\right)+\widetilde{\operatorname{Re}} \, \hat{\Gamma}_{i j}^{f, \mathrm{r}}\left(m_{f, j}^{2}\right)=0,
\qquad
m_{f, j} \, \widetilde{\operatorname{Re}} \, \hat{\Gamma}_{i j}^{f, \mathrm{R}}\left(m_{f, j}^{2}\right)+\widetilde{\operatorname{Re}} \, \hat{\Gamma}_{i j}^{f, \mathrm{l}}\left(m_{f, j}^{2}\right)=0, 
\label{eq:AuxFermionsA}
\\
&
\widetilde{\operatorname{Re}} \, \Bigg\{\hat{\Gamma}_{i i}^{f, \mathrm{R}}\left(m_{f, i}^{2}\right)+\hat{\Gamma}_{i i}^{f, \mathrm{L}}\left(m_{f, i}^{2}\right)+2 \frac{\partial}{\partial p^{2}} \bigg[m_{f, i}^{2}\left(\hat{\Gamma}_{i i}^{f, \mathrm{R}}\left(p^{2}\right)+\hat{\Gamma}_{i i}^{f, \mathrm{L}}\left(p^{2}\right)\right)
\hs{25mm} \nonumber \\[-3mm]
&
\hs{55mm} +m_{f, i}\left(\hat{\Gamma}_{i i}^{f, \mathrm{r}}\left(p^{2}\right)+\hat{\Gamma}_{i i}^{f, \mathrm{l}}\left(p^{2}\right)\right)\bigg]\bigg|_{p^{2}=m_{f, i}^{2}}\Bigg\}=2.
\label{eq:AuxFermionsB}
\end{flalign}
\es

\n Then, inserting eqs. \ref{eq:GammaRenoDecompFermion} in the two eqs. \ref{eq:AuxFermionsA} for $i=j$, we obtain respectively:%
\fn{We omit the indices $i$. Note that the case $i \neq j$ is irrelevant for this discussion, as the expressions for $\delta Z_{ij}^{f,\mathrm{L}}$ and $\delta Z_{ij}^{f,\mathrm{R}}$ (with $i \neq j$, eqs. \ref{eq:reno-fermB-pre} and \ref{eq:reno-fermB}) directly follow from the insertion of eqs. \ref{eq:GammaRenoDecompFermion} in the two equations \ref{eq:AuxFermionsA} with $i \neq j$. In that case, then, the OSS conditions leave no freedom.}
\bs
\label{eq:apD:setMasses}
\bea
\delta m_{f}^{\mathrm{R}}
&=&
\dfrac{1}{2} \widetilde{\operatorname{Re}}
\bigg[
2 \, m_{f} \Sigma^{f,\mathrm{L}}(m_{f}^2) +  2 \, \Sigma^{f,\mathrm{r}}(m_{f}^2) + m_{f} \left(\delta Z^{f, \mathrm{L}} - \delta Z^{f, \mathrm{R}} \right)
\bigg], \label{eq:apD:my4} \\
\delta m_{f}^{\mathrm{L}}
&=&
\dfrac{1}{2} \widetilde{\operatorname{Re}}
\bigg[
2 \, m_{f} \Sigma^{f,\mathrm{R}}(m_{f}^2) +  2 \, \Sigma^{f,\mathrm{l}}(m_{f}^2) + m_{f} \left(\delta Z^{f, \mathrm{R}} - \delta Z^{f, \mathrm{L}} \right)
\bigg].
\label{eq:apD:my3}
\eea
\es

\n Now, the hermiticity of the up-to-one-loop action implies:
\be
\widetilde{\operatorname{Re}} \,\Gamma_{ij}^{\bar{f} f}(p) = \gamma^0  \left[ \widetilde{\operatorname{Re}} \, \Gamma_{ij}^{\bar{f} f} (p) \right]^{\dagger} \gamma^0, 
\ee
so that, in particular,
\begin{gather}
\widetilde{\operatorname{Re}} \,\Sigma_{ij}^{f,\mathrm{L}}(p^2) =
\left[
\widetilde{\operatorname{Re}} \,\Sigma_{ji}^{f,\mathrm{L}}(p^2)
\right]^*,
\qquad
\widetilde{\operatorname{Re}} \,\Sigma_{ij}^{f,\mathrm{R}}(p^2)
=
\left[
\widetilde{\operatorname{Re}} \,\Sigma_{ji}^{f,\mathrm{R}}(p^2)
\right]^*,
\no
\widetilde{\operatorname{Re}} \,\Sigma_{ij}^{f,\mathrm{l}}(p^2)
=
\left[
\widetilde{\operatorname{Re}} \,\Sigma_{ji}^{f,\mathrm{r}}(p^2)
\right]^*.
\label{eq:apD:30c}
\end{gather}
Hence, inserting eqs. \ref{eq:GammaRenoDecompFermion} in eq. \ref{eq:AuxFermionsB} (again for $i=j$), and subtracting the complex conjugate of eq. \ref{eq:apD:my4} to eq. \ref{eq:apD:my3} while using eqs. \ref{eq:apD:reldeltams} and \ref{eq:apD:30c}, we get:
\bs
\label{eq:apD:setFields}
\begin{flalign}
\dfrac{\delta Z^{f, \mathrm{L}} + \delta Z^{f, \mathrm{L} *}}{2} &=  -\widetilde{\operatorname{Re}} \, \Sigma^{f, \mathrm{L}}\left(m_{f}^{2}\right)-  m_{f} \frac{\partial}{\partial p^{2}} \widetilde{\operatorname{Re}}\bigg[m_{f}\left(\Sigma^{f, \mathrm{L}}(p^2)+\Sigma^{f, \mathrm{R}}(p^2)\right) \nonumber
\\[-4mm]
& \hs{65mm} + \Sigma^{f, \mathrm{l}}(p^2)+\Sigma^{f, \mathrm{r}}(p^2)\bigg]\bigg|_{p^{2}=m_{f}^{2}} , \\[-2mm]
\dfrac{\delta Z^{f, \mathrm{R}} + \delta Z^{f, \mathrm{R} *}}{2} &=  -\widetilde{\operatorname{Re}} \, \Sigma^{f, \mathrm{R}}\left(m_{f}^{2}\right)- m_{f} \frac{\partial}{\partial p^{2}} \widetilde{\operatorname{Re}}\bigg[m_{f}\left(\Sigma^{f, \mathrm{L}}(p^2)+\Sigma^{f, \mathrm{R}}(p^2)\right)
\nonumber \\[-4mm]
&  \hs{65mm} +\Sigma^{f, \mathrm{l}}(p^2)+\Sigma^{f, \mathrm{r}}(p^2)\bigg]\bigg|_{p^{2}=m_{f}^{2}} .
\end{flalign}
\es
There are thus 6 real degrees of freedom (dof)---two for each of the three complex counterterms $\delta m_{f}, \delta Z^{f, \mathrm{L}}$ and $\delta Z^{f, \mathrm{R}}$---, but the diagonal OSS conditions only imply the 4 relations of eqs. \ref{eq:apD:setMasses} and \ref{eq:apD:setFields}. Hence, there are two free dof, so that many solutions satisfy those conditions. Before considering them, note that, for the specific case of the non-renormalized diagonal one-loop 2-point function,
eq. \ref{eq:apD:box} implies:
\be
\text{CP violation in} \, \widetilde{\operatorname{Re}} \,  \Sigma^{\bar{f} f}
\quad
\Leftrightarrow
\quad
\widetilde{\operatorname{Re}} \, \Sigma^{f, \mathrm{l}}
\neq
\widetilde{\operatorname{Re}} \, \Sigma^{f, \mathrm{r}}.
\label{eq:my-CP-one-loop}
\ee

\n Now, among the many possible solutions, we restrict ourselves to two.
The first chooses $\mathrm{Im} \left[\delta Z^{f, \mathrm{L}}\right] = \mathrm{Im} \left[\delta Z^{f, \mathrm{R}}\right] = 0$.
In this case, eq. \ref{eq:my-CP-one-loop} together with eqs. \ref{eq:apD:setMasses} and \ref{eq:apD:setFields} leads to:
\be
\text{CP violation in} \, \widetilde{\operatorname{Re}} \,  \Sigma^{\bar{f} f}
\quad
\stackrel{\scriptsize\mathrm{Im} \big[\delta Z^{f, {\mathrm{L}/\mathrm{R}}}\big] = 0}{\Leftrightarrow}
\quad 
\delta m_f^{\mathrm{R}} \neq \delta m_f^{\mathrm{L}} .
\label{eq:apD:box2}
\ee
This relation shows that, whenever there is no CP violation in  one-loop fermionic diagonal 2-point functions, there is no need to introduce $\delta m_{f}^{\mathrm{L}}$ nor $\delta m_{f}^{\mathrm{R}}$: they would be the same, and both equal to $\delta m_{f}$, which is thus preferred. Conversely, if there is CP violation in $\widetilde{\operatorname{Re}} \, \Sigma^{\bar{f} f}$, the choice $\mathrm{Im} \left[\delta Z^{f, \mathrm{L}}\right] = \mathrm{Im} \left[\delta Z^{f, \mathrm{R}}\right] = 0$ forces   $\delta m_{f}^{\mathrm{L}}$  and  $\delta m_{f}^{\mathrm{R}}$ to be different (or, what is equivalent, forces $\delta m_{f}$ to be complex). This is the solution we adopted in eqs. \ref{eq:reno-ferm} for the case $i=j$.\fn{These follow from eqs. \ref{eq:apD:setMasses} and \ref{eq:apD:setFields} when imposing $\mathrm{Im} \left[\delta Z^{f, \mathrm{L}}\right] = \mathrm{Im} \left[\delta Z^{f, \mathrm{R}}\right] = 0$.}

\n Another simple solution uses one of the two free dofs to impose $\delta m_{f}^{\mathrm{L}} = \delta m_{f}^{\mathrm{R}}$. In this case, by subtracting eq. \ref{eq:apD:my4} to eq. \ref{eq:apD:my3}, taking the imaginary part of the result and using eqs. \ref{eq:apD:30c} and \ref{eq:my-CP-one-loop}, we get:
\be
\text{CP violation in} \, \widetilde{\operatorname{Re}} \,  \Sigma^{\bar{f} f}
\quad
\stackrel{\delta m_{f}^{\mathrm{L}} = \delta m_{f}^{\mathrm{R}}}{\Leftrightarrow}
\quad 
\mathrm{Im} \left[\delta Z^{f, \mathrm{L}} \right] \neq \mathrm{Im} \left[\delta Z^{f, \mathrm{R}}\right].
\label{eq:apD:casedmLdmR}
\ee
It is then clear that, in the context of OSS, CP violation in fermionic diagonal 2-point functions does not necessarily force the relation $\delta m_{f}^{\mathrm{L}} \neq \delta m_{f}^{\mathrm{R}}$. Only, by choosing $\delta m_{f}^{\mathrm{L}} = \delta m_{f}^{\mathrm{R}}$, it follows that $\delta Z^{f, \mathrm{L}}$ and $\delta Z^{f, \mathrm{R}}$ cannot be both real. This is the solution usually adopted in the cMSSM (cf. e.g. refs.~\cite{Heinemeyer:2010mm,Fritzsche:2011nr}).

\renewcommand{\thesection}{C}
\renewcommand{\thesubsection}{C.\arabic{subsection}}
\renewcommand{\thesubsubsection}{C.\arabic{subsection}.\arabic{subsubsection}}
\setcounter{equation}{0}
\setcounter{figure}{0}
\renewcommand{\theequation}{\thesection.\arabic{equation}}
\renewcommand{\thefigure}{\thesection.\arabic{figure}}
\section{Symmetry relations}
\label{app:Sym}

\n To our knowledge, the procedure of fixing counterterms through symmetry relations has been originally proposed in ref.~\cite{Kanemura:2004mg}, having been later worked out and used by several authors \cite{Krause:2016oke,Krause:2017mal,Altenkamp:2017ldc,Denner:2018opp}.
As mentioned in section \ref{sec:UTOL}, the theory can be renormalized in its symmetric form (before SSB); in that case, it is enough to consider one field counterterm for each multiplet in order to cancel UV divergences \cite{Denner:2019vbn}.
On the other hand, the theory can also be renormalized in the broken phase, where counterterms for mixing parameters and mixed fields counterterms in general appear.
The two methods are not independent whenever fields in the symmetry basis are related with fields in the physical basis through mixing parameters.

\n In this appendix, we start by showing that one can exploit such dependence to write the divergent parts of the counterterms for mixing parameters in terms of the divergent parts of field counterterms; this we do in section \ref{sec:proc-sym}, considering a simple example. Then, in section \ref{sec:C2-case}, we apply the same method to the specific case of the mixing parameters of the scalar sector of the C2HDM, and discuss the limit of CP conservation.

\n The symmetry relations as such only have to be true for the divergent parts, since the common ground between the renormalization in the two forms (symmetric and broken) was the elimination of UV divergences.
Yet, as we will see in section \ref{sec:dep-sec}, in the case of \textit{independent} counterterms for mixing parameters (which need to be fixed) one can decide to fix them by requiring that the symmetry relations also hold for the finite parts. This is the reason why (independent) counterterms for mixing parameters can be fixed through symmetry relations. In that final section, we discuss the generality of the method, and consider the consequences of symmetry relations on dependent counterterms.

\subsection{A simple example}
\label{sec:proc-sym}

\n Let us consider an example involving SM particles only. From the bare version of eq. \ref{eq:gauge-rot-tree}, we have:
\be
\begin{pmatrix}
B_{\mu(0)} \\
W^3_{\mu(0)}
\end{pmatrix}
=
\begin{pmatrix}
c_{{\text{w}}(0)} & -s_{{\text{w}}(0)} \\
s_{{\text{w}}(0)} & c_{{\text{w}}(0)} 
\end{pmatrix}
\begin{pmatrix}
A_{\mu(0)} \\
Z_{\mu(0)}
\end{pmatrix}.
\label{eq:sym-bare}
\ee
According to the previous paragraphs, from the point of view of the cancellation of divergences, $W_{\mu}^a$ and $B_{\mu}$ can be renormalized with a single counterterm each. Hence,%
\fn{We define the counterterm $\delta Z_{W}^{\prime}$ with a prime to distinguish it from the counterterm for the physical $W^{+}$ field.}
\be
W^a_{\mu(0)}
\big|_{\Delta_{\epsilon}}
=
\left(1 + \dfrac{1}{2} \delta Z_{W}^{\prime}\right)\Big|_{\Delta_{\epsilon}} W^a_{\mu},
\qquad
B_{\mu(0)}\big|_{\Delta_{\epsilon}}
=
\left(1 + \dfrac{1}{2} \delta Z_{B}\right)\Big|_{\Delta_{\epsilon}} B_{\mu},
\ee
so that, in particular,
\be
\left.
\begin{pmatrix}
B_{\mu(0)} \\
W^3_{\mu(0)}
\end{pmatrix}
\right|_{\Delta_{\epsilon}}
=
\left.
\begin{pmatrix}
1 + \dfrac{1}{2} \delta Z_{B} & 0 \\
0 & 1 + \dfrac{1}{2} Z_{W}^{\prime}
\end{pmatrix}
\right|_{\Delta_{\epsilon}}
\begin{pmatrix}
B_{\mu} \\
W^3_{\mu}
\end{pmatrix}.
\label{eq:sym-expa-left}
\ee
On the other hand, from eqs. \ref{eq:field-gauge-expa} and \ref{eq:unnec},
\be
\begin{pmatrix}
A_{\mu(0)} \\ Z_{\mu(0)}
\end{pmatrix}
=
\begin{pmatrix}
1 + \dfrac{1}{2} \delta Z_{AA} & \dfrac{1}{2} \delta Z_{AZ} \\
\dfrac{1}{2} \delta Z_{ZA} & 1 + \dfrac{1}{2} \delta Z_{ZZ}
\end{pmatrix}
\begin{pmatrix}
A_{\mu}
\\
Z_{\mu}
\end{pmatrix},
\quad 
s_{{\text{w}}(0)} = s_{\text{w}} + \delta s_{\text{w}}, 
\quad
c_{{\text{w}}(0)} = c_{\text{w}} + \delta c_{\text{w}},
\label{eq:sym-expa-right}
\ee
and the renormalized quantities are defined such that:
\be
\begin{pmatrix}
B_{\mu} \\
W^3_{\mu}
\end{pmatrix}
=
\begin{pmatrix}
c_{\text{w}} & -s_{\text{w}} \\
s_{\text{w}} & c_{\text{w}} 
\end{pmatrix}
\begin{pmatrix}
A_{\mu} \\
Z_{\mu}
\end{pmatrix}.
\label{eq:sym-reno}
\ee
As a consequence, eq. \ref{eq:sym-expa-left} becomes:
\be
\left.
\begin{pmatrix}
B_{\mu(0)} \\
W^3_{\mu(0)}
\end{pmatrix}
\right|_{\Delta_{\epsilon}}
=
\left.
\begin{pmatrix}
1 + \dfrac{1}{2} \delta Z_{B} & 0 \\
0 & 1 + \dfrac{1}{2} Z_{W}^{\prime}
\end{pmatrix}
\right|_{\Delta_{\epsilon}}
\begin{pmatrix}
c_{\text{w}} & -s_{\text{w}} \\
s_{\text{w}} & c_{\text{w}} 
\end{pmatrix}
\begin{pmatrix}
A_{\mu} \\
Z_{\mu}
\end{pmatrix}.
\label{eq:sym-left}
\ee
On the other hand, we can apply eq. \ref{eq:sym-expa-right} to the right-hand side of eq. \ref{eq:sym-bare} and consider the divergent parts only to get:
\be
\left.
\begin{pmatrix}
B_{\mu(0)} \\
W^3_{\mu(0)}
\end{pmatrix}
\right|_{\Delta_{\epsilon}}
=
\left.
\left[
\begin{pmatrix}
c_{\text{w}} + \delta c_{\text{w}} & -s_{\text{w}}  - \delta s_{\text{w}}\\
s_{\text{w}} + \delta s_{\text{w}} & c_{\text{w}} + \delta c_{\text{w}} 
\end{pmatrix}
\begin{pmatrix}
1 + \dfrac{1}{2} \delta Z_{AA} & \dfrac{1}{2} \delta Z_{AZ} \\
\dfrac{1}{2} \delta Z_{ZA} & 1 + \dfrac{1}{2} \delta Z_{ZZ}
\end{pmatrix}
\right]
\right|_{\Delta_{\epsilon}}
\begin{pmatrix}
A_{\mu}
\\
Z_{\mu}
\end{pmatrix}.
\label{eq:sym-right}
\ee
Finally, equating eqs. \ref{eq:sym-left} and \ref{eq:sym-right} to first order and noting that $\delta \theta_{\text{w}} =  - \delta c_{\text{w}}/s_{\text{w}}$, one concludes that:
\be
\delta \theta_{\text{w}}
\big|_{\Delta_{\epsilon}}
=
\dfrac{1}{4} \left(\delta Z_{AZ} - \delta Z_{Z A}\right)\big|_{\Delta_{\epsilon}}.
\label{eq:relWein}
\ee
This is the symmetry relation; it fixes the divergent part of the counterterm for the mixing parameter $\theta_{\text{w}}$ in terms of the divergent parts of the counterterms for the physical fields associated with $\theta_{\text{w}}$.
This relation is also valid in the SM, as well as in all its extensions that keep its gauge structure.%
\fn{The same is true for another relation that also follows from equating eqs. \ref{eq:sym-left} and \ref{eq:sym-right} to first order, namely,
\be
\delta Z_{ZZ}
\big\rvert_{\Delta_{\epsilon}}
=
\left.
\bigg[
\delta Z_{AA}
+
\cot\left(2 \, \theta_{\text{w}}\right)
\Big(
\delta Z_{AZ} + \delta Z_{ZA}
\Big)
\bigg]
\right\rvert_{\Delta_{\epsilon}}.
\ee
}

\n The procedure just used is generalizable to other mixing parameters (or sets of mixing parameters) of different models and can be summarized in four steps.
First, one considers the relation involving the mixing parameters at stake (i.e. transforming the fields in the symmetry basis into the mass basis)
in its bare form; in the above example, this is simply eq. \ref{eq:sym-bare}. 
Second, the bare fields in the symmetry basis are expanded---aiming at the elimination of divergences, so that only one field counterterm for multiplet is taken and only the divergent parts are considered---and, after that, the renormalized relation between the two bases is used; in the above example, the result of this step is eq. \ref{eq:sym-left}. 
In the third step, one expands instead the bare mixing parameters and bare physical fields and takes the divergent parts; in the above example, the result is eq. \ref{eq:sym-right}.
Finally, one equates the two previous steps to first order, which yields the divergent parts of the counterterms for the mixing parameters in terms of the divergent parts of field counterterms.

\subsection{The case of the C2HDM}
\label{sec:C2-case}

\n Applying this procedure to the scalar sector of the C2HDM, we get, for the counterterms for mixing parameters of the charged fields,%
\fn{
Besides, we also find the following relations for the field counterterms:
\bs
\begin{gather}
\mathrm{Im} \, \delta Z_{G^+H^+}
\big\rvert_{\Delta_{\epsilon}}
=
\mathrm{Im} \, \delta Z_{H^+G^+}
\big\rvert_{\Delta_{\epsilon}},
\\
\delta Z_{H^+H^+}\big\rvert_{\Delta_{\epsilon}}
=
\left.
\bigg[
\delta Z_{G^+G^+}
+
\cot\left(2 \beta\right)
\mathrm{Re}
\Big(
\delta Z_{G^+H^+} + \delta Z_{H^+G^+}
\Big)
\bigg]
\right\rvert_{\Delta_{\epsilon}}.
\label{eq:dZHPHP}
\end{gather}
\es
Had we started with a parameterization of $X$ that included an overall phase, the symmetry relations would have forced the divergent part of its counterterm to be zero. This means that such phase is really not needed, which explains why we ignored it in the first place (cf. eq. \ref{eq:charged-param-original} and note \ref{note:overall}).
}
\bs
\label{eq:charged-div-rel}
\bea
\label{eq:dbeta-div-rel}
{\delta \beta}
\big\rvert_{\Delta_{\epsilon}}
&=& \dfrac{1}{4}
\operatorname{Re}
\left.\Big[\delta Z_{G^+H^+} - \delta Z_{H^+G^+}\Big]
\right|_{\Delta_{\epsilon}},
\\
\label{eq:dzetaa-div-rel}
\delta \zeta_a
\big|_{\Delta_{\epsilon}}
&=&
-\dfrac{1}{2} \cot(2 \beta) \operatorname{Im} \left[\delta Z_{G^+H^+} \right]\big|_{\Delta_{\epsilon}},
\eea
\es
and, for those of the neutral fields,
%
%
\bs
\label{eq:dalphas}
\begin{flalign}
&
\delta \alpha_0
\big|_{\Delta_{\epsilon}}
=
\dfrac{1}{4} \sec(\alpha_2) \sec(\alpha_3)  \big( \delta Z_{G_0h_3} - \delta Z_{h_3G_0} \big)
\Big|_{\Delta_{\epsilon}},
\\[3mm]
&
\delta \alpha_1
\big|_{\Delta_{\epsilon}}
=
\dfrac{1}{4} \sec(\alpha_2)  \Big[  \cos(\alpha_3) \left( \delta Z_{h_1h_2} - \delta Z_{h_2h_1} \right)  +  \sin(\alpha_3) \left( \delta Z_{h_3h_1} -\delta Z_{h_1h_3} \right)  \Big]
\Big|_{\Delta_{\epsilon}},
\\[3mm]
&
\delta \alpha_2
\big|_{\Delta_{\epsilon}}
=
\dfrac{1}{4}  \sin(\alpha_3) \Big[ \delta Z_{h_1h_2} - \delta Z_{h_2h_1} + \cot(\alpha_3) \left( \delta Z_{h_1h_3} - \delta Z_{h_3h_1} \right)  \Big]
\Big|_{\Delta_{\epsilon}},
\\[3mm]
&
\delta \alpha_3
\big|_{\Delta_{\epsilon}}
=
\dfrac{1}{4} \bigg[\delta Z_{h_2h_3} - \delta Z_{h_3h_2} -   \cos(\alpha_3) \tan(\alpha_2)  \left( \delta Z_{h_1h_2} - \delta Z_{h_2h_1} \right) \nonumber\\[-3mm]
& \hs{55mm} + \sin(\alpha_3) \tan(\alpha_2) \left( \delta Z_{h_1h_3} - \delta Z_{h_3h_1} \right)  \bigg]
\bigg|_{\Delta_{\epsilon}},
\\[3mm]
&
\delta \alpha_4
\big|_{\Delta_{\epsilon}}
=
\dfrac{1}{4} \bigg[\delta Z_{h_1G_0} -\delta Z_{G_0h_1} + \sec(\alpha_3) \tan(\alpha_2) \left( \delta Z_{G_0h_3} - \delta Z_{h_3G_0} \right)  \bigg]
\bigg|_{\Delta_{\epsilon}},
\\[3mm]
&
\delta \alpha_5
\big|_{\Delta_{\epsilon}}
=
\dfrac{1}{4} \bigg[\delta Z_{h_2G_0} -\delta Z_{G_0h_2} + \tan(\alpha_3) \left( \delta Z_{G_0h_3} - \delta Z_{h_3G_0} \right)
\bigg]
\bigg|_{\Delta_{\epsilon}}.
\end{flalign}
\es
%
Note that the expressions for the divergent parts of $\delta \alpha_0$, $\delta \alpha_4$ and $\delta \alpha_5$ depend only on the divergent parts of fields counterterms for the mixing between $G_0$ and other neutral scalar fields. This means that, should there be no such mixing at one-loop, the divergent parts of the corresponding mixed fields counterterms would vanish---which would then imply the vanishing of the divergent parts of $\delta \alpha_0$, $\delta \alpha_4$ and $\delta \alpha_5$. In other words, if $G_0$ did not mix with the fields $h_1$, $h_2$ and $h_3$ at one-loop, the counterterms $\delta \alpha_0$, $\delta \alpha_4$ and $\delta \alpha_5$ would not be needed. This is consistent with description of the theory considered solely at tree-level, where the angles $\alpha_0$, $\alpha_4$ and $\alpha_5$ are not introduced. However, since $G_0$ does mix with the remaining neutral scalar fields at one-loop, the divergent parts of $\delta \alpha_0$, $\delta \alpha_4$ and $\delta \alpha_5$ are in general not zero, so that these counterterms must be considered---which in turn means that $\alpha_{0}$ $\alpha_{4}$ and $\alpha_{5}$ have to be included at tree-level when aiming at the one-loop renormalization of the theory.

\n It is also interesting to consider the limit of CP conservation; one possible limit is \cite{ElKaffas:2006gdt}:
%
%
\be
\alpha_1 \stackrel{\mathrm{CP}}{=} \alpha + \dfrac{\pi}{2},
\qquad
\alpha_2 \stackrel{\mathrm{CP}}{=} 0,
\qquad
\alpha_3 \stackrel{\mathrm{CP}}{=} 0
\qquad
(\text{with}
\ \
\alpha_0 = \beta,
\quad
\alpha_4 = 0,
\quad
\alpha_5 = 0),
\ee
in which case \cite{ElKaffas:2006gdt}:
%
\be
h_1 \stackrel{\mathrm{CP}}{=} h, 
\qquad
h_2 \stackrel{\mathrm{CP}}{=} -H,
\qquad
h_3 \stackrel{\mathrm{CP}}{=} A_0,
\ee
with $h$ and $H$ being CP-even and $A_0$ CP-odd. Then, given that there would be no one-loop mixing between fields with different CP values, eqs. \ref{eq:dalphas} would become:
\begin{flalign}
&
\delta \alpha_0
\big|_{\Delta_{\epsilon}}
\stackrel{\mathrm{CP}}{=}
\dfrac{1}{4} \big( \delta Z_{G_0A_0} - \delta Z_{A_0G_0} \big)
\big|_{\Delta_{\epsilon}},
\\[3mm]
&
\delta \alpha_1
\big|_{\Delta_{\epsilon}}
\stackrel{\mathrm{CP}}{=}
\dfrac{1}{4} \left( \delta Z_{Hh} - \delta Z_{hH} \right)
\big|_{\Delta_{\epsilon}},
\\[3mm]
&
\delta \alpha_2\big|_{\Delta_{\epsilon}}
\stackrel{\mathrm{CP}}{=}
\delta \alpha_3\big|_{\Delta_{\epsilon}}
\stackrel{\mathrm{CP}}{=}
\delta \alpha_4\big|_{\Delta_{\epsilon}}
\stackrel{\mathrm{CP}}{=}
\delta \alpha_5\big|_{\Delta_{\epsilon}}
\stackrel{\mathrm{CP}}{=}
0.
\end{flalign}
This is consistent with the symmetry relations in a 2HDM with CP conservation (cf. e.g. ref.~\cite{Krause:2016oke}).
Moreover, still in the context of CP conservation, $Z_{G^+H^+}$ would be real, which would imply that
$
\delta \zeta_a
\big|_{\Delta_{\epsilon}}
$
would vanish; consequently, if CP were a symmetry of the theory, there would also be no need to consider $\delta \zeta_a$.

\subsection{Dependent and independent mixing parameters}
\label{sec:dep-sec}

\n The symmetry relations have interesting consequences both when the counterterm for the mixing parameter at stake is dependend and when it is independent.
In the first case, one can use the symmetry relations to establish relations between the divergent parts of different independent counterterms. For example, given that \cite{Denner:1991kt,Fontes:PhD}:
\be
\delta c_{\text{w}} = \dfrac{\delta m_{\mathrm{W}}^2}{2 m_{\mathrm{W}} m_{\mathrm{Z}}} - \dfrac{m_{\mathrm{W}} \, \delta m_{\mathrm{Z}}^2}{2 m_{\mathrm{Z}}^3},
\label{eq:cwCT}
\ee
one can use eq. \ref{eq:relWein} to conclude that:

\be
\dfrac{1}{4} \left(\delta Z_{A Z} - \delta Z_{Z A}\right)\Big|_{\Delta_{\epsilon}}
=
\left.
\left(\dfrac{m_{\mathrm{W}} \, \delta m_{\mathrm{Z}}^2}{2 s_{\text{w}} m_{\mathrm{Z}}^3} -  \dfrac{\delta m_{\mathrm{W}}^2}{2 s_{\text{w}} m_{\mathrm{W}} m_{\mathrm{Z}}} \right)
\right|_{\Delta_{\epsilon}}.
\label{eq:eq:my-rel-thetaw}
\ee
Note that the same relation is in general not valid for the finite parts.%
\fn{This is the reason why the counterterms involved are still independent (even if their divergent parts are related).
Again, eq. \ref{eq:eq:my-rel-thetaw} is also valid in the SM, as well as in all extensions of this model that do not modify its gauge structure.}

\n In the case of counterterms for independent mixing parameters, the finite parts are \textit{a priori} not fixed. In fact, even if we are able to determine their divergent parts, the finite parts are free, and must be fixed by renormalization conditions.
But here is the trick: we can choose as renormalization conditions the symmetry relations \textit{including finite parts}.
Said otherwise, we can decide that the way through which we fix the independent counterterms is requiring that the symmetry relations hold not only for the divergent parts, but also for the finite parts \cite{Denner:2018opp}.
In particular, we can decide that the relations involved in eqs. \ref{eq:charged-div-rel} and \ref{eq:dalphas} also hold for the finite parts---whenever the counterterm for mixing parameters at stake is independent (which depends on the combination $C_i$).
This fixes the counterterms involved, thus justifying the expressions presented in section \ref{sec:RenoMixingParameters}.%
\fn{As discussed in section \ref{sec:gauge-dep}, the independent counterterms for the mixing parameters are calculated in the Feynman gauge.}

\n The prescription we just described---of promoting symmetry relations for independent mi-xing parameter counterterms to renormalization conditions---is completely general, not being restricted to the C2HDM. In fact, just as the procedure presented in section \ref{sec:proc-sym} is easily generalizable to other models, so is this prescription to fix independent counterterms for mixing parameters. Actually, not only it is easily generalizable, as it is very simple, leading to gauge independent observables (as discussed in section \ref{sec:gauge-dep}). 

\renewcommand{\thesection}{D}
\renewcommand{\thesubsection}{D.\arabic{subsection}}
\renewcommand{\thesubsubsection}{D.\arabic{subsection}.\arabic{subsubsection}}
\setcounter{equation}{0}
\setcounter{figure}{0}
\renewcommand{\theequation}{\thesection.\arabic{equation}}
\renewcommand{\thefigure}{\thesection.\arabic{figure}}
\section{FeynMaster 2: an ideal tool for renormalizing the C2HDM}
\label{app:FM}


\n In this appendix, we introduce \FMTS and describe its application to the renorma-lization of the C2HDM. We start by presenting the main differences regarding the previous version of the software. We then summarize its usefulness in the renormalization of models. Finally, we explain how it can be exploited to renormalize a model such as the C2HDM, providing illustrations of the different steps involved.

\subsection{The new version}

\n An overall description of \FM \, was already presented in the Introduction. 
The second version of the program, \FMT, has been very recently made available online in the webpage:
\begin{center}
\url{https://porthos.tecnico.ulisboa.pt/FeynMaster/}.
\end{center}
There, the new software can be downloaded, and a complete and auto-sufficient manual can be found.
\FMTS has significant improvements over the previous version of the program.
First, it is much more practical, since a) a model is now specified by a single model file, b) the file controlling the \FMS run now contains just the data relevant for the run, and c) the presence of \ts{QGRAF} has virtually vanished.
The last point means that there is no longer such thing as a \ts{QGRAF} style or convention, which in turn implies that there is a single set of conventions (the \FR \, ones) used throughout the whole program.
A second major improvement in \FMTS is that the calculated results can now be automatically stored. This implies that a certain process can be calculated once and for all, which considerably simplifies the manipulation of results and the use of \FMS as a whole (this improvement is especially important for renormalization, as shall be seen).
In the third place, the renormalization is significantly faster; for example, the generation and impression of the total set of Feynman rules (tree-level and counterterms) for the SM is now completed in just 5 minutes in normal laptops.
Finally, several functions were corrected, improved or simply created.

\subsection{An ideal tool for renormalizing models}

\n Having seen this, and given the list of tasks presented in the Introduction, it is easy to conclude that \FMTS is an extremely useful tool in the renormalization of a model. Such usefulness is manifest, first and foremost, in the fact that the different steps required for the renormalization of a model---definition of the model, generation of Feynman rules, generation of counterterms, one-loop calculations and renormalization conditions---can all be performed with \FMT.
One starts by writing the model, with the desired conventions. This allows to automatically generate the Feynman rules not only for the tree-level interactions, but also for the counterterms. With the former, one-loop processes can be automatically calculated and stored, which allows to fix all the counterterms in a desired subtraction scheme, thus completing the renormalization of the model.

\n But this is not all. One can then study different one-loop processes (meanwhile renormalized), disposing not only of all the tools of \FC, but also of some specific functions of \FMT. These can also be combined with the numerical interface with \t{\ts{Fortran}}, thus allowing a more complete analysis. In the end of the day, renormalization and the study of NLO processes has become quite simple.

\n A constant element in the chain of processes just described should be stressed: \textit{flexibility}.
This property begins in the moment when, in the writing of the model, the user defines the conventions. Not only he or she has the freedom to define those conventions at will, but---and what is perhaps more relevant---they are kept throughout the whole process: in Feynman rules, one-loop calculations, counterterms, etc. Indeed, by combining all the required steps in a single software, \FMTS assures that the conventions do not change when changing from one step to the other.%
\fn{This tends to be a problem when one needs to combine different softwares (e.g., one software for the generation of Feynman rules, and another one for the one-loop calculations), since different softwares usually define different quantities and variables in different ways.}

\n Flexibility is present in many other ways. Intermediate and final results can be manipulated: not only due to the intuitive structure of \FMT, but also to the user-friendly character of \FC \, and \t{\ts{Mathematica}}. Notebooks for processes are automatically written, in which all of the results already calculated (Feynman rules, counterterms, one-loop amplitudes, etc.) are immediately available. The writing of Feynman diagrams and analytical expressions in \LaTeX \hs{0.1mm} is immediate. The treatment of Dirac structures and the calculation of (one-loop) decay widths and cross sections is remarkably simple given the functions available in \FMT. The numerical interface includes the beginning of a \t{\ts{Fortran}} main file (adapted to the process at stake), as well as a template of a makefile.

\n These examples suffice to make our point: \FMTS is very convenient for renormalizing a model and to study it at NLO. We now illustrate these features in the particular case of the C2HDM.

\subsection{The C2HDM}

\n We organize this section in three parts. In a first one, we briefly describe some particularities concerning the implementation of the C2HDM as a model for \FMT. Then, we explain how to calculate the counterterms. In the end, we show how to use \FMTS to study processes at NLO.

\subsubsection{The model}

\n The C2HDM as a model for \FMTS is available online in the webpage of \FMTS (in the Type II version, cf. section \ref{section:Yukawa}). 
Here, we do not describe in detail all the different components of the model (the manual can be checked for details). Rather, we stress some aspects which may be less obvious.

\n First, the model for the C2HDM is an extension of the model for the SM, in the sense that it was built upon the model file for the SM. This implies, in particular, that it also contains the $\eta$'s from ref.~\cite{Romao:2012pq}, which are very useful to compare different conventions.
Now, the theory (the Lagrangian, with its particles and parameters) is originally written as a bare theory. And since we intend to renormalize it, we need to write it in a general basis. The bare quantities are identified with the sum of a renormalized quantity and its corresponding counterterm, as usual. The renormalized quantities obey eqs. \ref{eq:key} to \ref{eq:X-reno}, which are enforced through the restrictions file.
%
%
%
%
%
Counterterms such as $\delta \lambda_1$ will show up because the model is written in such a way that most of the dependent parameters are renormalized.%
\fn{Recall note \ref{eq:note-dep-CT} and eq. \ref{eq:unnec}. The circumstance that we renormalize $m_{11}^2$, $m_{22}^2$ and $m_{12 \text{I}}^2$ for convenience implies that \FMS will generate counterterms for these variables, in such a way that it does not know their dependences---i.e. does not know how they depend on other counterterms. In that case, counterterms for the linear terms in $h_1$, $h_2$, $h_3$ and $G_0$ are apparently generated. Only, they are zero (which can be proven by replacing the bare values of $m_{11}^2$, $m_{22}^2$ and $m_{12 \text{I}}^2$ for their expressions, and renormalizing the latter). This is a consequence of the fact that we are using the FJTS. For more details, see ref. \cite{Fontes:PhD}.
}
Although we did not need to renormalize them (they are dependent, so that one could rewrite them in terms of independent parameters, and renormalize the latter), we decided to renormalize them for convenience. This is indeed very convenient: although it introduces several unnecessary counterterms, it renders the expressions considerably more compact.%
\fn{
Actually, if no unnecessary counterterms were used, the renormalization would probably be unfeasible: as it is, it takes about 50 minutes in our laptops to generate the total set of Feynman rules for the tree-level as well as for the counterterm interactions. It would take much longer if unnecessary counterterms were avoided.} 

\begin{figure}[htb]
\centering
\includegraphics[width=0.85\textwidth]{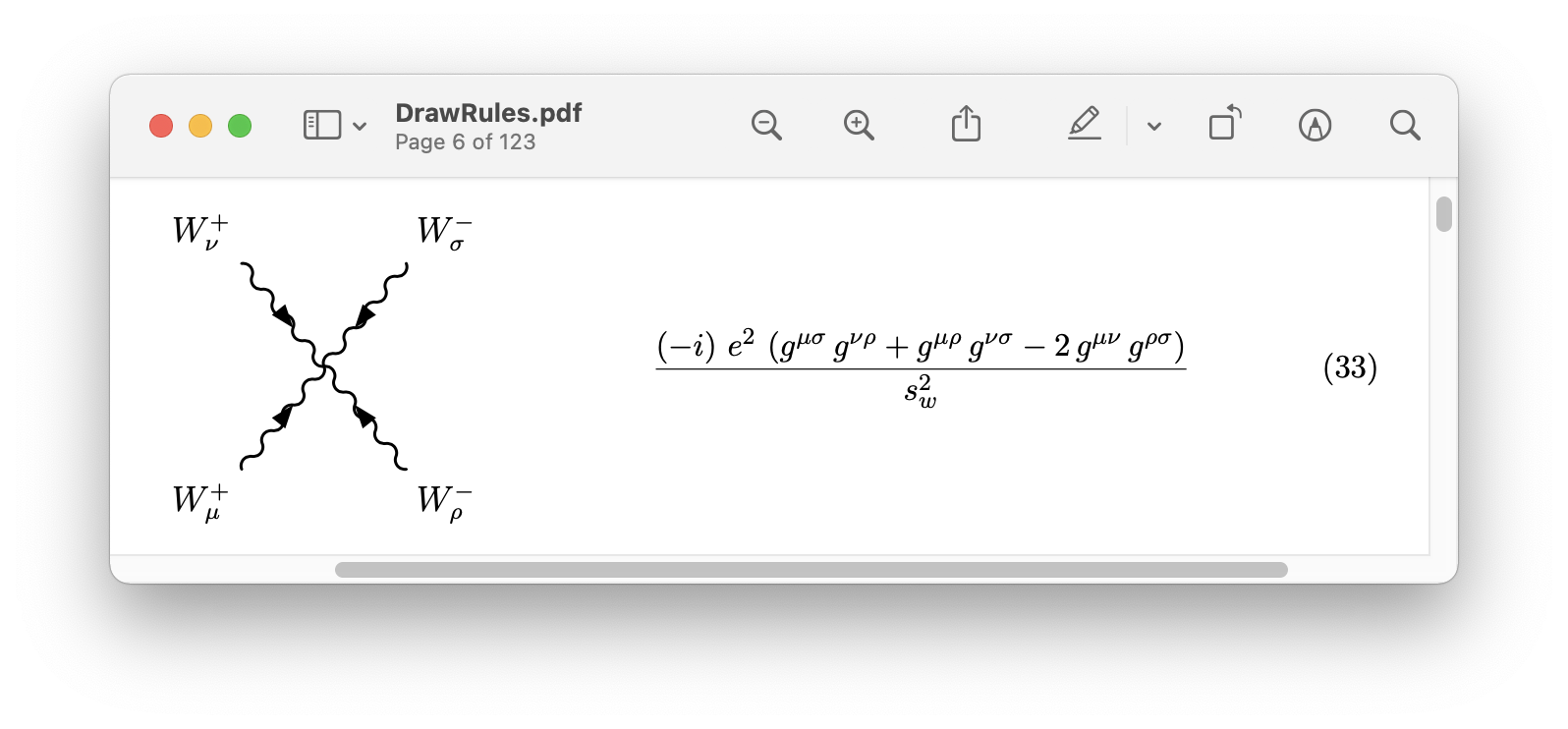}
\vs{-8mm}
\caption{Excerpt from the \t{DrawRules.pdf} file, containing the Feynman diagrams and rules for the tree-level interactions.}
\label{fig:FM:rule-ext}
\end{figure}

\begin{figure}[htb]
\centering
\includegraphics[width=0.85\textwidth]{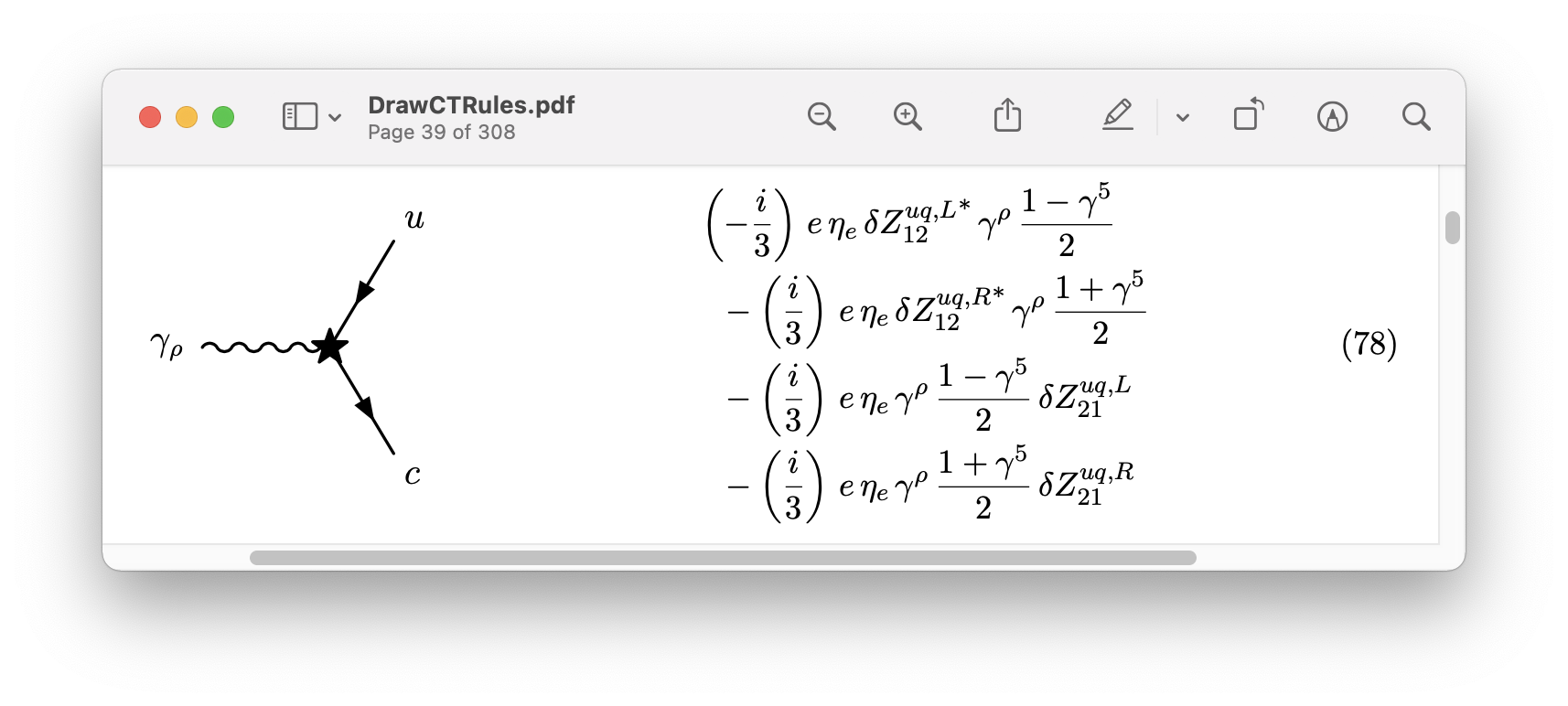}
\vs{-8mm}
\caption{Excerpt from the \t{DrawCTRules.pdf} file, containing the Feynman diagrams and rules for the counterterm interactions ($\eta_e$ is one of the $\eta$'s from ref.~\cite{Romao:2012pq}).}
\label{fig:FM:rule-CT-ext}
\end{figure}

\begin{figure}[htb]
\centering
\includegraphics[width=0.95\textwidth]{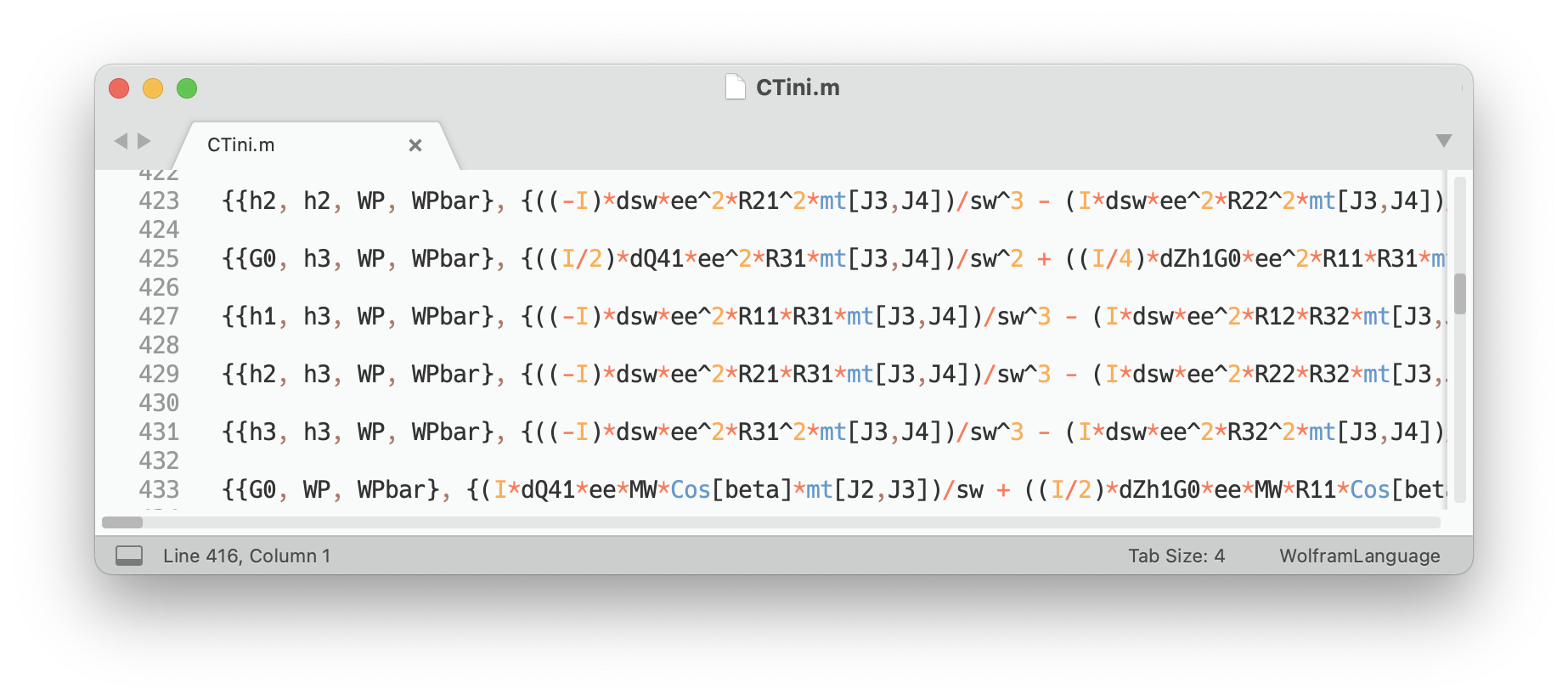}
\vs{-8mm}
\caption{Excerpt from the \t{CTini.m} file, containing the Feynman rules for the counterterms, written in a \FC-readable format.}
\label{fig:FM:rule-CT-int}
\end{figure}

\n By running \FMTS with the variables \t{FRinterLogic} and \t{RenoLogic} (of the \t{Control.m} file) set to \t{True}, the Feynman diagrams and rules for the tree-level and counterterm interactions are automatically generated. In figures \ref{fig:FM:rule-ext} and \ref{fig:FM:rule-CT-ext},
we present excerpts of the files containing the diagrams and rules written in \LaTeX. The corresponding files with the internal rules (i.e. the rules written in a \FC-readable format, to be used in the calculations) are also automatically generated. We show in figure \ref{fig:FM:rule-CT-int} 
an excerpt of one of them, \t{CTini.m}, which is the one concerned with the counterterms.
There, each non-empty line begins with the interaction at stake, containing after it the total set of counterterms contributing to it. Note the presence of dependent counterterms, such as $\delta Q_{41}$ and $\delta s_{\mathrm{w}}$ (represented by \t{dQ41} and \t{dsw}, respectively). Note
also the presence of renormalized parameters such as \t{R21}, corresponding to the entry $21$ of the matrix $R$.%
\fn{We used eqs. \ref{eq:Qsimple1} and \ref{eq:Qsimple2} to replace the elements of matrix $Q$ by those of matrix $R$. We could also have used eq. \ref{matrixR} to rewrite the entries of the matrix $R$, but the Feynman rules would then become longer.}
%

\begin{figure}[htb]
\centering
\includegraphics[width=0.95\textwidth]{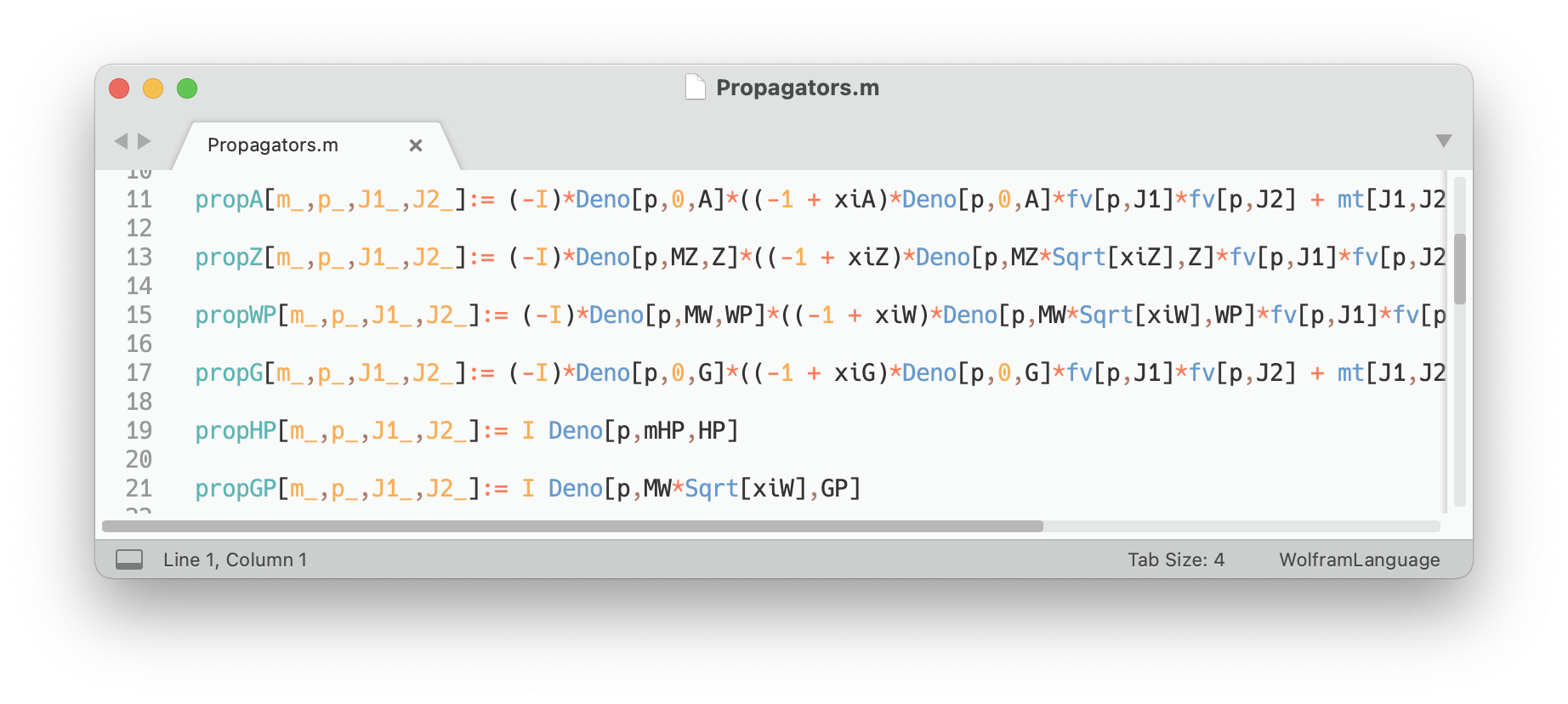}
\vs{-8mm}
\caption{Excerpt from the modified \t{Propagators.m} file of the C2HDM, containing the Feynman rules for the propagators written in a \FC-readable format.}
\label{fig:FM:rule-int}
\end{figure}

\n Finally, we have set the option \t{M\$PrMassFL} of the \FMS model file (defined in the \t{Extra.fr} file) to \t{False}. This means that \FMS does not extract the mass of a certain propagator from the term in the Lagrangian which is bilinear in the corresponding field, but rather writes a simplified version of the propagator.%
\fn{Otherwise, the masses in the propagators of the scalar particles would be written in terms of parameters in the potential. For more details, cf. the manual.}
But it also means that the gauge boson propagators are written in the Feynman gauge. In order to study gauge-dependences, however, they must be written in an arbitrary $R_{\xi}$ gauge. As the totality of propagators that depend on the gauge exist also in the SM, we can copy them---assuming the user has already generated the Feynman rules for SM---from the file \t{Propagators.m} of the SM (which lies inside the folder \t{FeynmanRules}, contained in the \FMS output of the SM), and paste it in the equivalent file for the C2HDM. The final version should look like figure \ref{fig:FM:rule-int}

\subsubsection{Calculation of counterterms}

\begin{figure}[htb]
\centering
\includegraphics[width=0.5\textwidth]{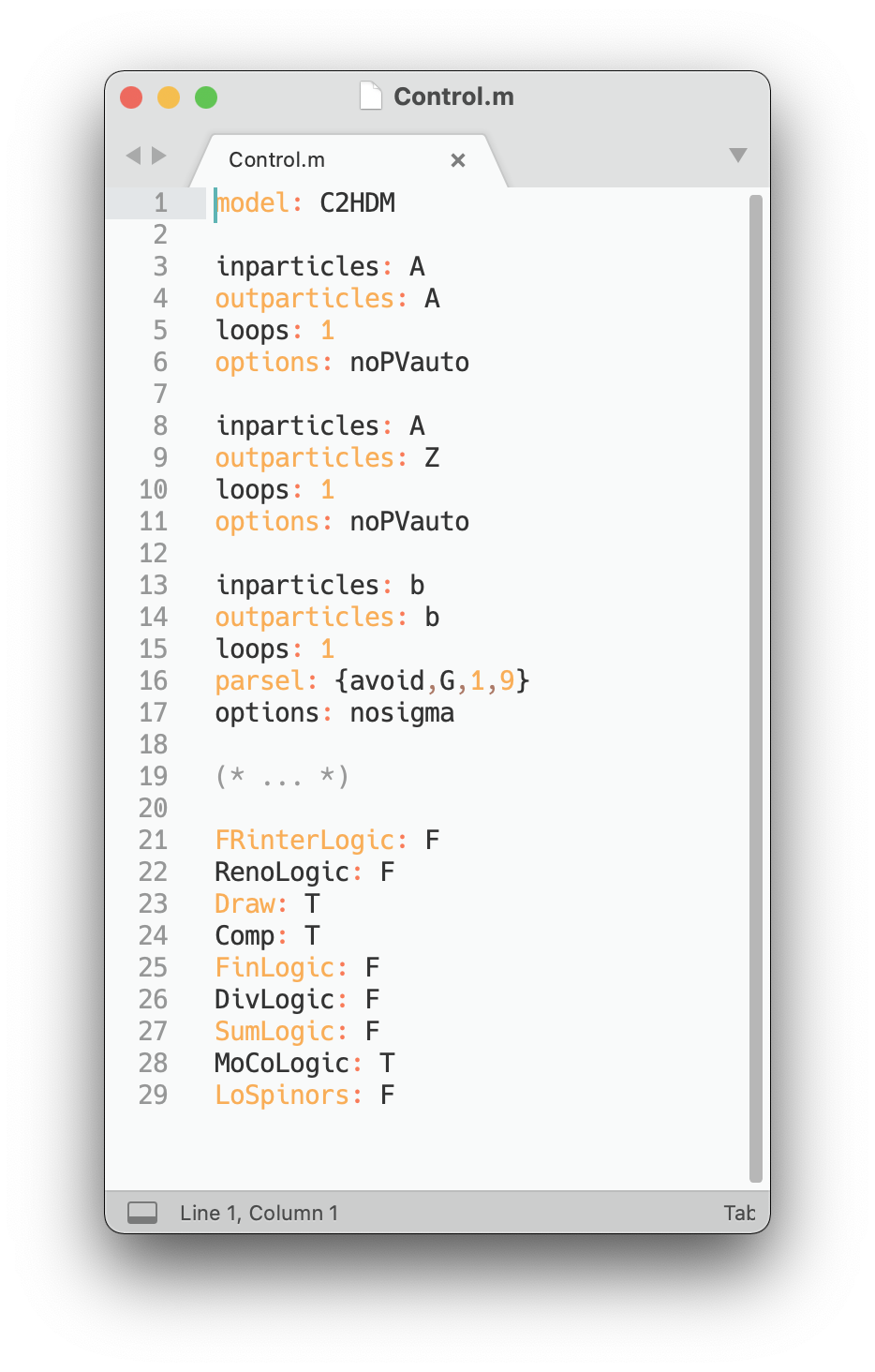}
\vs{-8mm}
\caption{Sketch of the \t{Control.m} file used to generate all the one-loop GFs required to renormalize the model. See text for details.}
\label{fig:FM:Control}
\end{figure}

\n As we just saw, the running of \FMS already generated the totality of Feynman rules for the counterterm interactions. That is, for each process, the total set of counterterms contributing to it is already known. Now, those counterterms need to be calculated, which can be done in different ways, i.e. through different subtraction schemes. Here, we adopt the schemes defined in section \ref{sec:calculation-CTs}. We thus need to calculate several GFs: mostly 2-point functions, but also some 3-point functions (due to $\overline{\text{MS}}$ renormalization). This can be done in a single \FMS run, as illustrated in figure \ref{fig:FM:Control}.
Some clarifications are in order, concerning this figure.

\n First, we only show some of the GFs involved in section \ref{sec:calculation-CTs}; the remaining ones should be added in the place of the ellipses ``\t{(* ...\,\,\,*)}''. 
Second, in the calculation of $\Sigma^{\bar{b} b}$, we exclude diagrams with gluons via \t{\{avoid,G,1,9\}}; the same applies for all GFs with external fermions.
Third, note the options \t{noPVauto} and \t{nosigma}.
The former must be employed in some GFs to prevent meaningless divergences.%
\fn{The conditions of section \ref{sec:calculation-CTs} predict that some counterterms are calculated taking the limit of zero external momentum in some 2-point functions. Now, as explained in the manual of \FMT, loop integrals are performed with the function \t{OneLoopTID}, which employs the \ts{FeynCalc} function \t{TID}. By default, \t{OneLoopTID} uses \t{TID} with the option \t{PaVeAutoReduce -> True}, which simplifies some special cases of Passarino-Veltman functions. But it turns out that, in a general $R_{\xi}$ gauge, that option causes spurious divergences to appear in some amplitudes when the limit of zero external momentum is taken. We verified this phenomenon for the cases $\Sigma^{AA}$, $\Sigma^{AZ}$, $\Sigma^{G_0G_0}$ and $\Sigma^{G^{+}G^{+}}$. This can be solved by making \t{PaVeAutoReduce -> False}, which in turn can be obtained by selecting the option \t{noPVauto} in the \t{Control.m} file. Finally, note that nothing of this happens in the Feynman gauge: all the counterterms are well-behaved in the limit of zero external momentum in that case.} 
The latter assures two things: on the one hand, that no one-loop corrections to the external legs are included (such corrections are not needed for renormalizing purposes);\fn{Actually, this aspect is only relevant in the case of GFs with more than two external particles.} on the other hand, that reducible diagrams with broad tadpoles are included, as is required by the FJTS (recall section \ref{sec:FJTS-C2HDM}). We thus prove what we claimed in that section (\ref{sec:FJTS-C2HDM}), namely, that the inclusion of broad tadpoles in \FMS is trivial.

\n Once all the GFs required in section \ref{sec:calculation-CTs} are calculated, the total set of counterterms can be directly determined in an analytical form through the results derived in that section.

\subsubsection{NLO processes}

\n As suggested above, the usefulness of \FMS does not end when the counterterms are determined. After that, multiple tasks are enabled by the tools provided in \FM. We now briefly illustrate three of them.

\n A first one consists in verifying the finiteness of a certain one-loop process. Once in the possession of the expressions for the counterterms, this task is almost immediate. One starts by selecting the process at stake in \t{Control.m} with the options \t{Comp} and \t{SumLogic} set to True. By running \FM, the total divergent part of the process is calculated, and a notebook for the process is automatically generated. One then opens the latter, having thus immediate access not only to the total divergent part (through the variable \t{resDtot}), but also to the total counterterms contributing to the process at stake (through the variable \t{CT}\textit{process}, where \textit{process} corresponds to the names of the incoming and the outgoing particles joined together). The expressions for the counterterms can be imported and replaced inside \t{CT}\textit{process}. The divergent part of the resulting expression can be obtained with the function \t{GetDiv} and compared with \t{resDtot}.%
\fn{As a matter of fact, the counterterm should be compared with $i$ times \t{resDtot}, since the amplitude calculated corresponds to $\mathcal{M}$, and not to $i \mathcal{M}$.}
If the process is finite, the difference must be zero.%
\fn{The comparison can be made analytically or numerically. While the former is more robust, it can take some time. The numerical method, on the other hand, requires a software to generate points in the parameter space of the model.}

\begin{figure}[htb]
\centering
\includegraphics[width=0.9\textwidth]{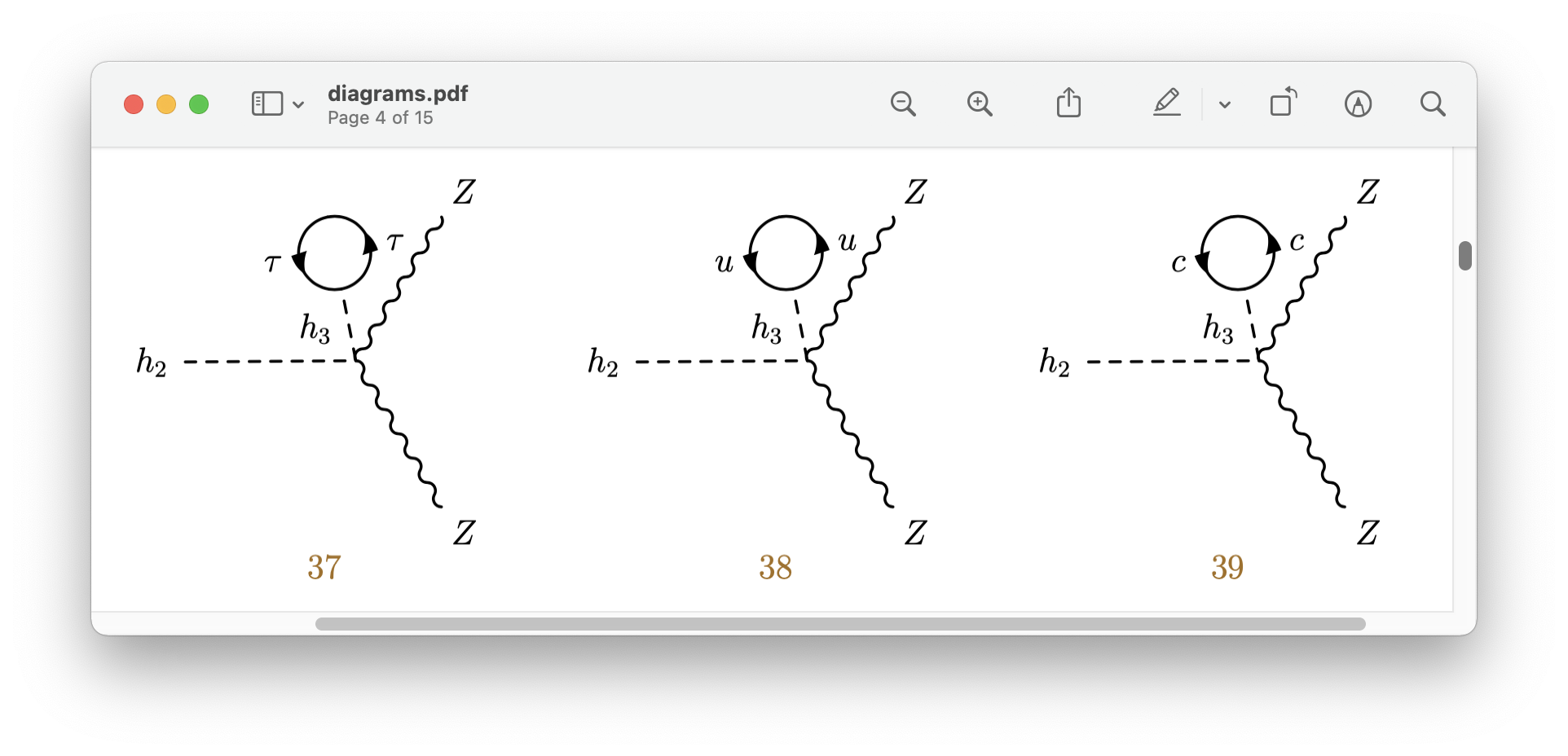}
\vs{-6mm}
\caption{Excerpt from \t{diagrams.pdf} for $h_2 \to ZZ$ at one-loop (in the FJTS).}
\label{fig:Feynman-diagram}
\end{figure}

\n A second task that becomes quite simple when using \FMS is the calculation of a decay width at up-to-one-loop level. This requires the calculation of the tree-level amplitude, as well as of the renormalized one-loop one. The user can start with the former (choosing the process at stake at tree-level in \t{Control.m} with \t{Comp} set to True). Then, in a similar way, he or she calculates the non-renormalized one-loop amplitude (if the variable \t{Draw} in \t{Control.m} is set to True, \FMS automatically draws the Feynman diagrams, as in figure \ref{fig:Feynman-diagram}).
Here, as in the previous paragraph, the total counterterm should be added to obtain the (finite) renormalized one-loop amplitude. 
After summing up the two amplitudes (tree-level and renormalized one-loop), one will in general have a complicated expression. In this case, although the \FMS  function \t{DecayWidth} can be directly applied to it, it will probably take quite some time; it is thus convenient to rewrite the expression in terms of form factors. This can be done automatically using another \FMS function, \t{FacToDecay}, which yields a two-element list: the first one is precisely the expression rewritten in terms of form factors; the second one is a list of replacements, associating to each form factor the corresponding analytical expression (written using the kinematics of the process). The function \t{DecayWidth} can now be easily applied to the first element of the output of \t{FacToDecay}.%
\fn{The user must in general define which parameters are complex.
Note, however, that the form factors used in \t{FacToDecay} are by default set as complex.}
In this way, one obtains in a couple of steps a simple formula for the NLO decay width, as well as a list with the different form factors contained in that formula. 

\n A final example concerns the numerical interface of \FMT, enabled by yet another \FMS function, \t{FCtoFT}. This function converts the expression it was applied to into \t{\ts{Fortran}} format and generates several files to run a \t{\ts{Fortran}} program (cf. the manual for details). \t{FCtoFT} is especially convenient in cases where the user has a simple expression written in terms of form factors and a list of replacement rules for each of them---precisely the case described in the previous paragraph. The reason is that \t{FCtoFT} can be used with two arguments, respectively: a simple expression in terms of form factors, and the list of replacement rules for them. In summary, the combination of functions \t{FacToDecay}, \t{DecayWidth} and \t{FCtoFT} is very powerful. 

%
%
%

\vspace*{1cm}
\bibliographystyle{JHEP}
\bibliography{MyReferences}

\end{document}